\documentclass[%
 reprint,
 superscriptaddress,
 amsmath,amssymb,
prb,
]{revtex4-2}
\usepackage[colorlinks=true,linktoc=page,citecolor=red,linkcolor=blue]{hyperref}
\usepackage{comment}
\usepackage{amsfonts}
\usepackage{float}
\usepackage{mathtools}
\usepackage{graphicx}
\usepackage{physics}
\usepackage{color}
\usepackage{tcolorbox}
\usepackage{xcolor}
\usepackage{amsthm}
\usepackage{mathdots}
\usepackage{tikz}
\usepackage{tikz-cd}
\usepackage{relsize}
\usepackage{hyperref}
\usepackage{caption}
\usepackage{subcaption}
\usepackage{dutchcal}
\usepackage{algpseudocode,algorithm}
\usepackage{CJK}
\usepackage{extarrows}

\usetikzlibrary{positioning}

\newcommand{\beq}{\begin{equation}\begin{aligned}}
\newcommand{\eeq}{\end{aligned}\end{equation}}

\usepackage{tikz-cd}
\usetikzlibrary{decorations.markings}
\usetikzlibrary{decorations.pathmorphing}
\tikzset{snake it/.style={decorate, decoration=snake}}
\usetikzlibrary{calc}
\usepackage{tikz-3dplot} 
\tdplotsetmaincoords{70}{120}  
\tdplotsetrotatedcoords{+100}{0}{0} 
\pgfdeclarelayer{bg}
\pgfsetlayers{bg,main}
\usepackage{tikzit}
\tikzstyle{pointoperator}=[fill=blue, draw=none, shape=circle, minimum size=.1 cm, inner sep=0 pt]
\tikzstyle{pointred}=[fill=red, draw=none, shape=circle, minimum size=.1 cm, inner sep=0 pt]
\tikzstyle{pointblack}=[fill=black, draw=none, shape=circle, minimum size=.1 cm, inner sep=0 pt]

\tikzstyle{dashedline}=[dashed, -, thick]
\tikzstyle{blueline}=[-, draw=blue, thick]
\tikzstyle{redline}=[-, draw=red, thick]
\tikzstyle{greenline}=[-, draw=green, thick]
\tikzstyle{arrowline}=[<-, thick]
\tikzstyle{bluearrowline}=[<-, draw=blue, thick]
\tikzstyle{normalline}=[-, thick]
\tikzstyle{greenfillline}=[-, thick, fill=green, fill opacity=.5]
\tikzstyle{grayfillline}=[-, fill opacity=.5, fill={rgb,255: red,128; green,128; blue,128}, thick]
\tikzstyle{redfillline}=[-, fill opacity=.5, fill=red, thick]
\tikzstyle{bluefillline}=[-, thick, fill=blue, fill opacity=.5]
\tikzstyle{bluedashedline}=[-, draw=blue, thick, dashed]
\tikzstyle{reddashedline}=[-, draw=red, thick, dashed]
\tikzstyle{grayfilldashedline}=[dashed, -, fill opacity=.5, fill={rgb,255: red,128; green,128; blue,128}, thick]

\def\red#1{{\color{black} #1}}

\usepackage[normalem]{ulem} 
\usepackage{soul}
\usepackage{titlesec}
\titleformat{\paragraph}[block]{\normalfont\bfseries\filcenter}{\theparagraph}{1em}{}
\begin{document}

\begin{CJK*}{UTF8}{}
\title{
Mixed-state Phases from Higher-order SSPTs with Kramers-Wannier Symmetry
}
 \author{Aswin Parayil Mana}
 \email{aswin.parayil.mana@phys.s.u-tokyo.ac.jp}
 \altaffiliation[Present address: ]{Department of Physics, The University of Tokyo, 7-3-1 Hongo, Bunkyo-ku, Tokyo 113-8654, Japan}
 \affiliation{C. N. Yang Institute for Theoretical Physics, State University of New York at Stony Brook, New York 11794, USA}
 \affiliation{Department of Physics and Astronomy, State University of New York at Stony Brook, New York 11794, USA}

\author{Zijian Song (\CJKfamily{bsmi}宋子健)}
 \affiliation{C. N. Yang Institute for Theoretical Physics, State University of New York at Stony Brook, New York 11794, USA}
 \affiliation{Department of Physics and Astronomy, State University of New York at Stony Brook, New York 11794, USA}

 \author{Fei Yan}
 \affiliation{Condensed Matter Physics and Materials Science Division, Brookhaven National Laboratory, Upton, New York 11973, USA}

 \author{Tzu-Chieh Wei (\CJKfamily{bsmi}魏子傑)}
 \affiliation{C. N. Yang Institute for Theoretical Physics, State University of New York at Stony Brook, New York 11794, USA}
 \affiliation{Department of Physics and Astronomy, State University of New York at Stony Brook, New York 11794, USA}

\date{\today}

\begin{abstract}
Mixed-state phases have recently attracted significant attention as a generalization beyond their pure-state counterparts. Prominent examples include mixed-state symmetry-protected topological (mSPT) phases and the strong-to-weak symmetry breaking (SWSSB) phases.  It has been shown recently that mSPT phases admit a holographic dual description in terms of higher-order subsystem SPT phases. In this work, we investigate the mixed-state phases obtained by tracing out the bulk degrees of freedom of higher-order subsystem SPT phases protected by non-invertible symmetries. We find that the resulting mixed-states exhibit the coexistence of the symmetry-protected topological order and  SWSSB. We also use the interface as a probe to characterize the mixed-state phases, and specifically, when there is no local modification to preserve the symmetries across the interface, the two sides of the interface are in distinct phases.
\end{abstract}

\maketitle
\end{CJK*}

\tableofcontents

\section{Introduction}
Symmetry-protected topological (SPT) phases have been extensively studied in the context of pure quantum states, with several foundational works establishing their defining properties and classification schemes~\cite{Chen:2011bcp, chen2013symmetry, chen2012symmetry, gu2014symmetry, kapustin2014symmetry}. In this framework, SPT phases are equivalence classes of gapped Hamiltonians with a fixed symmetry, where two Hamiltonians belong to the same phase if they can be connected by a symmetry-preserving adiabatic path without closing the energy gap. Equivalently, at the level of quantum states, two SPT states are in the same phase if they are related by a symmetry-respecting finite-depth local unitary (FDLU) circuit. However,  any SPT state can be mapped to a trivial product state by a non-symmetric FDLU.

Extending the notion of symmetry-protected topological phases to open quantum systems remains an open challenge. In such systems, unavoidable environmental couplings and thermal fluctuations generically drive the system into mixed-states, invalidating many of the assumptions underlying the pure-state classification. Recently, significant progress has been made towards extending symmetry-based classifications beyond strong symmetries, which are originally defined as symmetries of the ground state, to weak symmetries, which are symmetries of the density matrix characterizing the mixed-state system. In particular, extensive efforts have focused on strong-to-weak symmetry breaking (SWSSB) phases~\cite{Lessa:2024gcw,Guo:2025buy,huang2025hydrodynamics,song2025strong,zhang2025strongtoweak,gu2025spontaneous}, mixed-state symmetry-protected topological (mSPT) phases~\cite{ma2023average,Ma2025Topological,ma2025symmetry,xue2024tensor,guo2025locally,Lee2025symmetry,deGroot2022,aldossari2025tensor,Huang:2025asc} which are also known as averaged symmetry-protected topological (ASPT) phases~\footnote{As these two terms appear in the literature, so we will use mSPT and ASPT interchangeably.}, and finite-temperature topological phases~\cite{deGroot2022,ma2023average,Ma2025Topological,fan2024diagnostics,Lee2025symmetry,zhang2025fidelity,lee2023quantum,ma2025symmetry,xue2024tensor,guo2025locally,Guo2025design,chen2024symmetry,Lessa2025Mixed,wang2025intrinsic,wang2025anomaly,sala2024spontaneous,Lessa:2024gcw,huang2025hydrodynamics,zhang2025quantum,zhang2024fluctuation,ellison2025toward,sohal2025noisy,zhang2025strongtoweak,sala2025decoherence,song2025strong,Sun:2024iwm,sun2025anomalousmatrixproductoperator,sang2024mixed,Sang:2024vkl,gu2025spontaneous,guo2025anew,luo2025topological,Schafer-Nameki:2025fiy,qi2025symmetry,aldossari2025tensor,yang2025topological,ma2025circuit,sang2025mixedstatephaseslocalreversibility,lu2025holographic,Hauser:2026sgr,Huang:2024kzf}. Moreover, there are identifications of connection between measurement induced phase transitions and mixed-state phases~\cite{Wang:2025xwk,Eckstein:2025wce}.

At finite temperature, previous works have also shown that SPT phases can persist in certain scenarios, including in three-dimensional systems protected by higher-form symmetries \cite{roberts2017symmetry} and in two-dimensional fermionic systems \cite{viyuela2015symmetry}. Going beyond thermal states, mixed-states also arise from decoherence, dissipation, as well as tracing over part of the system. Recent proposals include defining mixed-state phases via symmetric finite-depth local quantum channels~\cite{ma2023average, Ma2025Topological}, as well as an alternative approach based on finite Markov length under local Lindbladian evolution~\cite{Sang:2024vkl}. Although each framework captures aspects of phase stability in dissipative settings, they are not equivalent. However, unlike in the pure-state case, a universally accepted definition and classification of mSPT phases remains elusive.

Non-invertible symmetries have recently attracted significant attention in both condensed matter and high-energy theory. A growing body of work has extended the notion of SPT phases to lattice systems with non-invertible symmetries, uncovering novel classes of topological phases that have no counterparts in the conventional group-symmetry setting~\cite{fechisin2023non, Seifnashri:2024dsd,li2024non, Inamura:2024jke,meng2024non, Choi:2024rjm,Jia:2024bng, pace2025spt,Li:2024gwx, jia2024weak,jia2024quantum, cao2025duality,Lu:2025rwd, Aksoy:2025rmg}. As a representative example, Ref.~\cite{Seifnashri:2024dsd}
studied lattice realizations of $\text{Rep}(D_8)$ SPT phases in $1+1$ dimensions, where the symmetry category includes both a $\mathbb{Z}_2\times\mathbb{Z}_2$ symmetry and the Kramers-Wannier (KW) duality, denoted by $\mathrm{D}^{(1)}$. 
In this setting, the $\mathbb{Z}_2 \times \mathbb{Z}_2$ cluster state, which is invariant under $\mathrm{D}^{(1)}$, was shown to split into three distinct non-invertible SPT phases.
Extensions to $2+1$ dimensions were explored in Ref.~\cite{ParayilMana:2025nxw}, where a 2D (we omit the $+1$ in $2+1$ for simplicity) cluster state invariant under a two-dimensional KW duality that gauges subsystem symmetries was analyzed. Owing to the presence of subsystem symmetries, this system supports a large family of SPT phases protected by non-invertible symmetries. Notably, certain subsets of these phases, when separated by a rectangular interface, differ by the presence of corner modes, thereby realizing higher-order subsystem SPT phases protected by non-invertible symmetry.

Recently, Ref.~\cite{Sun:2024iwm} investigated mixed-state SPT phases from a holographic perspective. Starting from a higher-order SPT in $(d+1)$ dimensions with subsystem symmetry $\mathcal{S}$ and global symmetry $\mathcal{G}$, tracing out the bulk degrees of freedom yields a mSPT in $d$ dimensions that exhibits a strong $\mathcal{S}$ symmetry (supported on the boundary) and a weak $\mathcal{G}$ symmetry. In this construction, a higher-order subsystem SPT in $(d+1)$ dimensions is holographically dual to a mSPT in $d$ dimensions. The higher-codimensional boundary modes originate from a mixed 't Hooft anomaly between the global $\mathcal{G}$ symmetry and the subsystem $\mathcal{S}$ symmetry in the parent theory. This anomaly descends to the boundary, giving rise to protected boundary modes in the mSPT and manifesting as a mixed 't Hooft anomaly between the weak $\mathcal{G}$ symmetry and the strong $\mathcal{S}$ symmetry. 

In this manuscript, we study the holographic dual description of higher-order SSPTs protected by non-invertible symmetries. By tracing out the bulk degrees of freedom, we obtain a class of mixed-state phases. In contrast to Ref. \cite{Sun:2024iwm}, the resulting mixed-state here exhibits the coexistence of both an SPT order and a strong-to-weak symmetry breaking (SSB) structure. We refer to this phase as a doubled average SPT (DASPT)~\cite{Kuno:2025dqf,Guo2025strong}, reflecting the presence of two extremal points characterizing the mixed-state phase. This observation does not contradict Ref. \cite{Sun:2024iwm}, since the SSPTs considered in the present work are not higher-order with respect to invertible symmetries. Instead, their higher-order nature originates from the underlying non-invertible symmetry. \red{In terms of invertible symmetries, they belong to the same phase}. 

\red{Recently, a systematic understanding of 1D mixed-state phases with $\mathbb{Z}_2\times\mathbb{Z}_2$ symmetry from a purification perspective was developed in \cite{Guo:2026dxx}. In their construction, various mixed-state phases are obtained by starting from a 1D pure state coupled to an ancilla and subsequently tracing out the ancilla. Our construction differs from this approach: following \cite{Sun:2024iwm}, we instead obtain mixed-state phases from higher-order SSPTs by tracing out their bulk degrees of freedom.}

The rest of the sections are organized in the following way. In Section~\ref{sec:holographicReview}, we review the holographic duality between higher-order SSPTs and mSPTs given in~\cite{Sun:2024iwm} using an example. In Section~\ref{sec:1D-DASPT}, we study DASPTs by tracing out the bulk of $2$D cluster state. Further, in Section~\ref{sec:1DNDASPT}, we study the DASPTs obtained from tracing out the bulk of $2$D higher-order SSPTs protected by the non-invertible Kramers-Wannier (KW) symmetry. We give a general analysis of the interface as a probe for mixed-state SPTs in Section~\ref{sec:generalinterface}. An analysis of the interface between two different DASPTs obtained in Sections~\ref{sec:1D-DASPT} and~\ref{sec:1DNDASPT} is presented in Section~\ref{sec:Interfaceanalysis}.  
In Section~\ref{sec:conclusion}, we provide concluding remarks and future directions. We supply further materials in the Appendices.  
\section{Review of the holographic picture for mSPT phases}\label{sec:holographicReview}
In this section, we review the holographic picture of mSPTs, also referred to as ASPT phases~\cite{Sun:2024iwm}, in more detail than previously outlined in the Introduction. A mixed-state density matrix $\rho$ can be interpreted as the reduced density matrix of a pure state obtained by tracing out ancillary degrees of freedom. Within the holographic framework, the pure state associated with a 
$d$-dimensional mSPT is taken to be a 
$(d+1)$-dimensional subsystem symmetry-protected topological (SSPT) state. Under this construction, weak symmetries of the mSPT are promoted to global symmetries of the pure-state SSPT, while strong symmetries of the mSPT become subsystem symmetries of the SSPT. Consequently, the holographic dual of an mSPT with strong $\mathcal{S}$ symmetry and weak $\mathcal{G}$ symmetry is a higher-order SSPT characterized by subsystem $\mathcal{S}$ symmetry and global $\mathcal{G}$ symmetry. The SSPT wave function exhibits short-range correlations in both the bulk and side surfaces, while hosting gapless modes localized at the hinges. These hinge modes realize a mixed anomaly between the global symmetry $\mathcal{G}$ and the subsystem symmetry $\mathcal{S}$. This observation suggests that the reduced density matrix of the two-dimensional SSPT wave function effectively behaves as a one-dimensional mSPT, motivating the following duality:
\begin{align}
    &\text{SSPT wavefunction in } d+1\text{ dim }\nonumber\\
    &\hspace{2cm}\updownarrow\nonumber\\
    &\hspace{1cm}\text{mSPT in } d \text{ dim}\,.\nonumber 
\end{align}
These mSPTs can be probed using the fidelity strange correlator. Let $\rho=\ket{\Psi_{\text{SSPT}}}\bra{\Psi_{\text{SSPT}}}$ and $\rho_0=\ket{\Psi_0}\bra{\Psi_0}$, where $\ket{\Psi_0}$ is a trivial product state. By tracing out ancillary degrees of freedom, we obtain one-dimensional mSPTs $\xi[\rho]$ and $\xi[\rho_0]$. These states can be distinguished through the fidelity strange correlator
\begin{align*}
    \frac{F(\xi[\rho]\,,O_rO_{r'}^{\dagger}\xi[\rho_0]O_{r'}O_r^{\dagger})}{F(\xi[\rho],\xi[\rho_0])}\,,
\end{align*}
which exhibits power-law decay as a function of $|r-r'|$. Here, $\xi$ denotes the decoherence channel that prepares the mSPT from the SSPT wave function, and $\mathcal{O}$ is an operator charged under the strong symmetry $\mathcal{S}$. As shown in Ref.~\cite{Sun:2024iwm}, this fidelity strange correlator is lower bounded by the strange correlator computed directly from the SSPT wave function and the product state $\ket{\Psi_0}$.

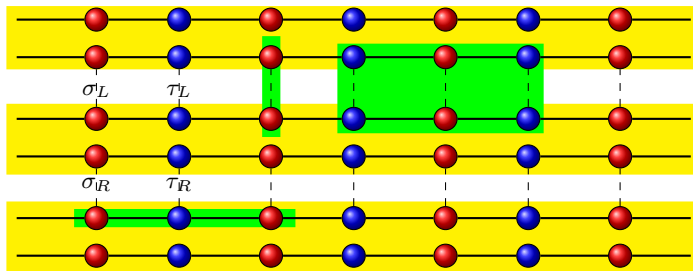
\begin{figure*}
    \centering
    \begin{tikzpicture}[auto,
    redBall/.style={circle, ball color = red,
    radius=0.1em},blueBall/.style={circle, ball color = blue,
    radius=0.4em}, decoration={markings,
  mark=between positions 0 and 1 step 6pt
  with { \draw [fill] (0,0) circle [radius=0.8pt];}}]
  \node[] (b100) at (0,0) {};
     \node[] (r100) at (-1.1,0) {};
     \node[] (b200) at (0,0.5) {};
     \node[] (r200) at (-1.1,0.5) {};
      \filldraw[yellow, very thick] (0,-0.2) rectangle (9.2,0.7);
      \filldraw[yellow, very thick] (0,1.1) rectangle (9.2,2);
      \filldraw[yellow, very thick] (0,2.5) rectangle (9.2,3.3);
     \filldraw[green, very thick] (0.9,0.4) rectangle (3.8,0.6);
     \filldraw[green, very thick] (3.4,1.6) rectangle (3.6,2.9);
     \filldraw[green, very thick] (4.4,1.65) rectangle (7.1,2.8);
  \foreach \x [remember=\x as \lastx (initially 0)] in {1,...,3}{\node [draw, blueBall, right =2cm of b1\lastx0] (b1\x0) {}; \foreach \y [remember=\y as \lasty (initially 0)] in {1,...,2}
    {\node [draw, blueBall, above =1cm of b1\x\lasty] (b1\x\y){};
    }
}

\foreach \x [remember=\x as \lastx (initially 0)] in {1,...,3}{\node [draw, blueBall, right =2cm of b2\lastx0] (b2\x0) {}; \foreach \y [remember=\y as \lasty (initially 0)] in {1,...,2}
    {\node [draw, blueBall, above =1cm of b2\x\lasty] (b2\x\y){};
    }
}

\foreach \x [remember=\x as \lastx (initially 0)] in {1,...,4}{\node [draw, redBall, right =2cm of r1\lastx0] (r1\x0) {}; \foreach \y [remember=\y as \lasty (initially 0)] in {1,2}
    {\node [draw, redBall, above =1cm of r1\x\lasty] (r1\x\y){};
    }
}

\foreach \x [remember=\x as \lastx (initially 0)] in {1,...,4}{\node [draw, redBall, right =2cm of r2\lastx0] (r2\x0) {}; \foreach \y [remember=\y as \lasty (initially 0)] in {1,2}
    {\node [draw, redBall, above =1cm of r2\x\lasty] (r2\x\y){};
    }
}
\node[below=0.01cm  of r111] (a) {$\sigma_R$};
\node[below=0.01cm  of b111] (b) {$\tau_R$};
\node[above=0.01cm  of r211] (a) {$\sigma_L$};
\node[above=0.01cm  of b211] (b) {$\tau_L$};
\node[right=2cm  of b130] (b140) {};
\node[left=2cm  of b111] (b101) {};
\node[right=2cm  of b131] (b141) {};
\node[left=2cm  of b112] (b102) {};
\node[right=2cm  of b132] (b142) {};
\draw[thick] (b100)--(r110)--(b110)--(r120)--(b120)--(r130)--(b130)--(r140)--(b140);
\draw[thick] (b101)--(r111)--(b111)--(r121)--(b121)--(r131)--(b131)--(r141)--(b141);
\draw[thick] (b102)--(r112)--(b112)--(r122)--(b122)--(r132)--(b132)--(r142)--(b142);
\node[right=2cm  of b230] (b240) {};
\node[left=2cm  of b211] (b201) {};
\node[right=2cm  of b231] (b241) {};
\node[left=2cm  of b212] (b202) {};
\node[right=2cm  of b232] (b242) {};
\draw[thick] (b200)--(r210)--(b210)--(r220)--(b220)--(r230)--(b230)--(r240)--(b240);
\draw[thick] (b201)--(r211)--(b211)--(r221)--(b221)--(r231)--(b231)--(r241)--(b241);
\draw[thick] (b202)--(r212)--(b212)--(r222)--(b222)--(r232)--(b232)--(r242)--(b242);
\draw[dashed] (r210)--(r111);
\draw[dashed] (b210)--(b111);
\draw[dashed] (r220)--(r121);
\draw[dashed] (b220)--(b121);
\draw[dashed] (r230)--(r131);
\draw[dashed] (b230)--(b131);
\draw[dashed] (r240)--(r141);
\draw[dashed] (r211)--(r112);
\draw[dashed] (b211)--(b112);
\draw[dashed] (r221)--(r122);
\draw[dashed] (b221)--(b122);
\draw[dashed] (r231)--(r132);
\draw[dashed] (b231)--(b132);
\draw[dashed] (r241)--(r142);
  \end{tikzpicture}
    \caption{Arrangement of qubits for the 2D SSPT Hamiltonian in \eqref{eq:H2DSSPT}. Each unit cell contains two types ($L$ and $R$) of $\sigma$ and $\tau$ spins. The green regions specify the different interactions in the SSPT Hamiltonian \eqref{eq:H2DSSPT}. Each yellow region contains two $L$ and $R$ spin chains.}
    \label{fig:HSSPT}
\end{figure*}

We now illustrate the holographic duality between SSPT phases and mSPTs using an explicit example introduced in Ref.~\cite{Sun:2024iwm}. Consider a two-dimensional lattice composed of two sets of spin chains oriented along the 
$x$-direction at each fixed value of 
$y$, which we label as the left ($L$) and right ($R$) chains. Along each row and at a given 
$x$-coordinate, there are two spins denoted by 
$\sigma$ and 
$\tau$. Consequently, each unit cell at position (x,y) contains four spins:
$\sigma_L$,
$\tau_L$, $\sigma_R$, and $\tau_R$. See Figure~\ref{fig:HSSPT} for an illustration.

We consider the following Hamiltonian,
\begin{widetext}
\begin{align}
    \mathrm{H}&=-\sum_{\alpha=L,R}\left[\sigma_{\alpha}^z(x,y)\tau_{\alpha}^x(x,y)\sigma_{\alpha}^z(x+1,y)+J\sigma^z_L(x,y)\sigma^z_R(x,y+1)\right.\nonumber\\
    &\left.+\tau^z_L(x,y)\sigma^x_L(x+1,y)\tau^z_L(x+1,y)\tau^z_R(x,y+1)\sigma^x_R(x+1,y+1)\tau^z_R(x+1,y+1)\right]
    \label{eq:H2DSSPT}
\end{align}
\end{widetext}
This Hamiltonian possesses subsystem symmetries acting independently along each 
$x$-row,  
\begin{align}
   U_s(y)=\prod_x\tau^x_R(x,y)\tau^x_L(x,y)\,. 
\end{align}
as well as a global $\mathbb{Z}_2$ symmetry given by
\begin{align}
    U_g=\prod_{x,y}\sigma^x_L(x,y)\sigma^x_R(x,y)\,.
\end{align}
In the presence of an open, smooth boundary at 
$x=0$, each 
$y$-row along the edge hosts two sets of dangling spin-$\frac{1}{2}$ degrees of freedom, labeled by 
$L$ and 
$R$. These spins can be paired into on-site singlets that are invariant under both the subsystem 
$\mathbb{Z}_2$ symmetry and the global $\mathbb{Z}_2$ symmetry, thereby fully gapping the edge.

For a system with open boundaries at both 
$x=0$ and 
$y=0$, the edge at 
$y=0$ contains an 
$R$-spin chain that is decoupled from the bulk building blocks. One may introduce edge transverse-field terms 
$\tau^x_R$ and 
$\sigma^x_R$ along 
$y=0$, which gap out this 
$R$-spin chain without breaking any symmetries. However, at the corner 
$x=y=0$, where the two edges intersect, there remain residual Pauli operators 
$\tilde{X}_L$, 
$\tilde{Y}_L$, and 
$\tilde{Z}_L$ that are decoupled from both the bulk and the gapped edges. These remaining spin-
$\frac{1}{2}$ degrees of freedom transform projectively under the combined action of the subsystem 
$\mathbb{Z}_2$ symmetry and the global  $\mathbb{Z}_2$ symmetry, giving rise to a protected corner mode.
  
We now impose periodic boundary conditions and trace out all spin chains except for a single one. The resulting reduced density matrix takes the form
\begin{align}
    \rho=\frac{1}{2^N}\sum_{\{\mathcal{D}\}}\ket{\Psi_{\mathcal{D}}}\bra{\Psi_{\mathcal{D}}},
\end{align}
where 
$N$ denotes the number of sites along the $x$-direction, and the sum runs over all configurations $\{\mathcal{D}\}$. The pure states $\ket{\Psi_{\mathcal{D}}}$ are given by
\begin{align}
    \ket{\Psi_{\mathcal{D}}}=\bigotimes_{j}\ket{\sigma^z_j=h_j}\otimes\ket{\tau^x_j=h_jh_{j+1}}\,,
\end{align}
with $h_j=\pm 1$ labeling classical d.o.f. , and $\sigma$ and $\tau$ are the remaining d.o.f. in the one-dimensional boundary.

The reduced density matrix 
$\rho$ is invariant under a strong $\mathbb{Z}_2$ symmetry,
\begin{align}
    U_s=\prod_j\tau^x_{j},
\end{align}
which originates from a subsystem symmetry inherited from the parent SSPT, as well as a weak $\mathbb{Z}_2$ symmetry,
\begin{align}
    U_g=\prod_j\sigma_j^x\,.
\end{align}
The state 
$\rho$ exhibits a nonvanishing, constant fidelity strange correlator with a trivial product state when evaluated using the charged operator $O=\tau^z$ associated with the strong symmetry, indicating $\rho$ is a non-trivial mSPT. 

In the remaining sections, we will focus on examples of higher-order SSPTs with non-invertible symmetries. When only invertible symmetries are present, the phases we consider below are SSPTs but not higher-order SSPTs. Tracing out the bulk, the subsystem symmetries on the boundary become the strong global symmetry for the mixed-state. However, the resulting mixed-state is not solely an SPT but exhibits the coexistence of SPT and SWSSB, namely the DASPT. We emphasize that a 2D non-invertible symmetry does not necessarily give rise to a 1D non-invertible symmetry upon tracing out the bulk. Starting from different 2D non-invertible SPT states, one may in some cases observe the emergence of a 1D non-invertible symmetry after tracing out the bulk, while in other cases no such symmetry appears. Below, we analyze these distinct scenarios in more detail.

\section{Toy model of DASPTs} \label{sec:1D-DASPT}
In this section, we study a simple example of a 1D DASPT obtained by placing the 2D cluster state on a semi-infinite cylinder with Dirichlet boundary conditions and tracing out the bulk degrees of freedom. We refer the reader to Appendix~\ref{app:1Dexamples} for other examples of 1D mixed-state phases involving invertible symmetries obtained by decohering various pure states, which serve as an alternative method to obtain mixed-state phases. See also  Appendix~\ref{app:1Dnon-invertibleSWSSB} for an example of SWSSB of non-invertible symmetry. In Appendix~\ref{app:commutative}, we present a commutative diagram illustrating the relationship between tracing and decoherence in obtaining 1D mixed-states from 2D pure states.
\subsection{Overview of the 2D cluster state}

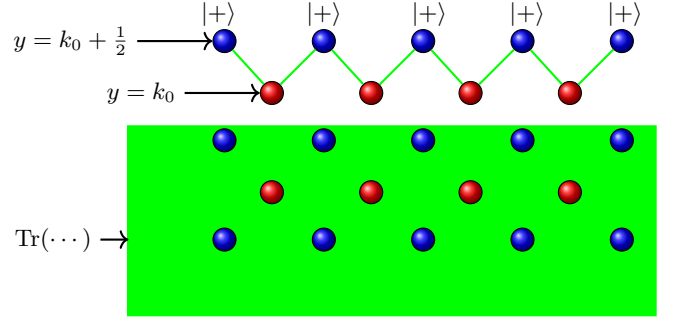
\begin{figure}
    \centering
    \begin{tikzpicture}[auto,
    redBall/.style={circle, ball color = red,
    radius=0.1em},blueBall/.style={circle, ball color = blue,
    radius=0.4em}, decoration={markings,
  mark=between positions 0 and 1 step 6pt
  with { \draw [fill] (0,0) circle [radius=0.8pt];}}]
   \node[] (b00) at (0,0) {};
     \node[] (r00) at (0.63,0.63) {};
     \filldraw[green, very thick] (0,-1) rectangle (7,1.5);
     \node[] (tr1) at (-1,0) {Tr($\cdots$)};
     \draw[->,thick] (tr1)--(0,0);
  \foreach \x [remember=\x as \lastx (initially 0)] in {1,...,5}{\node [draw, blueBall, right =of b\lastx0] (b\x0) {}; \foreach \y [remember=\y as \lasty (initially 0)] in {1,...,2}
    {\node [draw, blueBall, above =of b\x\lasty] (b\x\y){};
    }
}

\foreach \x [remember=\x as \lastx (initially 0)] in {1,...,4}{\node [draw, redBall, right =of r\lastx0] (r\x0) {}; \foreach \y [remember=\y as \lasty (initially 0)] in {1}
    {\node [draw, redBall, above =of r\x\lasty] (r\x\y){};
    }
}
\draw[thick,green] (b12)--(r11)--(b22)--(r21)--(b32)--(r31)--(b42)--(r41)--(b52);
\node[left=of b12] (b02) {$y=k_0+\frac{1}{2}$};
\node[left=of r11] (r01) {$y=k_0$};
\draw[->,thick] (b02)--(b12);
\draw[->,thick] (r01)--(r11);
\node[] (t0) at (1.2,3) {$\ket{+}$};
\node[] (t1) at (2.6,3) {$\ket{+}$};
\node[] (t2) at (4,3) {$\ket{+}$};
\node[] (t3) at (5.3,3) {$\ket{+}$};
\node[] (t4) at (6.6,3) {$\ket{+}$};
\end{tikzpicture}
    \caption{Bipartite lattice with open boundary condition in the $y$ direction and periodic boundary condition in the $x$ direction. The top boundary is at $y=k_0+\frac{1}{2}$. We trace out all the d.o.f. except those on region $\rm A$ defined to be the rows at $y=k_0$ and $y=k_0+\frac{1}{2}$. The green region contains the d.o.f. that we trace over. One can set the boundary row reference $k_0$ to zero without loss of generality.}
    \label{fig:regionA-open}
\end{figure}

Consider two mutually dual two-dimensional square lattices, as illustrated in Figure~\ref{fig:regionA-open}, distinguished by red and blue coloring, with degrees of freedom residing on their vertices. We denote the vertices and plaquettes of the red lattice by \(v_r\) and \(p_r\), and those of the blue lattice by \(v_b\) and \(p_b\), respectively. We identify \(p_b\) with \(v_r\) and \(p_r\) with \(v_b\). There are \(L_x\) vertices along the \(x\)-direction and \(L_y\) vertices along the \(y\)-direction. Vertices of the red sublattice are labeled by integer coordinates \((i,j)\), whereas vertices of the blue sublattice are labeled by half-integer coordinates \((i+\tfrac{1}{2}, j+\tfrac{1}{2})\), where \(i = 1, \dots, L_x\) and \(j = 1, \dots, L_y\). We denote the sets of vertices and plaquettes on the red and blue sublattices by \(\Delta_{v_r}\), \(\Delta_{p_r}\), \(\Delta_{v_b}\), and \(\Delta_{p_b}\), respectively. We use $\partial p$ to denote the vertices around the plaquette $p$.  The corresponding cluster Hamiltonian is given by 
\begin{align}
\begin{split}
    \bm{\mathrm{H}}_{\text{2D-cluster}}&=-\sum_{v_r}X_{v_r}\prod_{v_b\in \partial p_b}Z_{v_b}-\sum_{v_b}X_{v_b}\prod_{v_r\in \partial p_r}Z_{v_r} \, .
    \label{eq:2Dcluster}
\end{split}
\end{align}
The ground state of this Hamiltonian is the cluster state 
\begin{align}
    \ket{\text{2D-cluster}}=\prod_{p_b=v_r\in\Delta_{v_r}}\prod_{v_b\in\partial p_b}CZ_{v_r,v_b}\ket{+}^{\otimes\Delta_{v_r}}\ket{+}^{\otimes\Delta_{v_b}}\, ,
\end{align}
where $\ket{+}^{\otimes\Delta_{v_r}}$ and $\ket{+}^{\otimes\Delta_{v_b}}$ denote, respectively, the tensor product of $\ket{+}$ states over all vertex degrees of freedom associated with each of the two sublattices.

The cluster Hamiltonian \eqref{eq:2Dcluster} possesses subsystem symmetries generated by
\begin{align}
    \eta^x_{r,j}&=\prod_{i}X_{i,j}\, ,\quad \eta^y_{r,i}=\prod_{j}X_{i,j}\, ,\nonumber\\
    \eta^x_{b,j}&=\prod_{i}X_{i+\frac{1}{2},j+\frac{1}{2}}\, ,\quad \eta^y_{b,i}=\prod_{j}X_{i+\frac{1}{2},j+\frac{1}{2}}\, .
    \label{eq:subsystemsymmetries2Dclstr}
\end{align}
In addition to these symmetries, there is another symmetry for the Hamiltonian in Eq.~\eqref{eq:2Dcluster} obtained by swapping the following terms,
\begin{align}
    X_{v_r}\leftrightarrow \begin{array}{cc}
        Z_{v_b} & Z_{v_b}\\
        Z_{v_b} & Z_{v_b}
    \end{array},\qquad X_{v_b}\leftrightarrow \begin{array}{cc}
        Z_{v_r} & Z_{v_r}\\
        Z_{v_r} & Z_{v_r}
    \end{array}\, .
\end{align}
This transformation is the same as that of Kramers-Wannier duality, which is obtained by gauging the subsystem symmetries. An operator representation of this symmetry up to a half-lattice translation is given in~\cite{mana2024kennedy}. Here we define an operator with half-lattice translation included:
\begin{align}
    \mathbf{D}^{(2)}=\mathbf{T}^{-1}_{\frac{1}{2},\frac{1}{2}}\mathbf{D}^{(2)}_r\mathbf{D}^{(2)}_b\, .
    \label{eq:D^{(2)}}
\end{align}
The operator $\mathbf{D}^{(2)}_{r(b)}$ on the red(blue) sublattice is defined as 
\begin{align}
    \mathbf{D}^{(2)}_{r(b)}\equiv\mathbf{P}^{(2)}_{r(b)} \Tilde{\mathbf{D}}_{x;r(b)}^{(2)}\mathbf{H}^{\otimes(2)}_{r(b)}\Tilde{\mathbf{D}}_{y;r(b)}^{(2)} \mathbf{P}^{(2)}_{r(b)}\, ,
    \label{eq:Dr(b)def}
\end{align}
where $\mathbf{H}^{\otimes(2)}_{r(b)}$ denotes the product of Hadamard operators on the red (or blue) lattice and
\begin{subequations}
\begin{align}
\tilde{\mathbf{D}}_{x;r}^{(2)}&\equiv\prod_{j=1}^{L_y}\left(\left(\prod_{i=1}^{L_x-1}e^{i\frac{\pi}{4}X_{i,j}}e^{i\frac{\pi}{4}Z_{i,j}Z_{i+1j}}\right)\,e^{i\frac{\pi}{4}X_{L_x,j}}\right), \label{Dx-def}\\
\tilde{\mathbf{D}}_{y;r}^{(2)}&\equiv\prod_{i=1}^{L_x}\left(\left(\prod_{j=1}^{L_y-1}e^{i\frac{\pi}{4}X_{i,j}}e^{i\frac{\pi}{4}Z_{i,j}Z_{i,j+1}}\right)e^{i\frac{\pi}{4}X_{i,L_y}}\right),\label{Dy-def}\\
\mathbf{P}^{(2)}_{r}&\equiv\prod_{j=1}^{L_y}\frac{(1+\eta^x_{r,j})}{2}\prod_{i=1}^{L_x}\frac{(1+\eta^y_{r,i})}{2}\, ,
\end{align}
\label{eq:Dtilde2ddef}
\end{subequations}
with similar definitions for $\tilde{\mathbf{D}}_{x;b}^{(2)}$, $\tilde{\mathbf{D}}_{y;b}^{(2)}$ and $\mathbf{P}^{(2)}_{b}$.

The symmetry operator $\mathbf{D}^{(2)}$ is non-invertible and satisfies the following fusion rules
\begin{align}
\begin{split}
    &(\mathbf{D}^{(2)})^2\propto \mathbf{P}^{(2)}_{r}\mathbf{P}^{(2)}_{b}\, ,\\
    &\eta^{x}_{r,j}\mathbf{D}^{(2)}=\mathbf{D}^{(2)}\eta^{x}_{r,j}=\mathbf{D}^{(2)}\, ,\\
    &\eta^{y}_{r,i}\mathbf{D}^{(2)}=\mathbf{D}^{(2)}\eta^{y}_{r,i}=\mathbf{D}^{(2)}\, ,\\
    &\eta^x_{b,j+\frac{1}{2}}\mathbf{D}^{(2)}=\mathbf{D}^{(2)}\eta^x_{b,j+\frac{1}{2}}=\mathbf{D}^{(2)}\, ,\\
    &\eta^y_{b,i+\frac{1}{2}}\mathbf{D}^{(2)}=\mathbf{D}^{(2)}\eta^y_{b,i+\frac{1}{2}}=\mathbf{D}^{(2)}\,.
    \end{split}
\end{align}

\subsection{1D DASPT from 2D cluster state} \label{sec:DASPTfrom2D}

We now place the 2d cluster state on a semi-infinite cylinder, whose open boundary along the $y$-direction consists of the row of blue vertices at $y=k_0+\frac{1}{2}$ and the row of red vertices at $y=k_0$, as illustrated in Figure~\ref{fig:regionA-open}. Moreover, we consider a Dirichlet-type boundary condition where all the vertices in the row of blue vertices are in the $|+\rangle$ state. See Appendix~\ref{app:Extraexamples1} for other choices of boundary conditions.

Tracing out the bulk gives the following mixed-state
\begin{align}
    \rho_{\text{1D-DASPT}}^{\rm A}&=\prod_i\left(\frac{1+Z_{i,k_0}X_{i+\frac{1}{2},k_0+\frac{1}{2}}Z_{i+1,k_0}}{2}\right)\nonumber\\
    &\hspace{2cm}\times\frac{\left(1+\prod_iX_{i,k_0}\right)}{2^{|\Delta_{v_r}\cap \rm A|}}\,.
    \label{eq:rhoA2Dcluster}
\end{align}
The superscript $\rm A$ indicates the remaining sites after tracing over the rest.
This state has two strong symmetries: $\mathbb{Z}_2^{(r)}=\prod_iX_{i,k_0}$ and $\mathbb{Z}_2^{(b)}=\prod_iX_{i+\frac{1}{2},k_0+\frac{1}{2}}$. 

The two sets of stabilizers inherited from the 2D cluster state, $Z_{i,k_0}X_{i+\frac{1}{2},k_0+\frac{1}{2}}Z_{i+1,k_0}$ and $Z_{i-\frac{1}{2},k_0+\frac{1}{2}}X_{i,k_0}Z_{i+\frac{1}{2},k_0+\frac{1}{2}}$, act differently on the mixed-state $\rho_{\text{1D-DASPT}}^{\rm A}$. One set of stabilizers satisfies a strong symmetry condition while the other set satisfies a weak symmetry condition. Concretely we have
{\small
\begin{subequations}
   \begin{align}
    &Z_{i,k_0}X_{i+\frac{1}{2},k_0+\frac{1}{2}}Z_{i+1,k_0}\rho_{\text{1D-DASPT}}^{\rm A}=\rho_{\text{1D-DASPT}}^{\rm A}\,,\\
    &Z_{i-\frac{1}{2},k_0+\frac{1}{2}}X_{i,k_0}Z_{i+\frac{1}{2},k_0+\frac{1}{2}}\rho_{\text{1D-DASPT}}^{\rm A}Z_{i-\frac{1}{2},k_0+\frac{1}{2}}X_{i,k_0}Z_{i+\frac{1}{2},k_0+\frac{1}{2}}\nonumber\\
    &\hspace{3cm}=\rho_{\text{1D-DASPT}}^{\rm A}\, .
\end{align} 
\end{subequations}}

We note that this state can also be obtained by decohering the pure 1D cluster state, whose density matrix is
\begin{align}
    \rho_{\text{pSPT}}&=\prod_i\left(\frac{1+Z_{i,k_0}X_{i+\frac{1}{2},k_0+\frac{1}{2}}Z_{i+1,k_0}}{2}\right)\nonumber\\
    &\hspace{1cm}\prod_i\left(\frac{1+Z_{i-\frac{1}{2},k_0+\frac{1}{2}}X_{i,k_0}Z_{i+\frac{1}{2},k_0+\frac{1}{2}}}{2}\right)\,,
    \label{eq:pureSPT}
\end{align}
with decoherence channel 
\begin{align}
    \xi[\rho_{\text{pSPT}}]=\prod_i\xi_i[\rho_{\text{pSPT}}],
\end{align}
where 
{\small
\begin{align}
    \xi_i[\rho_{\text{pSPT}}]=\frac{1}{2}\left(\rho_{\text{pSPT}}+Z_{i,k_0}Z_{i+1,k_0}\rho_{\text{pSPT}} Z_{i,k_0}Z_{i+1,k_0}\right)\,.
\end{align}}

Using the fidelity correlator or R\'enyi-2 correlator defined in \cite{Lessa:2024gcw}, it can be checked that $\rho_{\text{1D-DASPT}}^{\rm A}$ is in the SWSSB phase with respect to the $\mathbb{Z}_2^{(r)}$ symmetry on the red sublattice. Namely we have
{\small
\begin{align}
&\frac{\text{Tr}\left(Z_{i,k_0}Z_{j,k_0}\rho_{\text{1D-DASPT}}^{\rm A}Z_{i,k_0}Z_{j,k_0}\rho_{\text{1D-DASPT}}^{\rm A}\right)}{\text{Tr}((\rho_{\text{1D-DASPT}}^{\rm A})^2)}=1\,,\\
&\text{Tr}\left(\sqrt{\sqrt{\rho_{\text{1D-DASPT}}^{\rm A}}Z_{i,k_0}Z_{j,k_0}\rho_{\text{1D-DASPT}}^{\rm A}Z_{i,k_0}Z_{j,k_0}\sqrt{\rho_{\text{1D-DASPT}}^{\rm A}}}\right)\nonumber\\
&\hspace{6.5cm}=1\,.
 \end{align}}
\red{ 
We also find the ordinary two-point correlator also evaluates to zero
\begin{align}
 \text{Tr}\left(\rho^{\rm A}_{\text{1D-DASPT}}Z_{i,k_0}Z_{j,k_0}\right)=0\,.
\end{align}
A summary of these correlators are provided in Table~\ref{tab:Renyi2andFidelityofthreestates}.
\begin{table*}[]
    \centering
    \begin{tabular}{c|c|c|c|c}
    \hline
       State & Order parameter $O_i$ & Tr$\left[\rho O_i O_j\right]$  & R\'enyi-2 correlator & Fidelity correlator \\
       \hline
        $\rho_{\text{1D-DASPT}}$ & $Z_{i,k_0}$ & 0 & 1 & 1\\
         $\rho_{\text{blue}}$ & $Y_{i,k_0}$ & 0 & 1 & 1\\
          $\rho_{\text{blue}^{k_0;k_1}}$ & $Y_{i,k_0}$ & 0 & 1 & 1\\
          \hline
    \end{tabular}
    \caption{\red{The Table lists the three DASPTs, their order parameter, ordinary two-point correlator, R\'enyi-2 correlator and the fidelity correlator with respect to the chosen order parameters used to exhibit SWSSB}.}
    \label{tab:Renyi2andFidelityofthreestates}
\end{table*}
We note that the definition of SWSSB using the fidelity correlator and the R\'enyi-2 correlator may give different results as discussed in Ref.~\cite{Lessa:2024gcw}. However, here both of them evaluate to unity. }

On the other hand, consider the following string order parameter for the strong $\mathbb{Z}_2^{(b)}$ symmetry
\begin{align}
\mathcal{O}_{i,l}^{(b)}:=Z_{i,k_0}X_{i+\frac{1}{2},k_0+\frac{1}{2}}X_{i+\frac{3}{2},k_0+\frac{1}{2}}...X_{i+l+\frac{1}{2},k_0+\frac{1}{2}}Z_{i+l+1,k_0}~,
\end{align}
where $i$ can be any red site in the row $y=k_0$. The expectation value of this string order parameter on $\rho_{\text{1D-DASPT}}^{\rm A}$ is computed to be
\begin{align}
\text{Tr}\left(\mathcal{O}_{i,l}^{(b)}\rho_{\text{1D-DASPT}}^{\rm A}\right)=1\,.
 \end{align}
This implies non-trivial edge correlations as discussed in \cite{Ma2025Topological}, as well as the standard symmetry localization at boundaries.
 
Therefore, we see that $\rho_{\text{1D-DASPT}}^{\rm A}$ exhibits both properties of ASPT and SWSSB,
 $\rho_{\text{1D-DASPT}}^{\rm A}$ is thus referred as a doubled average SPT (DASPT) state, which was recently numerically explored in~\cite{Guo2025strong,Kuno:2025dqf}. \red{The ASPT phase is protected by the strong $\mathbb{Z}_2^s$ symmetry and the other strong $\mathbb{Z}_2^{s\rightarrow w}$ symmetry that is spontaneously broken to weak symmetry.} In contrast, in the mixed-state examples in Ref.~\cite{Sun:2024iwm} by tracing over the bulk, only ASPT phases emerge. 
 
 \red{We note that SWSSB is stable under strongly symmetric finite-depth local channels using the criterion of the fidelity correlator as studied in Ref.~\cite{Lessa:2024gcw}. On the other hand, ASPTs remain in the same phase under any symmetric two-way connecting finite-depth local channel by the definition provided in Ref.~\cite{ma2023average}.  Since the symmetries we consider for the DASPT are all strong, we consider a strongly symmetric two-way connecting finite-depth local channel to take the state away from the fixed-point state. We expect that the above DASPT is also stable using the fidelity correlator criterion under such strongly symmetric two-way connecting finite-depth local channels, since it is the coexistence of SWSSB and ASPT. Moreover, Ref.[70] identifies a region for DASPT by numerics in their Figure 3(b).}

 We note that existence of symmetry breaking order and SPT order appeared in the pure ground states~\cite{Verresen:2017cwj,Zeng:2014qjk} and in non-equilibrium settings ~\cite{Morral-Yepes:2023vgo}.

In addition to the invertible $\mathbb{Z}_2^{(r)}$ and $\mathbb{Z}_2^{(b)}$ symmetries discussed above, there is an emergent strong non-invertible KW symmetry~\cite{mana2024kennedy,Seifnashri:2024dsd} in the one-dimensional system,
\begin{align}
    \mathrm{D}^{(1)}=\mathbf{T}^{-1}\mathrm{D}_r\mathrm{D}_b\, ,
    \label{eq:D1exp}
\end{align}
where $\mathrm{D}_r$ and $\mathrm{D}_b$ are non-invertible KW symmetry operators on the red row and blue row in the region $\rm A$. The exact expression for the KW symmetry operator $\mathrm{D}$ ( this could be $\mathrm{D}_r$ or $\mathrm{D}_b$) on a ring with $L$ sites is given by \cite{Seiberg:2023cdc,Chen:2023qst,Ho:2019hyv}
\begin{align}
    &\mathrm{D}=\red{e^{-\frac{2\pi i L}{8}}}\prod_{i=1}^{L-1}\left(e^{i\frac{\pi}{4}X_{i}}e^{i\frac{\pi}{4}Z_{i}Z_{i+1}}\right)e^{i\frac{\pi}{4}X_L}\nonumber\\&\qquad\qquad\times\frac{(1+\prod_{i=1}^LX_{i})}{2}\,. \label{eq:D1lattice}
\end{align}
$\mathrm{D}^{(1)}$ has the following action on the Pauli operators
\begin{subequations}
\begin{align}
&X_{i,k}\xleftrightarrow{\mathrm{D}^{(1)}}Z_{i-\frac{1}{2},k+\frac{1}{2}}Z_{i+\frac{1}{2},k+\frac{1}{2}}\,,\\
& X_{i+\frac{1}{2},k+\frac{1}{2}}\xleftrightarrow{\mathrm{D}^{(1)}}Z_{i,k}Z_{i+1,k}\,.
\end{align}
\end{subequations}
From the above action of $\mathrm{D}^{(1)}$ together with the stabilizer condition, it follows that $\mathrm{D}^{(1)}$ is a strong symmetry for $\rho_{\text{1D-DASPT}}^{\rm A}$.

A summary of the features of $\rho_{\text{1D-DASPT}}^{\rm A}$ is given in Table~\ref{tab:clusterstatetrace} and \ref{tab:clusterstatetrace+}.

\section{ DASPTs from non-invertible SSPTs} \label{sec:1DNDASPT}

In this section, we study the mixed-state phase by tracing out a different SSPT state with the non-invertible Kramers-Wannier symmetry. In particular, we first consider a non-invertible higher-order SSPT studied in \cite{ParayilMana:2025nxw}, denoted as the {\it blue state}. Later, we will also study another state denoted as {\it blue$^{k_0;k_1}$}. Our setup is the same as illustrated in 
Figure~\ref{fig:regionA-open}.

We start with 2D states that are in a different phase from the 2D cluster state due to the non-invertible symmetry $\mathbf{D}^{(2)}$. These states can be obtained by considering the Kennedy-Tasaki transformation, which maps SPT to SSB phases. Under KT, the non-invertible symmetry operator $\mathbf{D}^{(2)}$ is mapped to an operator proportional to  $\hat{\rm V}=\prod_{v_r,v_b}CZ_{v_r,v_b}$, while the subsystem symmetries are mapped to the subsystem symmetries on the dual theory. KT maps the cluster phase to a spontaneous symmetry breaking phase that consists of two copies of the plaquette Ising model. Hence, all the subsystem symmetries are spontaneously broken. However, there are various choices for the preserved symmetries. $\hat{\rm V}$ is one such choice, and on the dual theory side this gives rise to the cluster phase protected by the non-invertible symmetry $\mathbf{D}^{(2)}$, which we already discussed in the previous section. Another choice is preserving $\hat{\rm V}\prod_{v_r}X_{v_r}$, and this on the dual theory side gives the blue phase, which we describe below.

\subsection{Overview of 2D blue state}

The 2D blue state is in the same phase as the 2D cluster state if we only take into account the subsystem symmetries. However, including the non-invertible symmetry~\eqref{eq:D^{(2)}}, the cluster phase splits into different phases, and the blue state is in a different phase from the cluster phase protected by the non-invertible symmetry~\cite{ParayilMana:2025nxw}. The Hamiltonian of one such refined phase due to the splitting is described by the cluster Hamiltonian itself. On the other hand, the Hamiltonian for the phase associated with the blue state is also translation invariant and is given by
\begin{align}
\mathrm{H}_{\text{blue}}=&\sum_{v_r}\begin{array}{ccc}
         Z_{v_b} &  &Z_{v_b}\\
         & \boxed{X_{v_r}} & \\
         Z_{v_b} &  & Z_{v_b}
    \end{array}-\sum_{v_b}\begin{array}{ccc}
         Y_{v_r} &  &Y_{v_r}\\
         & \boxed{X_{v_b}} & \\
         Y_{v_r} &  & Y_{v_r}
    \end{array}\notag\\
    &-\sum_{v_b}\begin{array}{ccccc}
        Z_{v_b} & & & & Z_{v_b} \\
         & Z_{v_r} & &  Z_{v_r} & \\
         & & \boxed{X_{v_b}} & & \\
         & Z_{v_r} & &  Z_{v_r} & \\
         Z_{v_b} & & & & Z_{v_b} \\
    \end{array}\, .
    \label{eq:blueHam}
\end{align}
\red{We assume $L_x$ and $L_y$ are even.} The two phases differ as a higher-order SSPT with protected corner modes between their interface. 
We note that there are additional phases that are different from the cluster phase whose corresponding Hamiltonian is not translation invariant; see~\cite{ParayilMana:2025nxw} for a detailed analysis of such phases.

The Hamiltonian~(\ref{eq:blueHam}) has a unique ground state (which we have referred to as the blue state above) given by
\begin{align}
    \ket{\text{blue}}=&\prod_{ v_b} CZ_{v_b,v_b+(1,1)}CZ_{v_b,v_b+(-1,1)}\nonumber\\
    &\times\prod_{v_r}\prod_{v_b\in\partial (p_b=v_r)}CZ_{v_r,v_b}\ket{+}^{\otimes \Delta_{v_b}}\ket{-}^{\Delta_{v_r}}\, .
    \label{eq:blusstate}
\end{align}
The state $\ket{\text{blue}}$ is related to the 2D cluster state $\ket{\text{2D-cluster}}$ (the ground state of \eqref{eq:2Dcluster}) by the following finite-depth circuit 
\begin{align}
    \prod_{ v_r}Z_{v_r}\prod_{ v_b} CZ_{v_b,v_b+(1,1)}CZ_{v_b,v_b+(-1,1)}\, .
\end{align}
However, such a circuit does not respect the 2D KW symmetry, and, therefore, the cluster and blue states remain at different phases characterized by the KW symmetry.
\subsection{1D DASPT from 2D blue state}

We now consider the Hamiltonian $\mathrm{H}_{\text{blue}}$ on a cylinder, i.e., with open boundary conditions in the $y$ direction, with the top boundary on blue vertices at $y=k_0+\frac{1}{2}$. We consider the Dirichlet boundary condition where all the vertex d.o.f. at $k_0+\frac{1}{2}$ is set to $\ket{+}$. The analysis for other choices of boundary conditions is shown in Appendix~\ref{app:Extraexamples2}. Now, just as illustrated in Figure~\ref{fig:regionA-open}, we trace out all the 2D d.o.f. except for region $\rm A$ consisting of the top rows at $y=k_0$ and $y=k_0+\frac{1}{2}$,
{\small
\begin{widetext}
    \begin{align}
\rho_{\text{blue}}^{\mathrm{A}} &\equiv \text{Tr}_{\rm A^c}\Big(\ket{\text{blue}}\bra{\text{blue}}\Big)=\frac{2}{2^{|\Delta_{v_r}\cap \mathrm{A}|}}\sum_{\substack{\{i_l\}\in\Delta_{v_r}\cap \mathrm{A}\,,\\
    s \text{ even }}}\Big(Z_{i_1}\prod_{j_1\in \text{Nb}(i_1)}Z_{j_1}\Big)...\Big(Z_{i_s}\prod_{j_s\in \text{Nb}(i_s)}Z_{j_s}\Big)\\
    &\quad\prod CZ_{v_r,v_b}\ket{+}^{\otimes\Delta_{v_r}}\ket{+}^{\otimes\Delta_{v_b}}\bra{+}^{\otimes\Delta_{v_b}}\bra{+}^{\otimes\Delta_{v_r}}\prod CZ_{v_r,v_b}\Big(Z_{i_1}\prod_{j_1\in \text{Nb}(i_1)}Z_{j_1}\Big)...\Big(Z_{i_s}\prod_{j_s\in \text{Nb}(i_s)}Z_{j_s}\Big)\nonumber
\end{align} 
\end{widetext}}
where Nb($i_l$)  denotes the neighboring blue vertices to the red vertex $i_l$ in $\rm A$. This state has strong $\mathbb{Z}_2^{(r)}\times\mathbb{Z}_2^{(b)}$ 0-form symmetry and an emergent non-invertible symmetry $\mathrm{D}^{(1)}$~\cite{mana2024kennedy,Seifnashri:2024dsd}. Although not obvious, the non-invertible symmetry $\mathrm{D}^{(1)}$ is a strong symmetry for $\rho_{\text{blue}}^{\mathrm{A}}$, and we leave the detailed calculations to Appendix~\ref{app:strongD1onblue}. 

Additionally, we have the following strong and weak stabilizers
{\small
\begin{subequations}
   \begin{align}
    &Z_{i-\frac{1}{2},k_0+\frac{1}{2}}Y_{i,k_0}X_{i+\frac{1}{2},k_0+\frac{1}{2}}Y_{i+1,k_0}Z_{i+\frac{3}{2},k_0+\frac{1}{2}}\rho_{\text{blue}}^{\rm A}=-\rho_{\text{blue}}^{\rm A},\\
    &Z_{i,k}X_{i+\frac{1}{2},k_0+\frac{1}{2}}Z_{i+1,k_0}\rho_{\text{blue}}^{\rm A}Z_{i,k}X_{i+\frac{1}{2},k_0+\frac{1}{2}}Z_{i+1,k_0}\nonumber\\&\hspace{5cm}=\rho_{\text{blue}}^{\rm A}\,,\\
    &Z_{i-\frac{1}{2},k_0+\frac{1}{2}}X_{i,k_0}Z_{i+\frac{1}{2},k_0+\frac{1}{2}}\rho_{\text{blue}}^{\rm A}Z_{i-\frac{1}{2},k_0+\frac{1}{2}}X_{i,k_0}Z_{i+\frac{1}{2},k_0+\frac{1}{2}}\nonumber\\&\hspace{5cm}=\rho_{\text{blue}}^{\rm A}\, .
\end{align}
\end{subequations}}
From the above, we infer that 
{\small
\begin{align}
    \rho_{\text{blue}}^{\rm A}=&\prod_i\left(\frac{1-Z_{i-\frac{1}{2},k_0+\frac{1}{2}}Y_{i,k_0}X_{i+\frac{1}{2},k_0+\frac{1}{2}}Y_{i+1,k_0}Z_{i+\frac{3}{2},k_0+\frac{1}{2}}}{2}\right)\notag\\
    &\times\left(\frac{1+\prod_{i}X_{i,k_0}}{2^{|\Delta_{v_r}\cap A|}}\right).\label{eq:rhoAblue}
\end{align}
}
We note that this state is also a DASPT state, but with a different form from Eq.~(\ref{eq:rhoA2Dcluster}). We will study their interface in \ref{sec:1D-DASPT|blue}. A summary of properties of $\rho_{\text{blue}}^{\rm A}$ is given in Table~\ref{tab:bluestatetrace} and \ref{tab:bluestatetrace+}.

\red{We note that this state can also be obtained by decohering the pure 1D cluster state, whose density matrix is given in \eqref{eq:pureSPT}, 
with decoherence channel 
\begin{align}
    \tilde{\xi}[\rho_{\text{pSPT}}]=\prod_i\tilde{\xi}_i[\rho_{\text{pSPT}}],
\end{align}
where 
{\small
\begin{align}
    \tilde{\xi}_i[\rho_{\text{pSPT}}]=\frac{1}{2}\left(\rho_{\text{pSPT}}+Y_{i,k_0}Y_{i+1,k_0}\rho_{\text{pSPT}} Y_{i,k_0}Y_{i+1,k_0}\right)\,.
\end{align}}}

If we take Pauli-$Y$ operators as charged under the strong $\mathbb{Z}_2^{(r)}$ symmetry, we find the R\'enyi-2 correlator
\begin{align}
     &\frac{\text{Tr}\left(Y_{i,k_0}Y_{j,k_0}\rho_{\text{blue}}^{\rm A}Y_{i,k_0}Y_{j,k_0}\rho_{\text{blue}}^{\rm A}\right)}{\text{Tr}((\rho_{\text{blue}}^{\rm A})^2)}=1~,
\end{align}
and the fidelity correlator
    \begin{align}
    &\text{Tr}\left(\sqrt{\sqrt{\rho_{\text{blue}}^{\rm A}}Y_{i,k_0}Y_{j,k_0}\rho_{\text{blue}}^{\rm A}Y_{i,k_0}Y_{j,k_0}\sqrt{\rho_{\text{blue}}^{\rm A}}}\right)=1\,.
 \end{align}
 \red{The ordinary two-point correlator evaluates to zero
 \begin{align}
     \text{Tr}\left(\rho^{\rm A}_{\text{blue}}Y_{i,k_0}Y_{j,k_0}\right)=0\,.
 \end{align}}
 This confirms that the strong $\mathbb{Z}_2^{(r)}$ symmetry is spontaneously broken to a weak symmetry. \red{A summary of these correlators are provided in Table~\ref{tab:Renyi2andFidelityofthreestates}.}

 \red{We note that the state \eqref{eq:rhoAblue} can be obtained from \eqref{eq:rhoA2Dcluster} by conjugating with a unitary
 \begin{align}
     \rho^{\rm A}_{\text{blue}}=\mathcal{U}\rho^{\rm A}_{\text{1D-DASPT}}\mathcal{U}^{\dagger}\,.
 \end{align}
 The unitary $\mathcal{U}$ can be expressed as a product of local unitaries
 \begin{align}
     \mathcal{U}=\prod_{i=1}^{L_x}e^{i\pi/4\left(X_{i,k_0}+Z_{i-\frac{1}{2},k_0+\frac{1}{2}}Z_{i+\frac{1}{2},k_0+\frac{1}{2}}\right)}\,.
 \end{align}
 We note that the local gates $e^{i\pi/4\left(X_{i,k_0}+Z_{i-\frac{1}{2},k_0+\frac{1}{2}}Z_{i+\frac{1}{2},k_0+\frac{1}{2}}\right)}$ are $\mathbb{Z}_2\times\mathbb{Z}_2$  and $\mathrm{D}^{(1)}$ symmetric. Hence the unitary $\mathcal{U}$ acts as a finite-depth local channel without any additional ancillas. In other words, $\rho^{\rm A}_{\text{1D-DASPT}}$ and $\rho^{\rm A}_{\text{blue}}$ are in the same phase with respect to both invertible and non-invertible symmetries.}
 
 \red{Let us consider the one-parameter family of unitary gates defined as
 \begin{align}
     \mathcal{V}(\theta)=\prod_{i=1}^{L_x}e^{i\theta\left(X_{i,k_0}+Z_{i-\frac{1}{2},k_0+\frac{1}{2}}Z_{i+\frac{1}{2},k_0+\frac{1}{2}}\right)}\,,
 \end{align}
 and the one-parameter family of density matrices
 \begin{align}
     \rho(\theta)=\mathcal{V}(\theta)\rho^{\rm A}_{\text{1D-DASPT}}\mathcal{V}^{\dagger}(\theta)\,.
 \end{align}
 Note that $\rho(\pi/4)=\rho^{\rm A}_{\text{blue}}$. For every $\theta$, $\rho(\theta)$ has both strong $\mathbb{Z}_2\times\mathbb{Z}_2$ symmetry and strong $\rm D^{(1)}$ symmetry. Consider the order parameter as a function of $\theta$
 \begin{align}
     O_{i,k_0}(\theta)=\mathcal{V}(\theta)Z_{i,k_0}\mathcal{V}^{\dagger}(\theta)=\cos(2\theta)Z_{i,k_0}-\sin(2\theta)Y_{i,k_0}\,.
 \end{align}
 For far-separated points, we note that 
 \begin{align}
     O_{i,k_0}(\theta)O_{j,k_0}(\theta)\rho(\theta)O_{j,k_0}(\theta)O_{i,k_0}(\theta)=\rho(\theta)\,.
 \end{align}
 Hence, we have SWSSB for any parameter $\theta$, due to the fact that the fidelity correlator evaluates to 1 and the two-point correlator vanishes
 \begin{subequations}
 \begin{align}
     F(\rho(\theta),O_{i,k_0}(\theta)O_{j,k_0}(\theta)\rho(\theta)O_{j,k_0}(\theta)O_{i,k_0}(\theta))&=1\\
     \text{Tr}\left(\rho(\theta)O_{i,k_0}(\theta)O_{j,k_0}(\theta)\right)&=0\,.
 \end{align}
 \end{subequations}
 We can also define a string order parameter as a function of $\theta$
 \begin{align}
     S_{i,j}(\theta)=\mathcal{V}(\theta)Z_{i,k_0}\prod_{i=1}^{j-1}X_{i+\frac{1}{2},k_0+\frac{1}{2}}Z_{j,k_0}\mathcal{V}^{\dagger}(\theta)\,.
 \end{align}
The string order parameter evaluates to
\begin{align}
    \text{Tr}\left(\rho(\theta)S_{i,j}(\theta)\right)=1\,.
\end{align}
This implies that the whole parameter family of density matrices $\rho(\theta)$ exhibits the coexistence of SWSSB and ASPT and are DASPTs.
 }
\subsection{1D DASPT from 2D blue$^{k_0;k_1}$ state}
 Here we consider yet another choice of SSB phase that preserves $\prod CZ_{v_r,v_b}\times \prod_{k_1\leq j\leq k_0}\prod_{i}X_{v_r=(i,j)}$ and breaks all subsystem symmetries. The corresponding dual SSPT state is
{\small
\begin{align}
    &\ket{\text{blue}^{k_0;k_1}}\nonumber\\
    &=\prod_{\substack{v_b=(i+\frac{1}{2},j+\frac{1}{2})\\
    k_1\leq j \leq k_0-1}} CZ_{v_b,v_b+(1,1)}CZ_{v_b,v_b+(-1,1)}\nonumber\\
    &\times\prod_{\substack{v_b=(i+\frac{1}{2},j+\frac{1}{2})\\
   j=k_0}} CZ_{v_b,v_b+(1,0)}\prod_{\substack{v_b=(i+\frac{1}{2},j+\frac{1}{2})\\
   j=k_1-1}} CZ_{v_b,v_b+(1,0)}\nonumber\\
   &\times \prod_{v_r}\prod_{v_b\in\partial (p_b=v_r)}CZ_{v_r,v_b}\ket{+}^{\otimes \Delta_{v_b}}\ket{-}^{\otimes\Delta_{v_r}}\,.
   \label{eq:bluek0;k1}
\end{align}}
\red{Here also we assume $L_x$ and $L_y$ are even.} Now we place the $\text{blue}^{k_0;k_1}$ state on a geometry with open boundary condition in the $y$ direction with $y$ values ranging from $-\infty $ to $k_1$,$k_1+\frac{1}{2}$, $k_1+1$, $\dots$, $k_0-\frac{1}{2}$, $k_0$, $k_0+\frac{1}{2}$. We further impose the Dirichlet boundary condition where all the vertex d.o.f at $k_0+\frac{1}{2}$ are set to $\ket{+}$. We leave the analysis to Appendix~\ref{app:Extraexamples3} for other choices of boundary conditions. Taking $y=k_0$ and $y=k_0+\frac{1}{2}$ as the region $\rm A$ and the rest as $\rm A^c$ as before, after tracing out the region $\rm A^c$, we obtain the following mixed-state
\begin{align}
    &\text{Tr}_{\rm A^c}\left(\ket{\text{blue}^{k_0;k_1}}\bra{\text{blue}^{k_0;k_1}}\right)\notag\\
    =&\left[\prod_{\substack{v_b=(i+\frac{1}{2},k_0+\frac{1}{2})}} CZ_{v_b,v_b+(1,0)}\right]\rho^{\rm A}_{\text{blue}}\\&\times\left[\prod_{\substack{v_b=(i+\frac{1}{2},k_0+\frac{1}{2})}} CZ_{v_b,v_b+(1,0)}\right]\equiv\rho^{\rm A}_{\text{blue}^{k_0;k_1}}\,.\notag
\end{align}

This state has strong $\mathbb{Z}_2^{(r)}\times\mathbb{Z}_2^{(b)}$ 0-form symmetry and no strong/weak non-invertible symmetry $\mathrm{D}^{(1)}$. In addition, we have the strong and weak stabilizers
\begin{subequations}
\begin{align}
    &Y_{i,k_0}X_{i+\frac{1}{2},k_0+\frac{1}{2}}Y_{i+1,k_0}\rho^{\rm A}_{\text{blue}^{k_0;k_1}}=-\rho^{\rm A}_{\text{blue}^{k_0;k_1}}\\
    &\red{Z_{i-\frac{1}{2},k_0+\frac{1}{2}}X_{i,k_0}Z_{i+\frac{1}{2},k_0+\frac{1}{2}}\rho^{\rm A}_{\text{blue}^{k_0;k_1}}Z_{i-\frac{1}{2},k_0+\frac{1}{2}}X_{i,k_0}Z_{i+\frac{1}{2},k_0+\frac{1}{2}}}\nonumber\\
    &\hspace{3cm}=\rho^{\rm A}_{\text{blue}^{k_0;k_1}}\,,
\end{align}
\end{subequations}
From the above, we infer that this state can be explicitly written in terms of the following projectors
\begin{align}
    \rho^{\rm A}_{\text{blue}^{k_0;k_1}}&=\prod_i\left(\frac{1-Y_{i,k_0}X_{i+\frac{1}{2},k_0+\frac{1}{2}}Y_{i+1,k_0}}{2}\right)\nonumber\\
    &\hspace{2cm}\times\left(\frac{1+\prod_{i}X_{i,k_0}}{2^{|\Delta_{v_r}\cap \mathrm{A}|}}\right)\,.
    \label{eq:rhoAbluek0k1}
\end{align}
We note that this state is also a DASPT state different from $\rho_{\text{1D-DASPT}}^{\rm A}$. However, this state has different symmetries compared to $\rho^{\rm A}_{\text{1D-DASPT}}$ as the later possesses the strong non-invertible symmetry $\mathrm{D}^{(1)}$ but the former is not symmetric under $\mathrm{D}^{(1)}$. A summary of the properties of $\rho^{\rm A}_{\text{blue}^{k_0;k_1}}$ can be found in Table~\ref{tab:bluek0k1trace} and \ref{tab:bluek0k1trace+}.
\vspace{0.2cm}\\

Additionally, this state can also be obtained by decohering the following pure state 1D non-invertible SPT~\cite{Seifnashri:2024dsd}
\begin{align}
    \rho_{\text{pNSPT}}&=\prod_i\left(\frac{1-Y_{i,k_0}X_{i+\frac{1}{2},k_0+\frac{1}{2}}Y_{i+1,k_0}}{2}\right)\nonumber\\
    &\hspace{1cm}\prod_i\left(\frac{1-Z_{i-\frac{1}{2},k_0+\frac{1}{2}}X_{i,k_0}Z_{i+\frac{1}{2},k_0+\frac{1}{2}}}{2}\right)\,,
    \label{eq:pureNSPT}
\end{align}
with decoherence channel 
\begin{align}
    \bar{\xi}[\rho_{\text{pNSPT}}]=\prod_i\bar{\xi}_i[\rho_{\text{pNSPT}}],
\end{align}
where 
\begin{align}
    \bar{\xi}_i[\rho_{\text{pNSPT}}]=\frac{1}{2}\left(\rho_{\text{pNSPT}}+Y_{i,k_0}Y_{i+1,k_0}\rho_{\text{pNSPT}} Y_{i,k_0}Y_{i+1,k_0}\right)\,.
\end{align}

If we take Pauli-$Y$ operators as charged under the strong $\mathbb{Z}_2^{(r)}$ symmetry, we find the R\'enyi-2 correlator
\begin{align}
     &\frac{\text{Tr}\left(Y_{i,k_0}Y_{j,k_0}\rho_{\text{blue}^{k_0;k_1}}^{\rm A}Y_{i,k_0}Y_{j,k_0}\rho_{\text{blue}^{k_0;k_1}}^{\rm A}\right)}{\text{Tr}((\rho_{\text{blue}^{k_0;k_1}}^{\rm A})^2)}=1~,
\end{align}
and the fidelity correlator
    \begin{align}
    &\text{Tr}\left(\sqrt{\sqrt{\rho_{\text{blue}^{k_0;k_1}}^{\rm A}}Y_{i,k_0}Y_{j,k_0}\rho_{\text{blue}^{k_0;k_1}}^{\rm A}Y_{i,k_0}Y_{j,k_0}\sqrt{\rho_{\text{blue}^{k_0;k_1}}^{\rm A}}}\right)\nonumber\\
    &\hspace{6cm}=1\,.
 \end{align}
 \red{The ordinary two-point correlator evaluates to zero
 \begin{align}
     \text{Tr}\left(\rho^{\rm A}_{\text{blue}^{k_0;k_1}}Y_{i,k_0}Y_{j,k_0}\right)=0\,.
 \end{align}}
 This confirms that the strong $\mathbb{Z}_2^{(r)}$ symmetry is spontaneously broken to a weak symmetry. \red{A summary of these correlators are provided in Table~\ref{tab:Renyi2andFidelityofthreestates}.}

  \red{We note that \eqref{eq:rhoAbluek0k1} is in the same phase as \eqref{eq:rhoA2Dcluster} with respect to the invertible symmetries ($\mathbb{Z}_2\times\mathbb{Z}_2$). We can construct a unitary that conjugates \eqref{eq:rhoA2Dcluster} to \eqref{eq:rhoAbluek0k1}
  \begin{align}
      \rho^{\rm A}_{\text{blue}^{k_0;k_1}}=\mathcal{U}\rho^{\rm A}_{\text{1D-DASPT}}\mathcal{U}^{\dagger}\,.
  \end{align}
  Explicitly, this unitary is given by
  \begin{align}
      \mathcal{U}=\mathcal{U}_2\mathcal{U}_1\,,
  \end{align}
  where
  \begin{subequations}
      \begin{align}
      \mathcal{U}_1&=\prod_{i=1}^{L_x} e^{-i\pi/4 X_{i,k_0}}\\\mathcal{U}_2&=\prod_{i=1}^{L_x/2}Z_{2i-\frac{1}{2},k_0+\frac{1}{2}} Z_{2i+\frac{1}{2},k+\frac{1}{2}}\,.
  \end{align}
  \end{subequations}
  Note that individual gates in both $\mathcal{U}_1$ and $\mathcal{U}_2$ are $\mathbb{Z}_2\times\mathbb{Z}_2$ symmetric.
  Conjugating with $\mathcal{U}_1$, convert all the $Z$ operator to $-Y$ operator. Conjugating with $\mathcal{U}_2$ add an extra negative sign for $X_{i+\frac{1}{2},k_0+\frac{1}{2}}$. This effectively implements a symmetric finite-depth local unitary connecting \eqref{eq:rhoA2Dcluster} and \eqref{eq:rhoAbluek0k1}. We also comment that $\rho^{\rm A}_{\text{1D-DASPT}}$ has $\mathrm{D}^{(1)}$ symmetry while $\rho^{\rm A}_{\text{blue}^{k0;k1}}$ does not. If one applies a finite-depth strongly $\mathrm{D}^{(1)}$ symmetric channel, this would extend the distinction between the two states to a finite region beyond the fixed point.
  }
\section{Interface as a probe for mixed-state SPTs}\label{sec:generalinterface}
Let us consider a $1+1$D bosonic mixed-state SPT $\rho$ defined on a lattice with integer sites. $\rho$ is protected by strong symmetry $\mathcal{S}$ and weak invertible symmetry $W$ 
\begin{align}
    \mathcal{S}\rho\propto \rho\,,\quad W\rho W^{\dagger}=\rho\,.
\end{align}
Let $\rho'$ be another mixed-state obtained from $\rho$ by a finite depth symmetric channel, i.e.,
\begin{align}
    \rho'=\xi[\rho]\,,\quad\text{with}\quad \xi[\rho]=\text{Tr}_{\rm a}\left(U\rho\otimes\ket{0}_a\bra{0}U^{\dagger}\right)\,,
    \label{eq:unitaryinsidetrace}
\end{align}
and $U=\prod_{t=1}^{T}\mathfrak{U}_t$ with $\mathfrak{U}_t=\prod_{i}\mathcal{U}_i$ where $\mathcal{U}_i$ are local unitary operators supported near site $i$. Moreover, the local unitary operators need to commute with the strong and weak symmetry for $\xi$ to be a symmetric finite depth quantum channel, i.e., 
\begin{align}
    [\mathcal{S}\otimes I_a,\mathcal{U}_i]=0\,,\quad[W\otimes W_a,\mathcal{U}_i]=0\,,
\end{align}
where $W_a\ket{0}_a=\lambda_a\ket{0}_a$ is a scalar acting on the product state.

Now, let us consider a subregion of the lattice denoted by $B$ where we apply the finite-depth symmetric quantum channel as shown in Figure~\ref{fig:interfacerhorhop}. We denote this quantum channel by $\xi_B$, indicating it is applied on region $B$. Let us denote $\tilde{\rho}=\xi_B[\rho]$. The state $\tilde{\rho}$ can be thought of as an interface between $\rho$ and $\rho'$ as shown in Figure~\ref{fig:interfacerhorhop}. Now, we verify the symmetry properties of $\tilde{\rho}$
\begin{align}
    \mathcal{S}\tilde{\rho}&=\mathcal{S}\xi_B[\rho]\nonumber\\
    &=\mathcal{S}\text{Tr}_a\left(U\rho\otimes\ket{0}_a\bra{0}U^{\dagger}\right)\nonumber\\
    &=\text{Tr}_a\left(\mathcal{S}\otimes I_a\, U\rho\otimes\ket{0}_a\bra{0}U^{\dagger}\right)\nonumber\\
    &=\text{Tr}_a\left( U\mathcal{S}\otimes I_a\rho\otimes\ket{0}_a\bra{0}U^{\dagger}\right)\nonumber\\
    &\propto\text{Tr}_a\left( U\rho\otimes\ket{0}_a\bra{0}U^{\dagger}\right)=\tilde{\rho}\,.
    \label{eq:unitaryinsidetraceB}
\end{align}
Note that the ancillas $a$ in the above set of equations are coupled to only qubits in region $B$. Moreover,
\begin{align}
    W\tilde{\rho}W^{\dagger}&=W\xi_B[\rho]W^{\dagger}\nonumber\\
    &=W\text{Tr}_a\left(U\rho\otimes\ket{0}_a\bra{0}U^{\dagger}\right)W^{\dagger}\nonumber\\
    &=\text{Tr}_a\left(W\otimes I_a\, U\rho\otimes\ket{0}_a\bra{0}U^{\dagger}W^{\dagger}\otimes I_a\right)\nonumber\\
    &=\text{Tr}_a\left(W\otimes W_a\, U\rho\otimes\ket{0}_a\bra{0}U^{\dagger}W^{\dagger}\otimes W_a^{\dagger}\right)\nonumber\\
    &=\text{Tr}_a\left(U W\otimes W_a \rho\otimes\ket{0}_a\bra{0}W^{\dagger}\otimes W_a^{\dagger}U^{\dagger}\right)\nonumber\\
    &=\text{Tr}_a\left(U\rho\otimes\ket{0}_a\bra{0}U^{\dagger}\right)=\tilde{\rho}\,,
\end{align}
provided $W_a$ is a unitary operator. This implies that any two mixed-states $\rho$ and $\rho'$ that belong to the same mixed-state SPT phase protected by strong $\mathcal{S}$ and weak invertible $W$ symmetry should have a symmetric locality-preserving interpolation between them. In other words, if there is no symmetric \red{locality-preserving} interpolation between the two mixed-state SPTs $\rho$ and $\rho'$, then they should be in different phases protected by strong $\mathcal{S}$ and weak invertible $W$ symmetry. The interpolation between the two states includes the interfaces between them. In summary,
\begin{tcolorbox}[
    colback=blue!5!white, % Light blue background
    colframe=blue!75!black, % Dark blue border
    arc=2mm,              % Rounded corners
    boxrule=0.5mm,        % Border thickness
    width=\linewidth      % Forces box to fit the column exactly
]$\rho$ and $\rho'$ are in the same mixed-state SPT phase with strong $\mathcal{S}$ and weak $W$$\Rightarrow$ Symmetric locality-preserving interpolation between $\rho$ and $\rho'$.
\end{tcolorbox}
\begin{figure*}
    \centering
    \input{interfacerhorhop.tikz}
    \caption{Illustration of the unitary $U$ inside the $\text{Tr}_a$ for symmetric finite depth local channel. The black and red dots represent physical and ancilla qubits, respectively. Green boxes represent the symmetric local unitary gates. Shaded region in the bottom line represent the region $B$ under consideration. $(a)$ represent a finite depth local unitary inside the $\text{Tr}_a$ in \eqref{eq:unitaryinsidetrace}. At $t=T$, we have the state $\rho'$. $(b)$ represent the finite depth local unitary inside the $\text{Tr}_a$ in \eqref{eq:unitaryinsidetraceB}. At $t=T$, we get the state $\tilde{\rho}$, which is an interpolation between $\rho$ and $\rho'$ (in the shaded region).}
    \label{fig:interfacerhorhop}
\end{figure*}
\section{Interfaces between different phases}\label{sec:Interfaceanalysis}
In this section, we study the interface between two different ASPTs. First, we consider the interface between the trivial state and the doubled ASPT state $\rho_{\text{1D-DASPT}}^{\rm A}$ obtained from tracing out the 2D cluster state. Then, we study the interface between $\rho_{\text{1D-DASPT}}^{\rm A}$ and $\rho_{\text{blue}}^{\rm A}$ obtained from tracing out the 2D cluster state and 2D blue state respectively. Finally, we investigate the interface between $\rho^{\rm A}_{\text{1D-DASPT}}$ and $\rho^{\rm A}_{\text{blue}^{k_0;k_1}}$ obatined from tracing out the 2D cluster state and 2D blue$^{k_0;k_1}$ state respectively. In Appendix~\ref{app:interfacetwoDASPTs}, we provide the analysis for the interface between $\rho^{\rm A}_{\text{1D-DASPT}}$ and another similar-looking DASPT. In Appendix~\ref{app:interfaceZ3SPT}, we perform an interface analysis between two different mixed-state SPTs protected by $\mathbb{Z}_3^{(s)}\times \mathbb{Z}_3^{(w)}$.

\subsection{Interface between trivial and $\rho^{\rm A}_{\text{1D-DASPT}}$ states: different phases w.r.t. invertible symmetries}
\begin{figure*}
    \centering
    \begin{tikzpicture}[auto,
    redBall/.style={circle, ball color = red,
    radius=0.1em},blueBall/.style={circle, ball color = blue,
    radius=0.4em}, decoration={markings,
  mark=between positions 0 and 1 step 6pt
  with { \draw [fill] (0,0) circle [radius=0.8pt];}}]
   \node[] (b00) at (0,0) {};
     \node[] (r00) at (0.63,0.63) {};
     \filldraw[green, very thick] (0,-1) rectangle (13,1.5);
     \node[] (tr) at (-1,0) {Tr($\cdots$)};
     \draw[->,thick] (tr)--(0,0);
  \foreach \x [remember=\x as \lastx (initially 0)] in {1,...,9}{\node [draw, blueBall, right =of b\lastx0] (b\x0) {}; \foreach \y [remember=\y as \lasty (initially 0)] in {1,...,2}
    {\node [draw, blueBall, above =of b\x\lasty] (b\x\y){};
    }
}

\foreach \x [remember=\x as \lastx (initially 0)] in {1,...,8}{\node [draw, redBall, right =of r\lastx0] (r\x0) {}; \foreach \y [remember=\y as \lasty (initially 0)] in {1,...,1}
    {\node [draw, redBall, above =of r\x\lasty] (r\x\y){};
    }
}
\draw[thick,green] (b12)--(r11)--(b22)--(r21)--(b32)--(r31)--(b42)--(r41)--(b52)--(r51)--(b62)--(r61)--(b72)--(r71)--(b82)--(r81)--(b92);
\node[left=of b12] (b02) {$y=k_0+\frac{1}{2}$};
\node[left=of r11] (r01) {$y=k_0$};
\draw[->,thick] (b02)--(b12);
\draw[->,thick] (r01)--(r11);
\node[above=of b32] (b33) {};
\node[below=of b30] (b31m) {$x=l+\frac{1}{2}$};
\node[above=of b72] (b73) {};
\node[below=of b70] (b71m) {$x=L_x+\frac{1}{2}$};
\draw[thick,brown] (b31m)--(b33);
\draw[thick,brown] (b71m)--(b73);
\node[above=of b52] (b53) {$\ket{\Psi_{\rm I}}$};
\node[above=of b22] (b23) {$\ket{\Psi_{\rm II}}$};
\node[above=of b82] (b83) {$\ket{\Psi_{\rm II}}$};
\node[] (t0) at (1.2,3) {$\ket{+}$};
\node[] (t1) at (2.6,3) {$\ket{+}$};
\node[] (t2) at (4,3) {$\ket{+}$};
\node[] (t3) at (5.3,3) {$\ket{+}$};
\node[] (t4) at (6.6,3) {$\ket{+}$};
\node[] (t5) at (8,3) {$\ket{+}$};
\node[] (t6) at (9.3,3) {$\ket{+}$};
\node[] (t7) at (10.6,3) {$\ket{+}$};
\node[] (t8) at (11.9,3) {$\ket{+}$};
\end{tikzpicture}
    \caption{Interface between two 2D states $\ket{\Psi_{\rm I}}$ and $\ket{\Psi_{\rm II}}$. We impose Dirichlet boundary conditions with the boundary state to be the product state $\ket{+}^{\Delta_{v_b}\cap \rm A}$. After tracing out the bulk d.o.f. inside the green region, we get interfaces between the corresponding 1D states. The left vertical line is at $x=l+\frac{1}{2}$ and the right vertical line is at $L_x+\frac{1}{2}$.}
    \label{fig:Interface-cluster-blue}
\end{figure*}
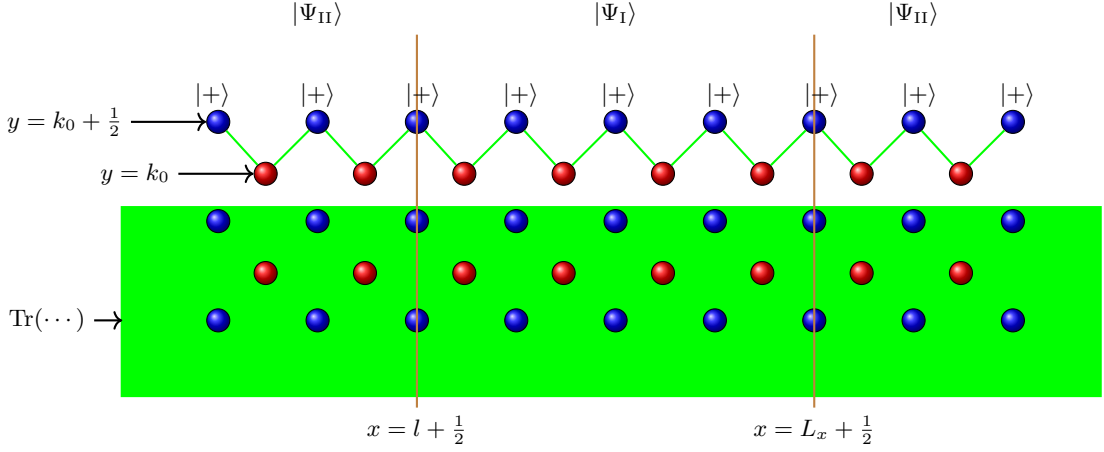

Let us consider the interface between the 2D trivial state and the 2D cluster state, subject to the open boundary condition along the $y$ direction with boundary at $y=k_0+\frac{1}{2}$. As before, all the vertex d.o.f. at $y=k_0+\frac{1}{2}$ is set to be $\ket{+}$ (see Figure~\ref{fig:Interface-cluster-blue}). Additionally, we set the blue vertex d.o.f. on the interface lines in the 2D bulk to $\ket{0}$, which is one choice of the interface boundary condition in the 2D bulk.
In this setting, after tracing out all bulk degrees of freedom except for the layers at $y = k_0$ and $y = k_0 + \tfrac{1}{2}$, the resulting state is 
\begin{align}
    &\rho^{\rm A}_{\text{trivial}|\text{1D-DASPT}}\notag\\
    =&\prod_{i=1}^l\left(\frac{1+X_{i,k_0}}{2}\right)\prod_{i=1}^{l-1}\left(\frac{1+X_{i+\frac{1}{2},k_0+\frac{1}{2}}}{2}\right)\notag\\
    &\prod_{i=l+1}^{L_x-1}\left(\frac{1+Z_{i,k_0}X_{i+\frac{1}{2},k_0+\frac{1}{2}}Z_{i+1,k_0}}{2}\right)\label{eq:trivial|1D-DASPT1b}\\
    &\left(\frac{1+X_{l+\frac{1}{2},k_0+\frac{1}{2}}Z_{l+1,k_0}}{2}\right)\left(\frac{1+Z_{L_x,k_0}X_{L_x+\frac{1}{2},k_0+\frac{1}{2}}}{2}\right)\notag\\
    &\left(\frac{1+Z_{l+\frac{1}{2},k_0+\frac{1}{2}}\prod_{i=l+1}^{L_x}X_{i,k_0}\,Z_{L_x+\frac{1}{2},k_0+\frac{1}{2}}}{2^{L_x-l}}\right)\notag.
\end{align}

This state explicitly breaks strong  $\mathbb{Z}_2^{(r)}$ symmetry to nothing but preserves the strong $\mathbb{Z}_2^{(b)}$ symmetry.
This is expected as the trivial state and the 1D-DASPT state are in different phases w.r.t. the invertible symmetries. 

Alternatively, we can consider the interface state purely from the 1D perspective, i.e., by connecting together the 1D density matrices for the trivial and DASPT phases at the interface without adding additional interface terms. 
\begin{align}
&\rho_{\text{trivial}|\text{1D-DASPT}}\notag\\=&\prod_{i=1}^l\left(\frac{1+X_{i,k_0}}{2}\right)\prod_{i=1}^{l-1}\left(\frac{1+X_{i+\frac{1}{2},k_0+\frac{1}{2}}}{2}\right)\notag\\
    &\prod_{i=l+1}^{L_x-1}\left(\frac{1+Z_{i,k_0}X_{i+\frac{1}{2},k_0+\frac{1}{2}}Z_{i+1,k_0}}{2}\right)\\
    &\left(\frac{1+Z_{l+\frac{1}{2},k_0+\frac{1}{2}}\prod_{i=l+1}^{L_x}X_{i,k_0}\,Z_{L_x+\frac{1}{2},k_0+\frac{1}{2}}}{2^{L_x-l}}\right)\,.\notag
    \label{eq:trivial|1D-DASPT2}
\end{align}
If we act with the $\mathbb{Z}_2^{(b)}$ generator $\prod_{i}X_{i+\frac{1}{2},k_0+\frac{1}{2}}$ on this state, we have
\begin{align}
    &\prod_{i}X_{i+\frac{1}{2},k_0+\frac{1}{2}}\rho_{\text{trivial}|\text{1D-DASPT}}\notag\\
    =&X_{l+\frac{1}{2},k_0+\frac{1}{2}}Z_{l+1,k_0}Z_{L_x,k_0}X_{L_x+\frac{1}{2},k_0+\frac{1}{2}}\rho_{\text{trivial}|\text{1D-DASPT}}\,.
\end{align}
To make the interface state strongly symmetric under the $\mathbb{Z}_2^{(b)}$ symmetry, we need to add the projector $X_{l+\frac{1}{2},k_0+\frac{1}{2}}Z_{l+1,k_0}=\pm 1$ and $Z_{L_x,k_0}X_{L_x+\frac{1}{2},k_0+\frac{1}{2}}=\pm 1$. If we choose the sign of these projectors to be both $+1$, then we recover the state~\eqref{eq:trivial|1D-DASPT1b}.  The other choices correspond to tuning one or both boundary blue vertices at the two vertical lines in Figure~\ref{fig:Interface-cluster-blue} to the state $\ket{-}$.  However, any such choice would explicitly break the $\mathbb{Z}_2^{(r)}$ symmetry ($\prod_iX_{i,k_0}$) to nothing. One could repeat this analysis for different choices of interface cuts (see Appendix~\ref{sec:otherinterfacecuts} for analysis on other choices of interface cuts). We find that it is not possible to construct an interface state that preserves all the symmetries. So any state that serves as an interface between the two states needs to explicitly break some symmetry. This is expected as the two states are in different mixed-state phases w.r.t. the invertible symmetries. The interface is therefore a useful tool to diagnose different phases in this context.

\subsection{Interface between $\rho^{\rm A}_{\text{1D-DASPT}}$ and $\rho^{\rm A}_{\text{blue}}$ states: same phase w.r.t. invertible symmetries}\label{sec:1D-DASPT|blue}
Now let us consider an interface between the 2D cluster state and the 2D blue state, subject to open boundary conditions along the $y$ direction. As in the previous case, the d.o.f. at $y=k_0+\frac{1}{2}$ is set to $\ket{+}$ (see Figure~\ref{fig:Interface-cluster-blue}). We also set the d.o.f on the blue vertices along the two bulk interface lines to the $\ket{+}$ state. Furthermore, we decorate the two interfaces with additional $\prod_{\substack{v_b=(l+\frac{1}{2},k+\frac{1}{2})\\k<k_0}}CZ_{v_b,v_b+(0,1)}$ and $\prod_{\substack{v_b=(L_x+\frac{1}{2},k+\frac{1}{2})\\k<k_0}}CZ_{v_b,v_b+(0,1)}$. This is allowed as we are not modifying the bulk of the cluster and the blue states. This additional decoration, as we shall see, in fact gives a symmetric interface state after we trace out the 2D bulk to produce a 1D mixed-state. In this setting, after tracing out all bulk degrees of freedom except for the layers at $y = k_0$ and $y = k_0 + \tfrac{1}{2}$, the resulting state is
{\small
\begin{align}
    &\rho_{\text{1D-DASPT}|\text{blue}}^{\rm A}\nonumber\\
    &=\prod_{1\leq i\leq l-1}\left(\frac{1+Z_{i,k_0}X_{i+\frac{1}{2},k_0+\frac{1}{2}}Z_{i+1,k_0}}{2}\right)\nonumber\\
    &\prod_{l+1\leq i\leq L_x-1}\left(\frac{1-Z_{i-\frac{1}{2},k_0+\frac{1}{2}}Y_{i,k_0}X_{i+\frac{1}{2},k_0+\frac{1}{2}}Y_{i+1,k_0}Z_{i+\frac{3}{2},k_0+\frac{1}{2}}}{2}\right)\nonumber\\
    &\left(\frac{1-Z_{l,k_0}Y_{l+\frac{1}{2},k_0+\frac{1}{2}}Y_{l+1,k_0}Z_{l+\frac{3}{2},k_0+\frac{1}{2}}}{2}\right)\nonumber\\
    &\left(\frac{1-Z_{L_x-\frac{1}{2},k_0+\frac{1}{2}}Y_{L_x,k_0}Y_{L_x+\frac{1}{2},k_0+\frac{1}{2}}Z_{1,k_0}}{2}\right)\left(\frac{1+\prod_{i}X_{i,k_0}}{2^{|\Delta_{v_r}\cap \rm A|}}\right)\,.
    \label{eq:1D-DASPT|blue(1)}
\end{align}}
This state has strong $\mathbb{Z}_2^{(r)}\times \mathbb{Z}_2^{(b)}$ and strong $\mathrm{D}^{(1)}$ symmetry. Recall that, individually, both states we get after tracing out the 2D cluster state and the 2D blue state have an emergent strong $\mathrm{D}^{(1)}$ symmetry.  After we insert the above interface between these two phases, the strong $\mathrm{D}^{(1)}$ symmetry is still maintained in this case. A summary of the features of the state $\rho_{\text{1D-DASPT}|\text{blue}}^{\rm A}$ is given in Table~\ref{tab:clusterblueinterface}.

Now suppose we consider the interface between $\rho_{\text{1D-DASPT}}^{\rm A}$ and $\rho_{\text{blue}}^{\rm A}$ without adding any additional terms on the interface, namely
{\small
\begin{align}
    &\rho_{\text{1D-DASPT}|\text{blue}}\nonumber\\
    =&\prod_{1\leq i\leq l-1}\left(\frac{1+Z_{i,k_0}X_{i+\frac{1}{2},k_0+\frac{1}{2}}Z_{i+1,k_0}}{2}\right)\nonumber\\
    &\hspace{-0.6cm}\prod_{l+1\leq i\leq L_x-1}\left(\frac{1-Z_{i-\frac{1}{2},k_0+\frac{1}{2}}Y_{i,k_0}X_{i+\frac{1}{2},k_0+\frac{1}{2}}Y_{i+1,k_0}Z_{i+\frac{3}{2},k_0+\frac{1}{2}}}{2}\right)\nonumber\\
    &\times\left(\frac{1+\prod_{i}X_{i,k_0}}{2^{|\Delta_{v_r}\cap \rm A|}}\right)\,.
    \label{eq:1D-DASPT|blue(2)}
\end{align}
}
In this state, we can act with the strong symmetry 
\begin{subequations}
\begin{align}
    &\prod_i X_{i,k_0}\,\rho_{\text{1D-DASPT}|\text{blue}}=\rho_{\text{1D-DASPT}|\text{blue}}\,,\\
    &\prod_{i}X_{i+\frac{1}{2},k_0+\frac{1}{2}}\rho_{\text{1D-DASPT}|\text{blue}}&\nonumber\\
    &=\mathbb{X}_{l+\frac{1}{2},k_0+\frac{1}{2}}\mathbb{X}_{L_x+\frac{1}{2},k_0+\frac{1}{2}}\rho_{\text{1D-DASPT}|\text{blue}}\,,
\end{align}
\end{subequations}
where 
\begin{align}
  \mathbb{X}_{l+\frac{1}{2},k_0+\frac{1}{2}}&\equiv Z_{l,k_0}Y_{l+\frac{1}{2},k_0+\frac{1}{2}}Y_{l+1,k_0}Z_{l+\frac{3}{2},k_0+\frac{1}{2}}\,,\nonumber\\
  \mathbb{X}_{L_x+\frac{1}{2},k_0+\frac{1}{2}}&\equiv Z_{L_x-\frac{1}{2},k_0+\frac{1}{2}}Y_{L_x,k_0}Y_{L_x+\frac{1}{2},k_0+\frac{1}{2}}Z_{1,k_0}\,.
\end{align}
To make the state $\rho_{\text{1D-DASPT}|\text{blue}}$ strongly symmetric under the invertible symmetries we need to add the projectors $(1\pm\mathbb{X}_{l+\frac{1}{2},k_0+\frac{1}{2}})/2$ and $(1\pm\mathbb{X}_{L_x+\frac{1}{2},k_0+\frac{1}{2}})/2$. We note that the additional projectors that appear in~\eqref{eq:1D-DASPT|blue(1)}, compared with ~\eqref{eq:1D-DASPT|blue(2)}, are exactly of the form $\mathbb{X}_{l+\frac{1}{2},k_0+\frac{1}{2}}=-1$ and $\mathbb{X}_{L_x+\frac{1}{2},k_0+\frac{1}{2}}
=-1$. The other sign choices come from modifying the states of the interface blue sites in Figure~\ref{fig:Interface-cluster-blue}. In any case, adding the projectors $(1-\mathbb{X}_{l+\frac{1}{2},k_0+\frac{1}{2}})/2$ and $(1-\mathbb{X}_{L_x+\frac{1}{2},k_0+\frac{1}{2}})/2$ make the interface state strongly symmetric under $\mathrm{D}^{(1)}$. For other choices of interface cuts, we can again find a symmetric interface, and leave the details to the  Appendix~\ref{app:Otherinterfacecuts2}. This maybe indicating that $\rho^{\rm A}_{\text{blue}}$ is in the same phase as that of $\rho^{\rm A}_{\text{1D-DASPT}}$ if both the strong $\mathbb{Z}_2^{(r)}\times\mathbb{Z}_2^{(b)}$ symmetries and the non-invertible $\mathrm{D}^{(1)}$ symmetry are considered. We leave further analysis for the future exploration.

\subsection{Interface between $\rho^{\rm A}_{\text{1D-DASPT}}$ and $\rho^{\rm A}_{\text{blue}^{k_0;k_1}}$ states: same phase w.r.t invertible symmetries}\label{sec:1D-DASPT|bluek0k1}
Now let us consider an interface between the 2D cluster state and the 2D blue$^{k_0;k_1}$ state, subject to open boundary conditions along the $y$ direction. As in the previous case, the d.o.f. at $y=k_0+\frac{1}{2}$ are set to $\ket{+}$ \red{except at $x=l+\frac{1}{2}$ and $x=L_x+\frac{1}{2}$. At $(x,y)=(l+\frac{1}{2},k_0+\frac{1}{2})$, we set it to $\exp(i\frac{\pi}{4}Z_{l+\frac{1}{2},k_0+\frac{1}{2}})\ket{+}\sim \ket{0}-i\,\ket{1}$, and at $(x,y)=(L_x+\frac{1}{2},k_0+\frac{1}{2})$, we set it to $\exp(i\frac{\pi}{4}(-1)^{L_x-l-1}Z_{L_x+\frac{1}{2},k_0+\frac{1}{2}})\ket{+}\sim \ket{0}\pm i\, \ket{1}$}. Like in the previous section, we also set the d.o.f on the blue vertices along the two bulk interface lines to the $\ket{+}$ state. Furthermore, we decorate the two interfaces with additional $\prod_{\substack{v_b=(l+\frac{1}{2},k+\frac{1}{2})\\k<k_0}}CZ_{v_b,v_b+(0,1)}$ and $\prod_{\substack{v_b=(L_x+\frac{1}{2},k+\frac{1}{2})\\k<k_0}}CZ_{v_b,v_b+(0,1)}$. In this setting, after tracing out all bulk degrees of freedom except for the layers at $y = k_0$ and $y = k_0 + \tfrac{1}{2}$, the resulting state is
{\small
\begin{align}
    &\rho_{\text{1D-DASPT}|\text{blue}^{k_0;k_1}}^{\rm A}\nonumber\\
    =&\prod_{1\leq i\leq l-1}\left(\frac{1+Z_{i,k_0}X_{i+\frac{1}{2},k_0+\frac{1}{2}}Z_{i+1,k_0}}{2}\right)\nonumber\\
    &\prod_{l+1\leq i\leq L_x-1}\left(\frac{1-Y_{i,k_0}X_{i+\frac{1}{2},k_0+\frac{1}{2}}Y_{i+1,k_0}}{2}\right)\nonumber\\
    &\qquad\left(\frac{1-Z_{l,k_0}X_{l+\frac{1}{2},k_0+\frac{1}{2}}Y_{l+1,k_0}}{2}\right)\nonumber\\
    &\left(\frac{1-\red{(-1)^{L_x-l-1}}Y_{L_x,k_0}X_{L_x+\frac{1}{2},k_0+\frac{1}{2}}Z_{1,k_0}}{2}\right)\nonumber\\
    &\times\left(\frac{1+\prod_{i}X_{i,k_0}}{2^{|\Delta_{v_r}\cap \rm A|}}\right)\,.
    \label{eq:1D-DASPT|blue(1)k0k1}
\end{align}
}
This state has strong $\mathbb{Z}_2^{(r)}\times \mathbb{Z}_2^{(b)}$  symmetry. \red{Note that, if we do not change the interface qubits from $|+\rangle$ to $\ket{0}\pm i\, \ket{1}$, then \eqref{eq:1D-DASPT|blue(1)k0k1} would not be symmetric under $\mathbb{Z}_2^{(r)}\times \mathbb{Z}_2^{(b)}$ as the $X$ operators at locations $(l+\frac{1}{2},k_0+\frac{1}{2})$ and $(L_x+\frac{1}{2},k_0+\frac{1}{2})$ would be replaced be $Y$}. A summary of the features of the state $\rho_{\text{1D-DASPT}|\text{blue}^{k_0;k_1}}^{\rm A}$ is given in Table~\ref{tab:clusterbluek0;k1interface}.

Now suppose we consider the interface between $\rho_{\text{1D-DASPT}}^{\rm A}$ and $\rho_{\text{blue}^{k_0;k_1}}^{\rm A}$ without adding any additional terms on the interface
\begin{align}
    &\rho_{\text{1D-DASPT}|\text{blue}^{k_0;k_1}}\nonumber\\
    =&\prod_{1\leq i\leq l-1}\left(\frac{1+Z_{i,k_0}X_{i+\frac{1}{2},k_0+\frac{1}{2}}Z_{i+1,k_0}}{2}\right)\nonumber\\
    &\prod_{l+1\leq i\leq L_x-1}\left(\frac{1-Y_{i,k_0}X_{i+\frac{1}{2},k_0+\frac{1}{2}}Y_{i+1,k_0}}{2}\right)\nonumber\\
    &\times\left(\frac{1+\prod_{i}X_{i,k_0}}{2^{|\Delta_{v_r}\cap \rm A|}}\right)\,.
    \label{eq:1D-DASPT|blue(2)k0k1}
\end{align}
On this state, we can act with the strong symmetry 
\begin{subequations}
\begin{align}
    &\prod_i X_{i,k_0}\,\rho_{\text{1D-DASPT}|\text{blue}^{k_0;k_1}}\nonumber\\
    &=\rho_{\text{1D-DASPT}|\text{blue}^{k_0;k_1}}\,,\\
    &\prod_{i}X_{i+\frac{1}{2},k_0+\frac{1}{2}}\rho_{\text{1D-DASPT}|\text{blue}^{k_0;k_1}}\nonumber\\
    &=\red{(-1)^{L_x-l-1}}\,\mathbb{X}_{l+\frac{1}{2},k_0+\frac{1}{2}}\mathbb{X}_{L_x+\frac{1}{2},k_0+\frac{1}{2}}\rho_{\text{1D-DASPT}|\text{blue}^{k_0;k_1}}\,,
\end{align}
\end{subequations}
where 
\begin{align}
  \mathbb{X}_{l+\frac{1}{2},k_0+\frac{1}{2}}&\equiv Z_{l,k_0}X_{l+\frac{1}{2},k_0+\frac{1}{2}}Y_{l+1,k_0}\,,\nonumber\\
  \mathbb{X}_{L_x+\frac{1}{2},k_0+\frac{1}{2}}&\equiv Y_{L_x,k_0}X_{L_x+\frac{1}{2},k_0+\frac{1}{2}}Z_{1,k_0}\,.
\end{align}
To make the state $\rho_{\text{1D-DASPT}|\text{blue}^{k_0;k_1}}$ strongly symmetric under the $\mathbb{Z}_2^{(r)}\times \mathbb{Z}_2^{(b)}$ symmetry, we can add the projectors $(1\pm \mathbb{X}_{l+\frac{1}{2},k_0+\frac{1}{2}})/2$ and $(1\pm \red{(-1)^{L_x-l-1}}\mathbb{X}_{L_x+\frac{1}{2},k_0+\frac{1}{2}})/2$. \red{From the previous discussion, we know that $\rho^{\rm A}_{\text{1D-DASPT}}$ and $\rho^{\rm A}_{\text{blue}^{k_0,k_1}}$ are in the same phase as far as the $\mathbb{Z}_2^{(r)}\times \mathbb{Z}_2^{(b)}$ strong symmetries are considered}. However, these two \red{states} are distinguished by the non-invertible $\mathrm{D}^{(1)}$ symmetry, since $\rho^{\rm A}_{\text{1D-DASPT}}$ has strong $\mathrm{D}^{(1)}$ symmetry while $\rho^{\rm A}_{\text{blue}^{k_0;k_1}}$ is not symmetric under $\mathrm{D}^{(1)}$. We leave the analysis of other interface cuts to the Appendix~\ref{app:Otherinterfacecuts3}.
\section{Conclusion}\label{sec:conclusion}

In this work, we have investigated a variety of 1D mixed-state phases obtained by tracing out the 2D bulk degrees of freedom.
\red{For the non-invertible higher-order SSPT examples and boundary conditions studied here, tracing
out the bulk yields DASPT states}, 
%When the bulk is a 2D higher-order SSPT, we obtain 1D DASPTs, 
which display a coexistence of mSPT order and SWSSB. A numerical study of DASPTs was carried out in~\cite{Guo2025strong,Guo2025strong}. In the specific examples we consider, these phases are characterized by a nonzero R\'enyi-2 (or fidelity) correlator signaling SWSSB, and a nonzero expectation value of string order parameters, indicating mixed-state SPT order. We note that the existence of both SPT and SSB order in pure states has been explored in~\cite{Verresen:2017cwj,Zeng:2014qjk}, and in non-equilibrium settings in~\cite{Morral-Yepes:2023vgo}.

To characterize and distinguish these 1D mSPT phases, we employed interfaces as diagnostic probes. We showed by explicit examples that when two mixed-states belong to the same phase, there exists an interface interpolation between them that preserves all the symmetries under consideration. When there is no local modification to preserve the symmetries
across the interface, the two sides are in distinct phases.

Furthermore, we found that there exists an interface interpolation between $\rho^{\rm A}_{\text{1D-DASPT}}$ and $\rho^{\rm A}_{\text{blue}}$ that is symmetric under the strong $\mathbb{Z}_2^{(r)} \times \mathbb{Z}_2^{(b)}$ symmetry as well as the strong non-invertible $\mathrm{D}^{(1)}$ symmetry. \red{We also explicitly constructed a finite-depth symmetric local channel (which is also a finite-depth circuit) connecting $\rho^{\rm A}_{\text{1D-DASPT}}$ and $\rho^{\rm A}_{\text{blue}}$.}

An important future direction is to establish, on general grounds, that interfaces provide a universal and robust probe for distinguishing mSPT phases. In this manuscript, we presented several concrete examples demonstrating how interfaces can be used to differentiate between distinct mSPT phases. It would be particularly valuable to connect this interface-based diagnostic to the formal definition of mSPT phases in terms of two-way symmetric finite-depth local quantum channels.
% This observation motivates an interesting open problem: if they are indeed in the same phase, how could we explicitly construct a symmetric \red{(also w.r.t $\rm D^{(1)})$} finite-depth local quantum channel that connects $\rho_{\text{1D-DASPT}}$ and $\rho_{\text{blue}}$? 
% Such a construction would provide a concrete realization of the equivalence between these phases and further clarify the role of the interface as a probe to distinguish between mSPT phases.\\

{\it Note:} While preparing this manuscript, a work appeared on the arXiv presenting a systematic understanding of 1D mixed-state phases with $\mathbb{Z}_2 \times \mathbb{Z}_2$ symmetry from an alternative purification perspective~\cite{Guo:2026dxx}.

\section*{Acknowledgment}

The authors thank Arkya Chatterjee, Yabo Li, Ruochen Ma, Shijun Sun, and Jian-Hao Zhang for helpful discussions. FY
thanks the Simons Center for Geometry and Physics at
Stony Brook University for great hospitality during various stages of this work. 
This work was supported by the U.S. Department of Energy, Office of Basic Energy Sciences, under Contract No. DE-SC0012704 (APM, FY \& TCW) and the National Science Foundation under Grant No. DRL I-TEST 2148467 (ZS). 

\bibliography{Ref}

\onecolumngrid
\appendix
\section{Extra examples}
\subsection{Mixed-state from 2D cluster state}\label{app:Extraexamples1}
In this section, we look at the states obtained by tracing out the 2D cluster state on a torus and on a cylinder with generic boundary conditions. The results are summarized in Table~\ref{tab:clusterstatetrace}.
\subsubsection{2D cluster state on a torus}
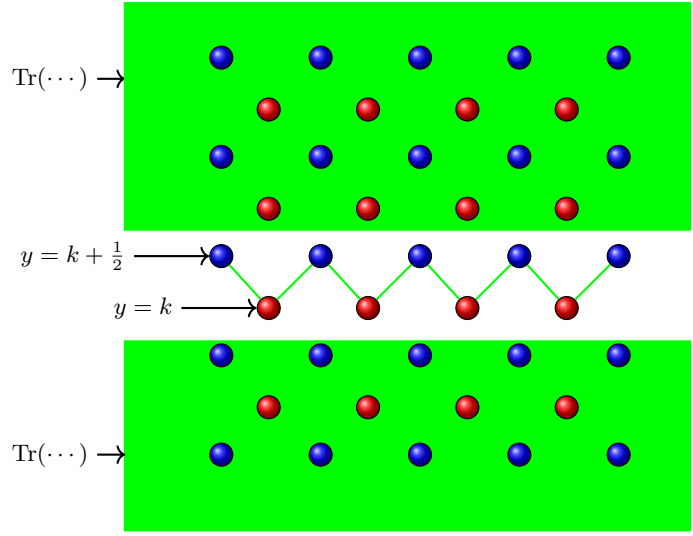
\begin{figure*}[ht]
    \centering
    \begin{tikzpicture}[auto,
    redBall/.style={circle, ball color = red,
    radius=0.1em},blueBall/.style={circle, ball color = blue,
    radius=0.4em}, decoration={markings,
  mark=between positions 0 and 1 step 6pt
  with { \draw [fill] (0,0) circle [radius=0.8pt];}}]
   \node[] (b00) at (0,0) {};
     \node[] (r00) at (0.63,0.63) {};
     \filldraw[green, very thick] (0,-1) rectangle (7.5,1.5);
     \node[] (tr) at (-1,0) {Tr($\cdots$)};
     \draw[->,thick] (tr)--(0,0);
     \node[] (tr2) at (-1,5) {Tr($\cdots$)};
     \draw[->,thick] (tr2)--(0,5);
     \filldraw[green, very thick] (0,3) rectangle (7.5,6);
  \foreach \x [remember=\x as \lastx (initially 0)] in {1,...,5}{\node [draw, blueBall, right =of b\lastx0] (b\x0) {}; \foreach \y [remember=\y as \lasty (initially 0)] in {1,...,4}
    {\node [draw, blueBall, above =of b\x\lasty] (b\x\y){};
    }
}

\foreach \x [remember=\x as \lastx (initially 0)] in {1,...,4}{\node [draw, redBall, right =of r\lastx0] (r\x0) {}; \foreach \y [remember=\y as \lasty (initially 0)] in {1,...,3}
    {\node [draw, redBall, above =of r\x\lasty] (r\x\y){};
    }
}
\draw[thick,green] (b12)--(r11)--(b22)--(r21)--(b32)--(r31)--(b42)--(r41)--(b52);
\node[left=of b12] (b02) {$y=k+\frac{1}{2}$};
\node[left=of r11] (r01) {$y=k$};
\draw[->,thick] (b02)--(b12);
\draw[->,thick] (r01)--(r11);
\end{tikzpicture}
    \caption{The vertices in the region $\rm A$ are at $y=k$ and $y=k+\frac{1}{2}$. The green line connecting red and blue vertices in region $\rm A$ indicates the $CZ$ gates that remain after tracing out $\rm A^c$.  The green region contains the d.o.f. that we trace over.}
    \label{fig:regionA}
\end{figure*}

We assume periodic boundary conditions for both the $x$ and $y$ directions for the cluster Hamiltonian. Now we trace out all the 2D d.o.f. except the rows with $y=k$ and $y=k+\frac{1}{2}$. Let us call this region $\rm A$ and the rest of the rows $\rm A^{c}$ (see Figure~\ref{fig:regionA}).
Tracing over the region $\rm A^c$ give

\begin{align}
    &\text{Tr}_{\rm A^c}\left(\ket{\text{2D-cluster}}\bra{\text{2D-cluster}}\right)\nonumber\\
    &=\frac{4}{2^{|(\Delta_{v_r}\cup\Delta_{v_b})\cap \mathrm{A}|}}\sum_{\substack{\{i_l\}\subset\Delta_{v_r}\cap \mathrm{A}\,,\quad\{j_l\}\subset\Delta_{v_b}\cap \mathrm{A}\\
    r,s \text{ even }}}Z_{i_1}...Z_{i_r}Z_{j_1}..Z_{j_s}\ket{\text{1D-cluster}}\bra{\text{1D-cluster}}Z_{i_1}...Z_{i_r}Z_{j_1}..Z_{j_s}\nonumber\\
    &=\frac{(1+\prod_{i}X_{i,k})}{2^{|\Delta_{v_r}\cap \mathrm{A}|}}\frac{(1+\prod_{i}X_{i+\frac{1}{2},k+\frac{1}{2}})}{2^{|\Delta_{v_b}\cap \mathrm{A}|}}\nonumber\\
    &\equiv \rho_{\text{1D-SWSSB}}^\mathrm{A}\, ,
    \label{eq:1D-SWSSB}
\end{align}

where $\{i_l\}$ and $\{j_l\}$ are subset of vertices in $\Delta_{v_r}\cap \mathrm{A}$ and $\Delta_{v_b}\cap \mathrm{A}$ respectively and the 1D cluster state is along the region $\mathrm{A}$
\begin{align}
    \ket{\text{1D-cluster}}=\prod_{i}CZ_{(i+\frac{1}{2},k+\frac{1}{2}),(i,k)}\red{CZ_{(i,k),(i+\frac{3}{2},k+\frac{1}{2})}}\ket{+}^{\otimes \Delta_{v_r}\cap \mathrm{A}}\ket{+}^{\otimes\Delta_{v_b}\cap \mathrm{A}}\, .
    \label{eq:1Dcluster}
\end{align}
We note that $\rho_{\text{1D-SWSSB}}^{\mathrm{A}}$ has strong $\mathbb{Z}_2^{(r)}\times\mathbb{Z}_2^{(b)}$ 0-form symmetry generated by $\eta^x_{r,k}$ and $\eta^x_{b,k}$, i.e,
\begin{subequations}
    \begin{align}
    \eta^x_{r,k}\rho_{\text{1D-SWSSB}}^{\mathrm{A}}&=\rho_{\text{1D-SWSSB}}^{\mathrm{A}}\,,\\ \eta^x_{b,k}\rho_{\text{1D-SWSSB}}^{\mathrm{A}}&=\rho_{\text{1D-SWSSB}}^{\mathrm{A}}\,.
\end{align}
\end{subequations}
This state has the following weak stabilizers
\begin{subequations}
   \begin{align}
    &Z_{i,k}X_{i+\frac{1}{2},k+\frac{1}{2}}Z_{i+1,k}\rho_{\text{1D-SWSSB}}^{\rm A}Z_{i,k}X_{i+\frac{1}{2},k+\frac{1}{2}}Z_{i+1,k}=\rho_{\text{1D-SWSSB}}^{\rm A}\,,\\
    &Z_{i-\frac{1}{2},k+\frac{1}{2}}X_{i,k}Z_{i+\frac{1}{2},k+\frac{1}{2}}\rho_{\text{1D-SWSSB}}^{\rm A}Z_{i-\frac{1}{2},k+\frac{1}{2}}X_{i,k}Z_{i+\frac{1}{2},k+\frac{1}{2}}=\rho_{\text{1D-SWSSB}}^{\rm A}\, .
\end{align} 
\end{subequations}
Interestingly, $\rho_{\text{1D-SWSSB}}^{\mathrm{A}}$ has weak non-invertible symmetry $\mathrm{D}^{(1)}$ defined in \eqref{eq:D1exp}.
Weak non-invertible symmetry satisfies
\begin{align}
    \mathrm{D}^{(1)}\rho_{\text{1D-SWSSB}}^{\rm A}(\mathrm{D}^{(1)})^{\dagger}=\rho_{\text{1D-SWSSB}}^{\rm A}\,.
\end{align}
We note that $Z_{i,k}$ and $Z_{i+\frac{1}{2},k+\frac{1}{2}}$ are good order parameters for the symmetries $\eta^x_{r,k}$ and $\eta^x_{b,k}$ respectively as they anti-commute with the symmetry operators. Using the fidelity correlator and R\'enyi-2 correlator defined in \cite{Lessa:2024gcw}, it can be easily verified that $\rho_{\text{1D-SWSSB}}^{\rm A}$ is SWSSB of the $\mathbb{Z}_2^{(r)}\times\mathbb{Z}_2^{(b)}$ symmetry. Concretely, for the R\'enyi-2 correlators we have:
\begin{align}
     &\frac{\text{Tr}\left(Z_{i,k}Z_{j,k}\rho_{\text{1D-SWSSB}}^{\rm A}Z_{i,k}Z_{j,k}\rho_{\text{1D-SWSSB}}^{\rm A}\right)}{\text{Tr}\left((\rho_{\text{1D-SWSSB}}^{\rm A})^2\right)}=1\,,\nonumber\\ &\frac{\text{Tr}\left(Z_{i+\frac{1}{2},k+\frac{1}{2}}Z_{j+\frac{1}{2},k+\frac{1}{2}}\rho_{\text{1D-SWSSB}}^{\rm A}Z_{i+\frac{1}{2},k+\frac{1}{2}}Z_{j+\frac{1}{2},k+\frac{1}{2}}\rho_{\text{1D-SWSSB}}^{\rm A}\right)}{\text{Tr}\left((\rho_{\text{1D-SWSSB}}^{\rm A})^2\right)}=1\,.
     \label{eq:Renyi2rho1DSWSSB}
     \end{align}
while for fidelity correlators we have:
     \begin{align}
     &\text{Tr}\left(\sqrt{\sqrt{\rho_{\text{1D-SWSSB}}^{\rm A}}Z_{i,k}Z_{j,k}\rho_{\text{1D-SWSSB}}^{\rm A}Z_{i,k}Z_{j,k}\sqrt{\rho_{\text{1D-SWSSB}}^{\rm A}}}\right)=1\,,\nonumber\\
     &\text{Tr}\left(\sqrt{\sqrt{\rho_{\text{1D-SWSSB}}^{\rm A}}Z_{i+\frac{1}{2},k+\frac{1}{2}}Z_{j+\frac{1}{2},k+\frac{1}{2}}\rho_{\text{1D-SWSSB}}^{\rm A}Z_{i+\frac{1}{2},k+\frac{1}{2}}Z_{j+\frac{1}{2},k+\frac{1}{2}}\sqrt{\rho_{\text{1D-SWSSB}}^{\rm A}}}\right)=1\,.
     \label{eq:fidelityrho1DSWSSB}
 \end{align}

\subsubsection{2D cluster state on a semi-infinite cylinder with generic boundary condition}

We consider the Hamiltonian $\mathrm{H}_{\text{2D-cluster}}$ on a cylinder, i.e, open boundary condition on the $y$ direction with top boundary ending at the row of blue vertices. The $y$ coordinate takes values from $-\infty$ to $1...,k_0,k_0+\frac{1}{2}$. Now we trace out all the 2D d.o.f. except for the top rows at $y=k_0$ and $y=k_0+\frac{1}{2}$ (we call the top rows again region $\rm A$). Tracing out $\rm A^c$ gives

\begin{align}
    &\qquad\text{Tr}_{\rm A^c}\left(\ket{\text{2D-cluster}}\bra{\text{2D-cluster}}\right)\nonumber\\&=\frac{2}{2^{|\Delta_{v_r}\cap \mathrm{A}|}}\sum_{\substack{\{i_l\}\in\Delta_{v_r}\cap \mathrm{A}\,,\\
    r \text{ even }}}Z_{i_1}...Z_{i_r}\prod CZ_{v_r,v_b}\ket{+}^{\otimes\Delta_{v_r}}\ket{\phi}^{\otimes\Delta_{v_b}}\bra{\phi}^{\otimes\Delta_{v_b}}\bra{+}^{\otimes\Delta_{v_r}}\prod CZ_{v_r,v_b} Z_{i_1}...Z_{i_r}\nonumber\\
    &\equiv \rho_{\text{1D-SWSSBr}}^{\mathrm{A}}\, ,
    \label{eq:1D-DASPT}
\end{align}
where the $\ket{\phi}^{\otimes\Delta_{v_b}}$ are dangling d.o.fs. The above state has strong $\mathbb{Z}_2^{(r)}=\prod_{i}X_{i,k_0}$ 0-form symmetry. In addition, we have the weak stabilizers
\begin{align}
    Z_{i-\frac{1}{2},k_0+\frac{1}{2}}X_{i,k_0}Z_{i+\frac{1}{2},k_0+\frac{1}{2}}\rho_{\text{1D-SWSSB}}^{\mathrm{A}}Z_{i-\frac{1}{2},k_0+\frac{1}{2}}X_{i,k_0}Z_{i+\frac{1}{2},k_0+\frac{1}{2}}=\rho_{\text{1D-SWSSB}}^{\mathrm{A}}\,.
\end{align}

$\rho_{\text{1D-SWSSB}}^{\rm A}$ is in the SWSSB phase regarding the $\mathbb{Z}_2^{(r)}$ symmetry. This can be checked using the fidelity correlator or the R\'enyi-2 correlator, both of which evaluate to 1 if we take $Z_{i,k_0}$ as the order parameter.
\subsubsection{Tracing the interface between 2D trivial state and 2D cluster state on a torus}
 Let us consider the two-dimensional state that constitutes an interface between the 2D trivial state and 2D cluster state, subject to periodic boundary conditions along the $y$ direction. We take the interface to lie along the lines $x=l+\frac{1}{2}$ and $x=L_x+\frac{1}{2}$. After tracing out the bulk degrees of freedom except for the layers at $y=k$ and $y=k+\frac{1}{2}$, the resulting reduced density matrix is
identical to the trivial state in one region and identical to $\rho_{\text{1D-SWSSB}}^{\rm A}$ in the other region. Equivalently, this can be thought of as $\rho_{\text{1D-SWSSB}}^{\rm A}$ with open boundary conditions. Otherwise, the strong $\mathbb{Z}_2^{(r)}$ symmetry is explicitly broken to weak symmetry and strong $\mathbb{Z}_2^{(b)}$ symmetry is broken to nothing unless we fine tune the state at $(x,y)=(l+\frac{1}{2},k+\frac{1}{2})$ and $(x,y)=(L_x+\frac{1}{2},k+\frac{1}{2})$. 
\subsubsection{Tracing the interface between 2D trivial state and 2D cluster state on a cylinder with generic open boundary condition}
Let us consider the two-dimensional state that constitutes an interface between the 2D trivial state and 2D cluster state, subject to an open boundary condition along the $y$ direction. We take the boundary to lie at $y=k_0$ and $y=k_0+\frac{1}{2}$. We take the interface to lie along the lines $x=l+\frac{1}{2}$ and $x=L_x+\frac{1}{2}$. After tracing out the bulk degrees of freedom except for the layers at $y=k_0$ and $y=k_0+\frac{1}{2}$, the resulting reduced density matrix is
identical to the trivial state in one region and identical to the $\rho_{\text{1D-SWSSB}_r}^{\rm A}$ in the other region. Equivalently, this can be thought of as $\rho_{\text{1D-SWSSB}_r}^{\rm A}$ with open boundary conditions. Otherwise, the strong $\mathbb{Z}_2^{(r)}$ symmetry is explicitly broken to a weak symmetry.
\subsection{Mixed-state from 2D blue state}\label{app:Extraexamples2}
In this section, we look at the states obtained by tracing out the 2D blue state on a torus and on a cylinder with generic boundary conditions. The results are summarized in Table~\ref{tab:bluestatetrace}.
\subsubsection{2D blue state on a torus}
In this section, we place the blue state on a torus and trace out all the bulk d.o.f. except at $y=k$ and $y=k+\frac{1}{2}$. As before, we denote these rows as $\rm A$ and the rest as $\rm A^c$. Tracing out $\rm A^c$ gives
    \begin{align}
    \text{Tr}_{\rm A^c}\left(\ket{\text{blue}}\bra{\text{blue}}\right)&=\frac{4}{2^{|\Delta_{v_b}\cap \mathrm{A}|+|\Delta_{v_r}\cap \mathrm{A}|}}\sum_{\substack{\{i_l\}\in\Delta_{v_r}\cap \mathrm{A}\,,\quad\{j_l\}\in\Delta_{v_b}\cap \mathrm{A}\\
    r,s \text{ even }}}Z_{i_1}...Z_{i_r}Z_{j_1}..Z_{j_s}\ket{\text{1D-cluster}}\bra{\text{1D-cluster}}Z_{i_1}...Z_{i_r}Z_{j_1}..Z_{j_s}\nonumber\\
    &= \rho_{\text{1D-SWSSB}}^{\mathrm{A}}\, ,
\end{align}
where the $\ket{\text{1D-cluster}}$ is given in \eqref{eq:1Dcluster}. Here, it seems that by tracing out the region $\rm A^c$, we lose information about the distinction between $\ket{\text{blue}}\bra{\text{blue}}$ and $\ket{\text{2D-cluster}}\bra{\text{2D-cluster}}$. Moreover, $\rho_{\text{1D-SWSSB}}^{\mathrm{A}}$, which appeared previously in~(\ref{eq:1D-SWSSB}), is in the  SWSSB phase regarding the $\mathbb{Z}_2^{(r)}\times\mathbb{Z}_2^{(b)}$ symmetries. We verified this by computing the R\'enyi-2 correlator and the fidelity correlators in \eqref{eq:Renyi2rho1DSWSSB} and \eqref{eq:fidelityrho1DSWSSB}.
\subsubsection{2D blue state on a semi-infinite cylinder with generic boundary condition}
We consider the Hamiltonian $\mathrm{H}_{\text{blue}}$ on a semi-infinite cylinder, i.e, with open boundary condition on the $y$ direction. Let the $y$ coordinate run from $-\infty$ to $...,k_0,k_0+\frac{1}{2}$ with the top boundary on blue vertices. Now we trace out all the 2D d.o.f. except the top row at $y=k_0$ and $y=k_0+\frac{1}{2}$ (we call the top rows again region $\rm A$). Tracing out $\rm A^c$ gives

\begin{subequations}
    \begin{align}
   &\text{Tr}_{\rm A^c}\left(\ket{\text{blue}}\bra{\text{blue}}\right)=\nonumber\\
    &\frac{2}{2^{|\Delta_{v_r}\cap \mathrm{A}|}}\sum_{\substack{\{i_l\}\in\Delta_{v_r}\cap \mathrm{A}\,,\\
    s \text{ even }}}\left(Z_{i_1}\prod_{j_1\in \text{Nb}(i_1)}Z_{j_1}\right)...\left(Z_{i_s}\prod_{j_s\in \text{Nb}(i_s)}Z_{j_s}\right)\prod CZ_{v_r,v_b}\ket{+}^{\otimes\Delta_{v_r}}\ket{\phi}^{\otimes\Delta_{v_b}}\\
    &\hspace{3cm}\bra{\phi}^{\otimes\Delta_{v_b}}\bra{+}^{\otimes\Delta_{v_r}}\prod CZ_{v_r,v_b}\left(Z_{i_1}\prod_{j_1\in \text{Nb}(i_1)}Z_{j_1}\right)...\left(Z_{i_s}\prod_{j_s\in \text{Nb}(i_s)}Z_{j_s}\right)\nonumber\\
    &\equiv \bar{\rho}_{\text{blue}}^{\mathrm{A}}\, ,
\end{align} 
\end{subequations}

where $\ket{\phi}$ is an arbitrary dangling d.o.f. on the blue vertices. Nb($i_l$) denotes the neighboring blue vertex to the red vertex $i_l$ in $\rm A$.
It has a strong $\mathbb{Z}_2^{(r)}=\prod_iX_{i,k_0}$ symmetry. Stabilizers satisfy the following weak symmetry condition
\begin{align}
    &Z_{i-\frac{1}{2},k_0+\frac{1}{2}}X_{i,k_0}Z_{i+\frac{1}{2},k_0+\frac{1}{2}}\bar{\rho}_{\text{blue}}^{\rm A}Z_{i-\frac{1}{2},k_0+\frac{1}{2}}X_{i,k_0}Z_{i+\frac{1}{2},k_0+\frac{1}{2}}=\bar{\rho}_{\text{blue}}^{\rm A}\, .
\end{align}
$\bar{\rho}_{\text{blue}}^{\rm A}$ exhibits SWSSB of $\mathbb{Z}_2^{(r)}$ symmetry when $\bra{\phi}Z\ket{\phi}\neq 0$.
\subsubsection{Tracing the interface between 2D cluster state and 2D blue state on a torus}
Now let us consider the two-dimensional state that constitutes an interface between the cluster state and the blue state, subject to periodic boundary conditions along the $y$ direction. We take the interface to lie along the lines $x = l + \tfrac{1}{2}$ and $x = L_x + \tfrac{1}{2}$. In this setting, after tracing out all bulk degrees of freedom except for the layers at $y = k$ and $y = k + \tfrac{1}{2}$, the resulting reduced density matrix is given by $\rho^{\rm A}_{\text{1D-SWSSB}}$. This outcome is anticipated, since tracing out all bulk degrees of freedom except those at $y = k$ and $y = k + \tfrac{1}{2}$ yields the same reduced state, regardless of whether the initial state is the cluster state or the blue state with periodic boundary conditions.
\subsubsection{Tracing the interface between 2D cluster state and 2D blue state on a cylinder with generic open boundary condition}
Let us consider the two-dimensional state that constitutes an interface between the 2D cluster state and 2D blue state, subject to an open boundary condition along the $y$ direction. We take the boundary to lie at $y=k_0$ and $y=k_0+\frac{1}{2}$. We take the interface to lie along the lines $x=l+\frac{1}{2}$ and $x=L_x+\frac{1}{2}$. After tracing out the bulk degrees of freedom except for the layers at $y=k_0$ and $y=k_0+\frac{1}{2}$, the resulting reduced density matrix has SWSSB of $\mathbb{Z}_2^{(r)}$ symmetry. 
\subsection{Mixed-state from 2D blue\texorpdfstring{$^{k_0;k_1}$}{Lg} state}\label{app:Extraexamples3}
In this section, we look at the states obtained by tracing out the 2D blue$^{k_0;k_1}$ state on a torus and on cylinder with generic boundary conditions. The results are summarized in Table~\ref{tab:bluek0k1trace}.
\subsubsection{2D blue\texorpdfstring{$^{k_0;k_1}$}{Lg} state on a torus}
    Now we trace out all the 2D d.o.f. except the row at $y=k_0$ and $y=k_0+\frac{1}{2}$. Tracing out $\rm A^c$ gives
\begin{subequations}    
\begin{align}
    &\text{Tr}_{A^c}\left(\ket{\text{blue}^{k_0;k_1}}\bra{\text{blue}^{k_0;k_1}}\right)\nonumber\\
    &=\frac{4}{2^{|\Delta_{v_b}\cap \mathrm{A}|+|\Delta_{v_r}\cap \mathrm{A}|}}\sum_{\substack{\{k_l\}\in\Delta_{v_b}\cap \mathrm{A}\,,\\
t\text{ even }\\
\{i_l\}\in\Delta_{v_r}\cap \mathrm{A}\,,\\
s\text{ even }}}Z^b_{k_1}...Z^b_{k_t} \times Z_{i_1}^r...Z_{i_s}^r\nonumber\\
&\hspace{2cm} \prod_{\substack{v_b=(i+\frac{1}{2},k_0+\frac{1}{2})}} CZ_{v_b,v_b+(1,0)}\ket{\text{1D-cluster}}\bra{\text{1D-cluster}}\prod_{\substack{v_b=(i+\frac{1}{2},k_0+\frac{1}{2})}} CZ_{v_b,v_b+(1,0)}\nonumber\\
&\hspace{10cm} \times Z_{i_1}^r...Z_{i_s}^r\times Z^b_{k_1}...Z^b_{k_t}\,\\
&=\left[\prod_{\substack{v_b=(i+\frac{1}{2},k_0+\frac{1}{2})}} CZ_{v_b,v_b+(1,0)}\right]\rho^{\mathrm{A}}_{\text{1D-SWSSB}}\left[\prod_{\substack{v_b=(i+\frac{1}{2},k_0+\frac{1}{2})}} CZ_{v_b,v_b+(1,0)}\right]=\rho^{\mathrm{A}}_{\text{1D-SWSSB}}\,.
\end{align}
\end{subequations}

This state was studied in \eqref{eq:1D-SWSSB} and has strong $\mathbb{Z}_2^{(r)}\times\mathbb{Z}_2^{(b)}$ 0-form symmetry and weak emergent non-invertible $\mathrm{D}^{(1)}$ symmetry. We verified that it is in the SWSSB phase regarding the $\mathbb{Z}_2^{(r)}\times\mathbb{Z}_2^{(b)}$ by computing the R\'enyi-2 correlator and the fidelity correlators in \eqref{eq:Renyi2rho1DSWSSB} and \eqref{eq:fidelityrho1DSWSSB}.
\subsubsection{2D blue\texorpdfstring{$^{k_0;k_1}$}{Lg} on a semi-infinite cylinder with generic boundary condition}
Now we consider the state $\text{blue}^{k_0;k_1}$ with open boundary condition in $y$ direction, with $y$ values ranging from $-\infty$, ..., $k_1$, $k_1+\frac{1}{2}$, ...., $k_0$, $k_0+\frac{1}{2}$. We take $y=k_0$ and $y=k_0+\frac{1}{2}$ as the region $\rm A$ and the rest as $\rm A^c$. Tracing out the region $\rm A^c$, we obtain the state

\begin{align}
    \text{Tr}_{\rm A^c}\left(\ket{\text{blue}^{k_0;k_1}}\bra{\text{blue}^{k_0;k_1}}\right)=\left[\prod_{\substack{v_b=(i+\frac{1}{2},k_0+\frac{1}{2})}} CZ_{v_b,v_b+(1,0)}\right]\bar{\rho}^{\rm A}_{\text{blue}}\left[\prod_{\substack{v_b=(i+\frac{1}{2},k_0+\frac{1}{2})}} CZ_{v_b,v_b+(1,0)}\right]\equiv\bar{\rho}^{\rm A}_{\text{blue}^{k_0;k_1}}\,.
\end{align}

with arbitrary dangling d.o.f. $\ket{\phi}$ on the vertices at $y=k_0+\frac{1}{2}$ at the boundary. 

This state has the weak stabilizers
\begin{align}
Z_{i-\frac{1}{2},k_0+\frac{1}{2}}X_{i,k_0}Z_{i+\frac{1}{2},k_0+\frac{1}{2}}\bar{\rho}^{\rm A}_{\text{blue}^{k_0;k_1}}Z_{i-\frac{1}{2},k_0+\frac{1}{2}}X_{i,k_0}Z_{i+\frac{1}{2},k_0+\frac{1}{2}}=\bar{\rho}^{\rm A}_{\text{blue}^{k_0;k_1}}\,,
\end{align}
$\bar{\rho}^{\rm A}_{\text{blue}^{k_0;k_1}}$ is in the SWSSB phase regarding the $\mathbb{Z}_2^{(r)}$ symmetry when $\bra{\phi}Z\ket{\phi}\neq 0$.
\begin{table*}[h!]
    \centering
    \resizebox{0.95\columnwidth}{!}{
\begin{tabular}{|c|c|c|c|}
\hline
Boundary condition in 2D & Strong symmetries & Weak symmetries & Phase\\
\hline
   Periodic in $x$ and $y$  & \begin{tabular}{@{}c@{}}$\mathbb{Z}_2^{(r)}=\prod_{i}X_{i,k}$\,,\\ $\mathbb{Z}_2^{(b)}=\prod_{i}X_{i+\frac{1}{2},k+\frac{1}{2}}$\end{tabular} & \begin{tabular}{@{}c@{}}$\mathrm{D}^{(1)}$\\
   $Z_{i,k}X_{i+\frac{1}{2},k+\frac{1}{2}}Z_{i+1,k}$\\
   $Z_{i-\frac{1}{2},k+\frac{1}{2}}X_{i,k}Z_{i+\frac{1}{2},k+\frac{1}{2}}$
   \end{tabular}& SWSSB of $\mathbb{Z}_2^{(r)}$ and $\mathbb{Z}_2^{(b)}$\\
\hline
   \begin{tabular}{@{}c@{}}
   Periodic in $x$ and\\
 open in $y$ with boundary at $y=k_0+\frac{1}{2}=\frac{1}{2}$
   \end{tabular}  &    $\mathbb{Z}_2^{(r)}=\prod_{i}X_{i,0}$ & $Z_{i-\frac{1}{2},\frac{1}{2}}X_{i,0}Z_{i+\frac{1}{2},\frac{1}{2}}$ & SWSSB of $\mathbb{Z}_2^{(r)}$ \\
   \hline
   \begin{tabular}{@{}c@{}}
   Periodic in $y$ and
 open in $x$ with boundary at  \\ $x=l+\frac{1}{2}$ and $x=L_x+\frac{1}{2}$
   \end{tabular}  &    - &\begin{tabular}{@{}c@{}}$\mathbb{Z}_2^{(r)}=Z_{l+\frac{1}{2},k+\frac{1}{2}}\prod_{i=l+1}^{L_x}X_{i,k}Z_{L_x+\frac{1}{2},k+\frac{1}{2}}$\\
   $Z_{i-\frac{1}{2},k+\frac{1}{2}}X_{i,k}Z_{i+\frac{1}{2},k+\frac{1}{2}}$($i=l+1,...,L_x$)\\
   $Z_{i,k}X_{i+\frac{1}{2},k+\frac{1}{2}}Z_{i+1,k}$($i=l+1,...,L_x-1$)
   \end{tabular}&   \begin{tabular}{@{}c@{}}$\mathbb{Z}_2^{(r)}$ is explcitly\\ broken to weak symmetry\\
   $\mathbb{Z}_2^{(b)}$ is explicitly\\
   broken to nothing
   \end{tabular}\\ 
   \hline
   \begin{tabular}{@{}c@{}}
   Open in $x$ and
 open in $y$ with boundary at  \\ $y=k_0+\frac{1}{2}=\frac{1}{2}$ and $x=l+\frac{1}{2},L_x+\frac{1}{2}$
   \end{tabular}  &    - & \begin{tabular}{@{}c@{}}$\mathbb{Z}_2^{(r)}=Z_{l+\frac{1}{2},\frac{1}{2}}\prod_{i=l+1}^{L_x}X_{i,0}Z_{L_x+\frac{1}{2},\frac{1}{2}}$\\
   $Z_{i-\frac{1}{2},\frac{1}{2}}X_{i,0}Z_{i+\frac{1}{2},\frac{1}{2}}$($i=l+1,...,L_x$)
   \end{tabular}&   \begin{tabular}{@{}c@{}}$\mathbb{Z}_2^{(r)}$ is explcitly\\ broken to weak symmetry
   \end{tabular}\\
   \hline
\end{tabular}}
\caption{A table illustrating the properties of the state after tracing out the cluster state in the 2D with arbitrary dangling d.o.f. on boundaries.}
\label{tab:clusterstatetrace}
\end{table*}
\begin{scriptsize}
\begin{table*}[h!]
    \centering
    \resizebox{0.95\columnwidth}{!}{
\begin{tabular}{|c|c|c|c|}
\hline
Boundary condition in 2D & Strong symmetries & Weak symmetries & Phase\\
\hline
   \begin{tabular}{@{}c@{}}
   Periodic in $x$ and
 open in $y$ \\with boundary at $y=k_0+\frac{1}{2}=\frac{1}{2}$
   \end{tabular}  &    \begin{tabular}{@{}c@{}}$\mathbb{Z}_2^{(r)}=\prod_{i}X_{i,0}$\\
   $\mathbb{Z}_2^{(b)}=\prod_{i}X_{i+\frac{1}{2},\frac{1}{2}}$\\
   $\mathrm{D}^{(1)}$\\
   $Z_{i,0}X_{i+\frac{1}{2},\frac{1}{2}}Z_{i+1,0}$
   \end{tabular}& $Z_{i-\frac{1}{2},\frac{1}{2}}X_{i,0}Z_{i+\frac{1}{2},\frac{1}{2}}$ & \begin{tabular}{@{}c@{}}SWSSB of $\mathbb{Z}_2^{(r)}$\\
   Double ASPT \\
   \end{tabular}\\
   \hline
   \begin{tabular}{@{}c@{}}
   Periodic in $y$ and
 open in $x$\\ with boundary at $x=l+\frac{1}{2}$\\ and $x=L_x+\frac{1}{2}$
   \end{tabular}  &    $\mathbb{Z}_2^{(b)}=\prod_{i=l}^{L_x}X_{i+\frac{1}{2},k+\frac{1}{2}}$ &\begin{tabular}{@{}c@{}}$\mathbb{Z}_2^{(r)}=Z_{l+\frac{1}{2},k+\frac{1}{2}}\prod_{i=l+1}^{L_x}X_{i,k}Z_{L_x+\frac{1}{2},k+\frac{1}{2}}$\\
   $Z_{i-\frac{1}{2},k+\frac{1}{2}}X_{i,k}Z_{i+\frac{1}{2},k+\frac{1}{2}}$($i=l+1,...,L_x$)\\
   $Z_{i,k}X_{i+\frac{1}{2},k+\frac{1}{2}}Z_{i+1,k}$($i=l+1,...,L_x-1$)\\
   $X_{l+\frac{1}{2}}Z_{l+1,k}$\,,\quad$X_{L_x+\frac{1}{2},k+\frac{1}{2}}Z_{L_x,k}$
   \end{tabular}&   \begin{tabular}{@{}c@{}}$\mathbb{Z}_2^{(r)}$ is explcitly\\ broken to weak symmetry
\end{tabular}\\
   \hline
   \begin{tabular}{@{}c@{}}
   Open in $x$ and
 open in $y$\\ with boundary at $y=k_0+\frac{1}{2}=\frac{1}{2}$\\ and $x=l+\frac{1}{2},L_x+\frac{1}{2}$
   \end{tabular}  &    \begin{tabular}{@{}c@{}}$\mathbb{Z}_2^{(b)}=\prod_{i=l}^{L_x}X_{i+\frac{1}{2},\frac{1}{2}}$\\
   $Z_{i,0}X_{i+\frac{1}{2},\frac{1}{2}}Z_{i+1,0}$\\
   \qquad($i=l+1,...,L_x-1$)\\
   $X_{l+\frac{1}{2}}Z_{l+1,0}$\,,\quad$X_{L_x+\frac{1}{2},\frac{1}{2}}Z_{L_x,0}$
   \end{tabular}& \begin{tabular}{@{}c@{}}$\mathbb{Z}_2^{(r)}=Z_{l+\frac{1}{2},\frac{1}{2}}\prod_{i=l+1}^{L_x}X_{i,0}Z_{L_x+\frac{1}{2},\frac{1}{2}}$\\
   $Z_{i-\frac{1}{2},\frac{1}{2}}X_{i,0}Z_{i+\frac{1}{2},\frac{1}{2}}$($i=l+1,...,L_x$)
   \end{tabular}&   \begin{tabular}{@{}c@{}}mSPT 
   \end{tabular}\\
   \hline
\end{tabular}}
\caption{A table illustrating the properties of the state after tracing out the cluster state in the 2D with the dangling d.o.f. set to $\ket{+}$ state.}
\label{tab:clusterstatetrace+}
\end{table*}
\end{scriptsize}
\begin{table*}[h!]
    \centering
    \resizebox{0.95\columnwidth}{!}{
\begin{tabular}{|c|c|c|c|}
\hline
Boundary condition in 2D & Strong symmetries & Weak symmetries & Phase\\
\hline
   Periodic in $x$ and $y$  & \begin{tabular}{@{}c@{}}$\mathbb{Z}_2^{(r)}=\prod_{i}X_{i,k}$\,,\\ $\mathbb{Z}_2^{(b)}=\prod_{i}X_{i+\frac{1}{2},k+\frac{1}{2}}$\end{tabular} & \begin{tabular}{@{}c@{}}
   $Z_{i,k}X_{i+\frac{1}{2},k+\frac{1}{2}}Z_{i+1,k}$\\
   $Z_{i-\frac{1}{2},k+\frac{1}{2}}X_{i,k}Z_{i+\frac{1}{2},k+\frac{1}{2}}$\\
   $\mathrm{D}^{(1)}$
   \end{tabular} &\begin{tabular}{@{}c@{}} SWSSB of $\mathbb{Z}_2^{(b)}$ and $\mathbb{Z}_2^{(r)}$\\ 
   \end{tabular}\\
\hline
   \begin{tabular}{@{}c@{}}
   Periodic in $y$ and
 open in $x$\\ with boundary at $x=l+\frac{1}{2},L_x+\frac{1}{2}$
   \end{tabular}  &    - & \begin{tabular}{@{}c@{}}
   $Z_{i-\frac{1}{2},k+\frac{1}{2}}X_{i,k}Z_{i+\frac{1}{2},k+\frac{1}{2}}$($i=l+1,...,L_x$)\\
   $Z_{i,k}X_{i+\frac{1}{2},k+\frac{1}{2}}Z_{i+1,k}$($i=l+1,...,L_x-1$)\\
   $\mathbb{Z}_2^{(r)}=Z_{l+\frac{1}{2}}\prod_{i=l+1}^{L_x}X_{i,k}Z_{L_x+\frac{1}{2},k+\frac{1}{2}}$
   \end{tabular}& \begin{tabular}{@{}c@{}}
   $\mathbb{Z}_2^{(r)}$ is explcitly\\ broken to weak symmetry\\
   $\mathbb{Z}_2^{(b)}$ is explicitly\\
   broken to nothing
   \end{tabular}\\
   \hline
   \begin{tabular}{@{}c@{}}
   Periodic in $x$ and
 open in $y$\\ with boundary at $y=k_0+\frac{1}{2}=\frac{1}{2}$
   \end{tabular}  &    $\mathbb{Z}_2^{(r)}=\prod_{i}X_{i,0}$ & $Z_{i-\frac{1}{2},\frac{1}{2}}X_{i,0}Z_{i+\frac{1}{2},\frac{1}{2}}$ & SWSSB of $\mathbb{Z}_2^{(r)}$ \\
   \hline
   \begin{tabular}{@{}c@{}}
   Open in $x$ and
 open in $y$ with boundary\\ at $y=k_0+\frac{1}{2}=\frac{1}{2}$ and $x=l+\frac{1}{2},L_x+\frac{1}{2}$
   \end{tabular}  &    - & \begin{tabular}{@{}c@{}}$\mathbb{Z}_2^{(r)}=Z_{\frac{1}{2},\frac{1}{2}}\prod_{i=1}^{l-1}X_{i,0}Z_{l-\frac{1}{2},\frac{1}{2}}$\\
   $Z_{i-\frac{1}{2},\frac{1}{2}}X_{i,0}Z_{i+\frac{1}{2},\frac{1}{2}}$($i=l+1,...,L_x$)
   \end{tabular}&   \begin{tabular}{@{}c@{}}$\mathbb{Z}_2^{(r)}$ is explcitly\\ broken to weak symmetry
   \end{tabular}\\
   \hline
\end{tabular}}
\caption{A table illustrating the properties of the state after tracing out the blue state in the 2D with arbitrary d.o.f. at the boundary.}
\label{tab:bluestatetrace}
\end{table*}
\begin{table*}[h!]
    \centering
    \resizebox{0.95\columnwidth}{!}{
\begin{tabular}{|c|c|c|c|}
\hline
Boundary condition in 2D & Strong symmetries & Weak symmetries & Phase\\
\hline
   \begin{tabular}{@{}c@{}}
   Periodic in $x$ and open in $y$\\
  with boundary at $y=k_0+\frac{1}{2}=\frac{1}{2}$
   \end{tabular}  &    \begin{tabular}{@{}c@{}}$\mathbb{Z}_2^{(r)}=\prod_{i}X_{i,0}$\\
   $\mathbb{Z}_2^{(b)}=\prod_{i}X_{i+\frac{1}{2},\frac{1}{2}}$\\
   $\mathrm{D}^{(1)}$\\
   $Z_{i-\frac{1}{2},\frac{1}{2}}Y_{i,0}X_{i+\frac{1}{2},\frac{1}{2}}Y_{i+1,0}Z_{i+\frac{3}{2},\frac{1}{2}}$
   \end{tabular}& \begin{tabular}{@{}c@{}}$Z_{i-\frac{1}{2},\frac{1}{2}}X_{i,0}Z_{i+\frac{1}{2},\frac{1}{2}}$\\
   $Z_{i,0}X_{i+\frac{1}{2},\frac{1}{2}}Z_{i+1,0}$
   \end{tabular}& \begin{tabular}{@{}c@{}}SWSSB of $\mathbb{Z}_2^{(r)}$\\
   Double ASPT
   \end{tabular} \\
   \hline
   \begin{tabular}{@{}c@{}}
   Periodic in $y$ and
 open in $x$\\ with boundary at\\ $x=l+\frac{1}{2},L_x+\frac{1}{2}$
   \end{tabular}  &    
   - & \begin{tabular}{@{}c@{}}
   $Z_{i-\frac{1}{2},k+\frac{1}{2}}X_{i,k}Z_{i+\frac{1}{2},k+\frac{1}{2}}$ ($i=l+1,...,L_x$)\\
   $Z_{i,k}X_{i+\frac{1}{2},k+\frac{1}{2}}Z_{i+1,k}$ ($i=l+1,...,L_x-1$)\\
   $X_{l+\frac{1}{2},k+\frac{1}{2}}Z_{l+1,k}$\,,\quad$Z_{L_x,k}X_{L_x+\frac{1}{2},k+\frac{1}{2}}$\\
   $\mathbb{Z}_2^{(r)}=Z_{l+\frac{1}{2}}\prod_{i=l+1}^{L_x}X_{i,k}Z_{L_x+\frac{1}{2},k+\frac{1}{2}}$\\
   $\mathbb{Z}_2^{(b)}=\prod_{i=l}^{L_x}X_{i+\frac{1}{2},k+\frac{1}{2}}$
   \end{tabular}& \begin{tabular}{@{}c@{}}
   $\mathbb{Z}_2^{(r)}$ and $\mathbb{Z}_2^{(b)}$\\ is explicitly
   broken\\ to weak symmetry
   \end{tabular}\\
   \hline
   \begin{tabular}{@{}c@{}}
   Open in $x$ and
 open in $y$\\ with boundary at $y=k_0+\frac{1}{2}=\frac{1}{2}$\\ and $x=l+\frac{1}{2}, L_x+\frac{1}{2}$
   \end{tabular}  & 
   \begin{tabular}{@{}c@{}}
   $Z_{i-\frac{1}{2},\frac{1}{2}}Y_{i,0}X_{i+\frac{1}{2},\frac{1}{2}}Y_{i+1,0}Z_{i+\frac{3}{2},\frac{1}{2}}$\\
   $(i=l+1,...,L_x-1)$
   \end{tabular}& \begin{tabular}
   {@{}c@{}}
$\mathbb{Z}_2^{(r)}=Z_{l+\frac{1}{2},\frac{1}{2}}\prod_{i=l+1}^{L_x}X_{i,0}Z_{L_x+\frac{1}{2},\frac{1}{2}}$\\
$\mathbb{Z}_2^{(b)}=\prod_{i=l}^{L_x}X_{i+\frac{1}{2},\frac{1}{2}}$\\
$Z_{i,0}X_{i+\frac{1}{2},\frac{1}{2}}Z_{i+1,0}$($i=l+1,...,L_x-1$)\\
   $Z_{i-\frac{1}{2},\frac{1}{2}}X_{i,0}Z_{i+\frac{1}{2},\frac{1}{2}}$($i=l+1,...,L_x$)\\
   $X_{l+\frac{1}{2},\frac{1}{2}}Z_{l+1,0}$\,\quad$X_{L_x+\frac{1}{2},\frac{1}{2}}Z_{L_x,0}$ 
   \end{tabular}&   \begin{tabular}{@{}c@{}}$\mathbb{Z}_2^{(r)}$ and $\mathbb{Z}_2^{(b)}$\\ is explicitly
   broken\\ to weak symmetry
   \end{tabular}\\
   \hline
\end{tabular}}
\caption{ A table illustrating the properties of the state after tracing out the blue state in the 2D with the dangling d.o.f. set to $\ket{+}$ state.}
\label{tab:bluestatetrace+}
\end{table*}
\begin{table*}[h!]
    \centering
    \resizebox{0.95\columnwidth}{!}{
\begin{tabular}{|c|c|c|c|}
\hline
Boundary condition in 2D & Strong symmetries & Weak symmetries & Phase\\
\hline
Periodic in $x$ and $y$ &\begin{tabular}{@{}c@{}}$\mathbb{Z}_2^{(r)}=\prod_{i}X_{i,k}$\,,\\ $\mathbb{Z}_2^{(b)}=\prod_{i}X_{i+\frac{1}{2},k+\frac{1}{2}}$\end{tabular} & \begin{tabular}{@{}c@{}}$\mathrm{D}^{(1)}$\\
   $Z_{i,k}X_{i+\frac{1}{2},k+\frac{1}{2}}Z_{i+1,k}$\\
   $Z_{i-\frac{1}{2},k+\frac{1}{2}}X_{i,k}Z_{i+\frac{1}{2},k+\frac{1}{2}}$
   \end{tabular}& SWSSB of $\mathbb{Z}_2^{(r)}$ and $\mathbb{Z}_2^{(b)}$\\
\hline
\begin{tabular}{@{}c@{}}
Periodic in $x$ and open in $y$\\
with boundary at $y=k_0+\frac{1}{2}=\frac{1}{2}$ 
\end{tabular} & $\mathbb{Z}_2^{(r)}=\prod_{i}X_{i,0}$ & $Z_{i-\frac{1}{2},\frac{1}{2}}X_{i,0}Z_{i+\frac{1}{2},\frac{1}{2}}$ & SWSSB of $\mathbb{Z}_2^{(r)}$\\
\hline
\begin{tabular}{@{}c@{}}
Periodic in $y$ and open in $x$\\
with boundary at \\$x=l+\frac{1}{2},L_x+\frac{1}{2}$ 
\end{tabular}&- & \begin{tabular}{@{}c@{}}
   $Z_{i-\frac{1}{2},k+\frac{1}{2}}X_{i,k}Z_{i+\frac{1}{2},k+\frac{1}{2}}$\, ($i=l+1,...,L_x$)\\
   $Z_{i-\frac{1}{2},k+\frac{1}{2}}Z_{i,k}X_{i+\frac{1}{2},k+\frac{1}{2}}Z_{i+1,k}Z_{i+\frac{3}{2},k+\frac{1}{2}}$\,\\
   \hspace{4cm}($i=l+1,...,L_x-1$)\\
$\mathbb{Z}_2^{(r)}=Z_{l+\frac{1}{2},k+\frac{1}{2}}\prod_{i=l+1}^{L_x}X_{i,k}Z_{L_x+\frac{1}{2},k+\frac{1}{2}}$\\
   $\mathbb{Z}_2^{(b)}=Z_{l+\frac{1}{2},k+\frac{1}{2}}Z_{l+1,k}Y_{l+\frac{3}{2},k+\frac{1}{2}}\prod_{i=l+2}^{L_x-2}X_{i+\frac{1}{2},k+\frac{1}{2}}$\\
   $\times Y_{L_x-\frac{1}{2},k+\frac{1}{2}}Z_{L_x,k}Z_{L_x+\frac{1}{2},k+\frac{1}{2}}$
   \end{tabular}& \begin{tabular}{@{}c@{}}
   $\mathbb{Z}_2^{(r)}$ and $\mathbb{Z}_2^{(b)}$ is explicitly\\
   broken to weak symmetry
   \end{tabular}\\
\hline
\begin{tabular}{@{}c@{}}
Open in $x$ and open in $y$\\
with boundary at $y=k_0+\frac{1}{2}=\frac{1}{2}$\\ and  $x=l+\frac{1}{2},L_x+\frac{1}{2}$
\end{tabular}&- & \begin{tabular}{@{}c@{}}$\mathbb{Z}_2^{(r)}=Z_{\frac{1}{2},\frac{1}{2}}\prod_{i=1}^{l-1}X_{i,0}Z_{l-\frac{1}{2},\frac{1}{2}}$\\
   $Z_{i-\frac{1}{2},\frac{1}{2}}X_{i,0}Z_{i+\frac{1}{2},\frac{1}{2}}$($i=l+1,...,L_x$)
   \end{tabular}&   \begin{tabular}{@{}c@{}}$\mathbb{Z}_2^{(r)}$ is explicitly\\ broken to weak symmetry
   \end{tabular}\\
\hline
\end{tabular}}
\caption{A table illustrating the properties of the state after tracing out the blue$^{k_0;k_1}$ state in the 2D with arbitrary dangling d.o.f.}
\label{tab:bluek0k1trace}
\end{table*}
\begin{table*}
\resizebox{0.95\columnwidth}{!}{
    \begin{tabular}{|c|c|c|c|}
    \hline
       Boundary condition in 2D  & Strong symmetries & Weak symmetries & Phase \\
        \hline\begin{tabular}{@{}c@{}}
Periodic in $x$ and open in $y$\\
with boundary at $y=k_0+\frac{1}{2}=\frac{1}{2}$ 
\end{tabular} &\begin{tabular}{@{}c@{}}$\mathbb{Z}_2^{(r)}=\prod_{i}X_{i,0}$\\
   $\mathbb{Z}_2^{(b)}=\prod_{i}X_{i+\frac{1}{2},\frac{1}{2}}$\\
   $Y_{i,0}X_{i+\frac{1}{2},\frac{1}{2}}Y_{i+1,0}$
   \end{tabular}& \begin{tabular}{@{}c@{}}$Z_{i-\frac{1}{2},\frac{1}{2}}X_{i,0}Z_{i+\frac{1}{2},\frac{1}{2}}$\\
   $Z_{i-\frac{1}{2},\frac{1}{2}}Z_{i,0}X_{i+\frac{1}{2},\frac{1}{2}}Z_{i+1,0}Z_{i+\frac{3}{2},\frac{1}{2}}$
   \end{tabular}& \begin{tabular}{@{}c@{}}SWSSB of $\mathbb{Z}_2^{(r)}$\\
   Double ASPT
   \end{tabular} \\
\hline
\begin{tabular}{@{}c@{}}
Periodic in $y$ and open in $x$\\
with boundary at \\$x=l+\frac{1}{2},L_x+\frac{1}{2}$ 
\end{tabular}&- & \begin{tabular}{@{}c@{}}
   $Z_{i-\frac{1}{2},k+\frac{1}{2}}X_{i,k}Z_{i+\frac{1}{2},k+\frac{1}{2}}$\, ($i=l+1,...,L_x$)\\
   $Z_{i-\frac{1}{2},k+\frac{1}{2}}Z_{i,k}X_{i+\frac{1}{2},k+\frac{1}{2}}Z_{i+1,k}Z_{i+\frac{3}{2},k+\frac{1}{2}}$\,\\
   \hspace{4cm}($i=l+1,...,L_x-1$)\\
   $X_{l+\frac{1}{2},k+\frac{1}{2}}Z_{l+1,k}Z_{l+\frac{3}{2},k+\frac{1}{2}}$\\$X_{L_x+\frac{1}{2},k+\frac{1}{2}}Z_{L_x,k}Z_{L_x-\frac{1}{2},k+\frac{1}{2}}$\\
$\mathbb{Z}_2^{(r)}=Z_{l+\frac{1}{2},k+\frac{1}{2}}\prod_{i=l+1}^{L_x}X_{i,k}Z_{L_x+\frac{1}{2},k+\frac{1}{2}}$\\
   $\mathbb{Z}_2^{(b)}=Y_{l+\frac{1}{2},k+\frac{1}{2}}\prod_{i=l+1}^{L_x-1}X_{i+\frac{1}{2},k+\frac{1}{2}}
   Y_{L_x+\frac{1}{2},k+\frac{1}{2}}$
   \end{tabular}& \begin{tabular}{@{}c@{}}
   $\mathbb{Z}_2^{(r)}$ and $\mathbb{Z}_2^{(b)}$ is explicitly\\
   broken to weak symmetry
   \end{tabular}\\
\hline
\begin{tabular}{@{}c@{}}
Open in $x$ and open in $y$\\
with boundary at $y=k_0+\frac{1}{2}=\frac{1}{2}$\\ and $x=l+\frac{1}{2},L_x+\frac{1}{2}$ 
\end{tabular} &\begin{tabular}{@{}c@{}}
   $Y_{i,0}X_{i+\frac{1}{2},\frac{1}{2}}Y_{i+1,0}$\\
   $(i=l+1,...,L_x-1)$
   \end{tabular}
   &\begin{tabular}{@{}c@{}}
   $Z_{i-\frac{1}{2},\frac{1}{2}}X_{i,0}Z_{i+\frac{1}{2},\frac{1}{2}}$\, ($i=l+1,...,L_x$)\\
   $Z_{i-\frac{1}{2},\frac{1}{2}}Z_{i,0}X_{i+\frac{1}{2},\frac{1}{2}}Z_{i+1,0}Z_{i+\frac{3}{2},\frac{1}{2}}$\,\\
   \hspace{4cm}($i=l+1,...,L_x-1$)\\
   $X_{l+\frac{1}{2},\frac{1}{2}}Z_{l+1,0}Z_{l+\frac{3}{2},\frac{1}{2}}$\\$X_{L_x+\frac{1}{2},\frac{1}{2}}Z_{L_x,0}Z_{L_x-\frac{1}{2},\frac{1}{2}}$\\
$\mathbb{Z}_2^{(r)}=Z_{l+\frac{1}{2},\frac{1}{2}}\prod_{i=l+1}^{L_x}X_{i,0}Z_{L_x+\frac{1}{2},\frac{1}{2}}$\\
   $\mathbb{Z}_2^{(b)}=Y_{l+\frac{1}{2},\frac{1}{2}}\prod_{i=l+1}^{L_x-1}X_{i+\frac{1}{2},\frac{1}{2}}
   Y_{L_x+\frac{1}{2},\frac{1}{2}}$
   \end{tabular}& \begin{tabular}{@{}c@{}}
   $\mathbb{Z}_2^{(r)}$ and $\mathbb{Z}_2^{(b)}$ is explicitly\\
   broken to weak symmetry
   \end{tabular}\\
\hline
    \end{tabular}}
    \caption{A table illustrating the properties of the state after tracing out the blue$^{k_0;k_1}$ state in the 2D with dangling d.o.f. set to $\ket{+}$ state.}
\label{tab:bluek0k1trace+}
\end{table*}
\begin{table*}
\begin{tabular}{|c|c|c|c|}
\hline
   Boundary condition in 2D  & Strong symmetries & SWSSB & Weak symmetries \\
   \hline
    \begin{tabular}{@{}c@{}}
    Periodic in $y$ with\\
    interface at $x=l+\frac{1}{2},L_x+\frac{1}{2}$
    \end{tabular}&  $\mathbb{Z}_2^{(r)}\times\mathbb{Z}_2^{(b)}$ &  $\mathbb{Z}_2^{(r)}\times\mathbb{Z}_2^{(b)}$  & $\mathrm{D}^{(1)}$\\
    \hline
    \begin{tabular}{@{}c@{}}
    Open in $y$ with\\
    interface at $x=l+\frac{1}{2},L_x+\frac{1}{2}$
    \end{tabular}  & $\mathbb{Z}_2^{(r)}\times\mathbb{Z}_2^{(b)}$, $\mathrm{D}^{(1)}$ & $\mathbb{Z}_2^{(r)}$& \\
    \hline
\end{tabular}
\caption{A table illustrating the properties of the interface between 2D cluster state and 2D blue state. Vertex d.o.f. at the interface is set to $\ket{+}$. }
\label{tab:clusterblueinterface}
\end{table*}
\begin{table*}[h!]
\begin{tabular}{|c|c|c|c|}
\hline
   Boundary condition in 2D  & Strong symmetries& SWSSB & Weak symmetries \\
   \hline
    \begin{tabular}{@{}c@{}}
    Periodic in $y$ with\\
    interface at $x=l+\frac{1}{2},L_x+\frac{1}{2}$
    \end{tabular}&$\mathbb{Z}_2^{(r)}\times\mathbb{Z}_2^{(b)}$ & $\mathbb{Z}_2^{(r)}\times\mathbb{Z}_2^{(b)}$ &$\mathrm{D}^{(1)}$\\
    \hline
    \begin{tabular}{@{}c@{}}
    Open in $y$ with\\
    interface at $x=l+\frac{1}{2},L_x+\frac{1}{2}$
    \end{tabular} & $\mathbb{Z}_2^{(r)}\times\mathbb{Z}_2^{(b)}$& $\mathbb{Z}_2^{(r)}$ &\\
    \hline
\end{tabular}
\caption{A table illustrating the properties of interface between 2D cluster state and 2D blue$^{k_0;k_1}$ state. Vertex d.o.f. at the interface is set to $\ket{+}$.}
\label{tab:clusterbluek0;k1interface}
\end{table*}
\subsubsection{Tracing the interface between 2D cluster state and 2D blue$^{k_0;k_1}$ state on a torus}
Now let us consider the two-dimensional state that constitutes an interface between the cluster state and the blue$^{k_0;k_1}$ state, subject to periodic boundary conditions along the $y$ direction. We take the interface to lie along the lines $x = l + \tfrac{1}{2}$ and $x = L_x + \tfrac{1}{2}$. In this setting, after tracing out all bulk degrees of freedom except for the layers at $y = k_0$ and $y = k_0 + \tfrac{1}{2}$, the resulting reduced density matrix is given by

    \begin{align}
        \prod_{\substack{v_b=(i+\frac{1}{2},k_0+\frac{1}{2})\\
l\leq i< L_x}} CZ_{v_b,v_b+(1,0)}\rho^{\rm A}_{\text{1D-SWSSB}}\prod_{\substack{v_b=(i+\frac{1}{2},k_0+\frac{1}{2})\\
l\leq i< L_x}} CZ_{v_b,v_b+(1,0)}\,.
    \end{align}

 This outcome is anticipated, since tracing out all bulk degrees of freedom except those at $y = k_0$ and $y = k_0 + \tfrac{1}{2}$ yields the same reduced state, regardless of whether the initial state is the cluster state or the blue state with periodic boundary conditions and we have an extra decoration by the $CZ$ gate to account for the blue$^{k_0;k_1}$ state.
\subsubsection{Tracing the interface between 2D cluster state and 2D blue$^{k_0;k_1}$ state on a cylinder with generic open boundary condition}
Now let us consider an interface between the 2D cluster state and the 2D blue$^{k_0;k_1}$ state, subject to open boundary conditions along the $y$ direction. As in the previous case, the d.o.f. at $y=k_0+\frac{1}{2}$ are arbitrary and we call it $\ket{\phi}$. For this generic boundary condition, the resulting state after tracing the bulk d.o.f is in the SWSSB phase of $\mathbb{Z}_2^{(r)}$ symmetry. This is expected since tracing the two states individually results in SWSSB of $\mathbb{Z}_2^{(r)}$ and does not possess the non-invertible symmetry $\mathrm{D}^{(1)}$.
\section{A symmetric interface between two DASPTs}\label{app:interfacetwoDASPTs}
In this appendix, we consider an interface between two DASPTs and show that the interface can be made symmetric under $\mathbb{Z}_2^{(r)}\times\mathbb{Z}_2^{(b)}$ and $\mathrm{D}^{(1)}$. The two DASPTs are
\begin{subequations}
\begin{align}
\rho_{\text{1D-DASPT}}&=\prod_{i}\frac{1+Z_{i-1,k_0}X_{i+\frac{1}{2},k_0+\frac{1}{2}}Z_{i+1,k_0}}{2}\times \left(\frac{1+\prod_{i}X_{i,k_0}}{2^{L_x}}\right)\\
    \rho_{\text{1D-DASPT}'}&=\prod_{i}\frac{1+Z_{i-\frac{1}{2},k_0+\frac{1}{2}}X_{i,k_0}Z_{i+\frac{1}{2},k_0+\frac{1}{2}}}{2}\times \left(\frac{1+\prod_{i}X_{i+\frac{1}{2},k_0+\frac{1}{2}}}{2^{L_x}}\right)
    \,.
\end{align}
\end{subequations}
Now we consider an interface between the two, with the interface located at $(l+\frac{1}{2},k_0+\frac{1}{2})$ and $(L_x+\frac{1}{2},k_0+\frac{1}{2})$. As before, we assume periodic boundary conditions $L_x+1\equiv 1$. The interface state is obtained by truncating the two DASPTs
\begin{align}
    \rho_{\text{1D-DASPT}|\text{1D-DASPT}'}&=\prod_{i=1}^{l}\frac{1+Z_{i-\frac{1}{2},k_0+\frac{1}{2}}X_{i,k_0}Z_{i+\frac{1}{2},k_0+\frac{1}{2}}}{2}\times\frac{1+Z_{1,k_0}\prod_{i=1}^{l-1}X_{i+\frac{1}{2},k_0+\frac{1}{2}}Z_{l,k_0}}{2^{l-1}}\nonumber\\
    &\hspace{2cm}\prod_{i=l+1}^{L_x-1}\frac{1+Z_{i,k_0}X_{i+\frac{1}{2},k_0+\frac{1}{2}}Z_{i+1,k_0}}{2}\times\frac{1+Z_{l+\frac{1}{2},k_0+\frac{1}{2}}\prod_{i=l+1}^{L_x}X_{i,k_0}Z_{L_x+\frac{1}{2},k_0+\frac{1}{2}}}{2^{L_x-l}}\,.
\end{align}
Acting with the invertible symmetry operators
\begin{align}
    \prod_{i}X_{i,k_0}\rho_{\text{1D-DASPT}|\text{1D-DASPT}'}&=\rho_{\text{1D-DASPT}|\text{1D-DASPT}'}\\
    \prod_{i}X_{i+\frac{1}{2},k_0+\frac{1}{2}}\rho_{\text{1D-DASPT}|\text{1D-DASPT}'}&=Z_{l,k_0}X_{l+\frac{1}{2},k_0+\frac{1}{2}}Z_{l+1,k_0}\times Z_{L_x,k_0}X_{L_x+\frac{1}{2},k_0+\frac{1}{2}}Z_{1,k_0}\rho_{\text{1D-DASPT}|\text{1D-DASPT}'}\,.
\end{align}
To make the interface state symmetric under the invertible symmetries, we can add the projectors and define
\begin{align}
    \tilde{\rho}_{\text{1D-DASPT}|\text{1D-DASPT}'}=\frac{1+Z_{l,k_0}X_{l+\frac{1}{2},k_0+\frac{1}{2}}Z_{l+1,k_0}}{2}\times \frac{1+Z_{L_x,k_0}X_{L_x+\frac{1}{2},k_0+\frac{1}{2}}Z_{1,k_0}}{2}\rho_{\text{1D-DASPT}|\text{1D-DASPT}'}\,.
\end{align}
It is a straightforward exercise to check that $\tilde{\rho}_{\text{1D-DASPT}|\text{1D-DASPT}'}$ has strong $\mathrm{D}^{(1)}$ symmetry in addition to the strong $\mathbb{Z}_2^{(r)}\times\mathbb{Z}_2^{(b)}$. 
\section{Details on the strong symmetry of $\rho^{\rm A}_{\text{blue}}$}\label{app:strongD1onblue}
Let us consider the action of $\mathrm{D}^{(1)}$ on $\rho^{\rm A}_{\text{blue}}$

\begin{equation}
\begin{aligned}
    \mathrm{D}^{(1)}\rho^{\rm A}_{\text{blue}}&=\mathrm{D}^{(1)}\frac{2}{2^{|\Delta_{v_r}\cap \mathrm{A}|}}\sum_{\substack{\{i_l\}\in\Delta_{v_r}\cap \mathrm{A}\,,\\
    s \text{ even }}}\left(Z_{i_1}\prod_{j_1\in \text{Nb}(i_1)}Z_{j_1}\right)...\left(Z_{i_s}\prod_{j_s\in \text{Nb}(i_s)}Z_{j_s}\right)\prod CZ_{v_r,v_b}\ket{+}^{\otimes\Delta_{v_r}}\ket{+}^{\otimes\Delta_{v_b}}\\
    &\hspace{3cm}\bra{+}^{\otimes\Delta_{v_b}}\bra{+}^{\otimes\Delta_{v_r}}\prod CZ_{v_r,v_b}\left(Z_{i_1}\prod_{j_1\in \text{Nb}(i_1)}Z_{j_1}\right)...\left(Z_{i_s}\prod_{j_s\in \text{Nb}(i_s)}Z_{j_s}\right)\\
    &=\frac{2}{2^{|\Delta_{v_r}\cap \mathrm{A}|}}\sum_{\substack{\{i_l\}\in\Delta_{v_r}\cap \mathrm{A}\,,\\
    s \text{ even }}}X_{i_1}...X_{i_s}\mathrm{D}^{(1)}Z_{i_1}...Z_{i_s}\prod CZ_{v_r,v_b}\ket{+}^{\otimes\Delta_{v_r}}\ket{+}^{\otimes\Delta_{v_b}}\\
    &\hspace{3cm}\bra{+}^{\otimes\Delta_{v_b}}\bra{+}^{\otimes\Delta_{v_r}}\prod CZ_{v_r,v_b}\left(Z_{i_1}\prod_{j_1\in \text{Nb}(i_1)}Z_{j_1}\right)...\left(Z_{i_s}\prod_{j_s\in \text{Nb}(i_s)}Z_{j_s}\right)\\
    &=\frac{2}{2^{|\Delta_{v_r}\cap \mathrm{A}|}}\sum_{\substack{\{i_l\}\in\Delta_{v_r}\cap \mathrm{A}\,,\\
    s \text{ even }}}X_{i_1}...X_{i_s}X_{k_1}...X_{k_r}\mathrm{D}^{(1)}\prod CZ_{v_r,v_b}\ket{+}^{\otimes\Delta_{v_r}}\ket{+}^{\otimes\Delta_{v_b}}\\
    &\hspace{3cm}\bra{+}^{\otimes\Delta_{v_b}}\bra{+}^{\otimes\Delta_{v_r}}\prod CZ_{v_r,v_b}\left(Z_{i_1}\prod_{j_1\in \text{Nb}(i_1)}Z_{j_1}\right)...\left(Z_{i_s}\prod_{j_s\in \text{Nb}(i_s)}Z_{j_s}\right)\\
    &=\frac{2}{2^{|\Delta_{v_r}\cap \mathrm{A}|}}\sum_{\substack{\{i_l\}\in\Delta_{v_r}\cap \mathrm{A}\,,\\
    s \text{ even }}}X_{i_1}...X_{i_s}Z_{i_1}...Z_{i_s}\prod CZ_{v_r,v_b}\ket{+}^{\otimes\Delta_{v_r}}\ket{+}^{\otimes\Delta_{v_b}}\\
    &\hspace{3cm}\bra{+}^{\otimes\Delta_{v_b}}\bra{+}^{\otimes\Delta_{v_r}}\prod CZ_{v_r,v_b}\left(Z_{i_1}\prod_{j_1\in \text{Nb}(i_1)}Z_{j_1}\right)...\left(Z_{i_s}\prod_{j_s\in \text{Nb}(i_s)}Z_{j_s}\right)\\
    &=\frac{2}{2^{|\Delta_{v_r}\cap \mathrm{A}|}}\sum_{\substack{\{i_l\}\in\Delta_{v_r}\cap \mathrm{A}\,,\\
    s \text{ even }}}\left(Z_{i_1}\prod_{j_1\in \text{Nb}(i_1)}Z_{j_1}\right)...\left(Z_{i_s}\prod_{j_s\in \text{Nb}(i_s)}Z_{j_s}\right)\prod CZ_{v_r,v_b}\ket{+}^{\otimes\Delta_{v_r}}\ket{+}^{\otimes\Delta_{v_b}}\\
    &\hspace{3cm}\bra{+}^{\otimes\Delta_{v_b}}\bra{+}^{\otimes\Delta_{v_r}}\prod CZ_{v_r,v_b}\left(Z_{i_1}\prod_{j_1\in \text{Nb}(i_1)}Z_{j_1}\right)...\left(Z_{i_s}\prod_{j_s\in \text{Nb}(i_s)}Z_{j_s}\right)\\
    &=\rho^{\rm A}_{\text{blue}}
\end{aligned}
\end{equation}
In the third equality, we converted the product of $Z$ operators on the red sublattice to a product of $X$ operators on the blue sublattice. In the fourth line, we used the fact that $\mathrm{D}^{(1)}\ket{\text{cluster}}=\ket{\text{cluster}}$ and converted back the product of $X$ operators on the blue sublattice to a product of $Z$ operators on the red sublattice. Finally, in the fifth line, we converted back the product of $X$ operators on the red sublattice to a product of $Z$ operators on the blue sublattice.
\section{Other choices of interface cuts}
\subsection{Other choices of interface cuts between $\rho_{\text{trivial}}$ and $\rho_{\text{1D-DASPT}}$}\label{sec:otherinterfacecuts}
\subsubsection{Interface cut at $(l,k_0)$ and $(L_x,k_0)$}
Let us consider an interface cut passing through $(l,k_0)$ and $(L_x,k_0)$. The interface state is

\begin{align}
    \rho^{(1)}_{\text{trivial}|\text{1D-DASPT}}=\prod_{i=1}^{l-1}\frac{(1+X_{i,k_0})}{2}\prod_{i=0}^{l-1}\frac{(1+X_{i+\frac{1}{2},k_0+\frac{1}{2}})}{2}\prod_{i=l}^{L_x-1}\frac{(1+Z_{i,k_0}X_{i+\frac{1}{2},k_0+\frac{1}{2}}Z_{i+1,k_0})}{2}\left(\frac{1+\prod_{i=l}^{L_x}X_{i,k_0}}{2^{L_x-l+1}}\right)\,.
\end{align}
Now we apply the symmetries on this state
\begin{subequations}
\begin{align}
    \prod_{i}X_{i,k_0}\rho^{(1)}_{\text{trivial}|\text{1D-DASPT}}&=\rho^{(1)}_{\text{trivial}|\text{1D-DASPT}}\\
    \prod_{i}X_{i+\frac{1}{2},k_0+\frac{1}{2}}\rho^{(1)}_{\text{trivial}|\text{1D-DASPT}}&=Z_{l,k_0}Z_{L_x,k_0}\rho^{(1)}_{\text{trivial}|\text{1D-DASPT}}\,.
\end{align}
\end{subequations}
Now if we add the projectors $Z_{l,k_0}=\pm 1$ and $Z_{L_x,k_0}=\pm 1$, then that would explicitly break the symmetry $\prod_{i}X_{i,k_0}$to nothing.
\subsubsection{Interface cut at $(l+\frac{1}{2},k_0+\frac{1}{2})$ and $(L_x,k_0)$}
Now let us consider an interface cut at $(l+\frac{1}{2},k_0+\frac{1}{2})$ and $(L_x,k_0)$. The interface state is
\begin{align}
    \rho^{(2)}_{\text{trivial}|\text{1D-DASPT}}&=\prod_{i=1}^{l}\frac{(1+X_{i,k_0})}{2}\prod_{i=0}^{l-1}\frac{(1+X_{i+\frac{1}{2},k_0+\frac{1}{2}})}{2}\prod_{i=l+1}^{L_x-1}\frac{(1+Z_{i,k_0}X_{i+\frac{1}{2},k_0+\frac{1}{2}}Z_{i+1,k_0})}{2}\nonumber\\
    &\hspace{7cm}\left(\frac{1+Z_{l+\frac{1}{2},k_0+\frac{1}{2}}\prod_{i=l+1}^{L_x}X_{i,k_0}}{2^{L_x-l}}\right)\,.
\end{align}
Now we apply the symmetries on this state
\begin{subequations}
\begin{align}
    \prod_iX_{i,k_0}\rho^{(2)}_{\text{trivial}|\text{1D-DASPT}}&=Z_{l+\frac{1}{2},k_0+\frac{1}{2}}\rho^{(2)}_{\text{trivial}|\text{1D-DASPT}}\\
    \prod_iX_{i+\frac{1}{2},k_0+\frac{1}{2}}\rho^{(2)}_{\text{trivial}|\text{1D-DASPT}}&=X_{l+\frac{1}{2},k_0+\frac{1}{2}}Z_{l+1,k_0}Z_{L_x,k_0}\rho^{(2)}_{\text{trivial}|\text{1D-DASPT}}\,.
\end{align}
\end{subequations}
It is clear that if we try to preserve one of the symmetries by adding a projector onto the interface state, then the other symmetry is explicitly broken to nothing. Interface cut at $(l,k_0)$ and $(L_x+\frac{1}{2},k_0+\frac{1}{2})$ is similar to this case.
\subsection{Other choices of interface cuts between $\rho_{\text{1D-DASPT}}$ and $\rho_{\text{blue}}$}\label{app:Otherinterfacecuts2}
\subsubsection{Interface cut at $(l,k_0)$ and $(L_x,k_0)$}
Let us consider an interface cut passing through $(l,k_0)$ and $(L_x,k_0)$. The interface state is
\begin{align}
    \rho^{(1)}_{\text{1D-DASPT}|\text{blue}}=\prod_{i=0}^{l-1}\frac{(1+Z_{i,k_0}X_{i+\frac{1}{2},k_0+\frac{1}{2}}Z_{i+1,k_0})}{2}\prod_{i=l+1}^{L_x-2}\frac{(1-Z_{i-\frac{1}{2},k_0+\frac{1}{2}}Y_{i,k_0}X_{i+\frac{1}{2},k_0+\frac{1}{2}}Y_{i+1,k_0}Z_{i+\frac{3}{2},k_0+\frac{1}{2}})}{2}\frac{(1+\prod_iX_{i,k_0})}{2^{L_x}}
\end{align}
Now we apply the symmetries on this state
\begin{subequations}
\begin{align}
    \prod_{i}X_{i,k_0}\rho^{(1)}_{\text{1D-DASPT}|\text{blue}}&=\rho^{(1)}_{\text{1D-DASPT}|\text{blue}}\\
    \begin{split}\prod_iX_{i+\frac{1}{2},k_0+\frac{1}{2}}\rho^{(1)}_{\text{1D-DASPT}|\text{blue}}&=Z_{l,k_0}Y_{l+\frac{1}{2},k_0+\frac{1}{2}}Y_{l+1,k_0}Z_{l+\frac{3}{2},k_0+\frac{1}{2}}\times\\
    &\hspace{3cm}Z_{L_x-\frac{3}{2},k_0+\frac{1}{2}}Y_{L_x-1,k_0}Y_{L_x-\frac{1}{2},k_0+\frac{1}{2}}Z_{L_x,k_0}\rho^{(1)}_{\text{1D-DASPT}|\text{blue}}\,.
    \end{split}
\end{align}
\end{subequations}
To make the state $\rho^{(1)}_{\text{1D-DASPT}|\text{blue}}$ symmetric under the invertible symmetries, we need to add the projectors $Z_{l,k_0}Y_{l+\frac{1}{2},k_0+\frac{1}{2}}Y_{l+1,k_0}Z_{l+\frac{3}{2},k_0+\frac{1}{2}}=\pm 1$ and $Z_{L_x-\frac{3}{2},k_0+\frac{1}{2}}Y_{L_x-1,k_0}Y_{L_x-\frac{1}{2},k_0+\frac{1}{2}}Z_{L_x,k_0}=\pm 1$. The state with  $Z_{l,k_0}Y_{l+\frac{1}{2},k_0+\frac{1}{2}}Y_{l+1,k_0}Z_{l+\frac{3}{2},k_0+\frac{1}{2}}=- 1$ and $Z_{L_x-\frac{3}{2},k_0+\frac{1}{2}}Y_{L_x-1,k_0}Y_{L_x-\frac{1}{2},k_0+\frac{1}{2}}Z_{L_x,k_0}=- 1$ has a strong non-invertible symmetry $\mathrm{D}^{(1)}$. \red{We know that $\rho_{\text{1D-DASPT}}$ and $\rho_{\text{blue}}$ are in the same phase with both invertible and non-invertible symmetries.} 

\subsubsection{Interface cut at $(l+\frac{1}{2},k_0+\frac{1}{2})$ and $(L_x,k_0)$}
Now let us consider an interface cut passing through $(l+\frac{1}{2},k_0+\frac{1}{2})$ and $(L_x,k_0)$. The interface state is
\begin{align}
    \begin{split}\rho^{(2)}_{\text{1D-DASPT}|\text{blue}}&=\prod_{i=1}^{l-1}\frac{(1+Z_{i,k_0}X_{i+\frac{1}{2},k_0+\frac{1}{2}}Z_{i+1,k_0})}{2}\prod_{i=l+1}^{L_x-2}\frac{(1-Z_{i-\frac{1}{2},k_0+\frac{1}{2}}Y_{i,k_0}X_{i+\frac{1}{2},k_0+\frac{1}{2}}Y_{i+1,k_0}Z_{i+\frac{3}{2},k_0+\frac{1}{2}})}{2}\\
&\hspace{3cm}\times\frac{(1+\prod_{i}X_{i,k_0})}{2^{L_x}}
\end{split}
\end{align}
Now we apply the symmetries on this state
\begin{subequations}
\begin{align}
    \prod_iX_{i,k_0}\rho^{(2)}_{\text{1D-DASPT}|\text{blue}}&=\rho^{(2)}_{\text{1D-DASPT}|\text{blue}}\\
    \begin{split}\prod_iX_{i+\frac{1}{2},k_0+\frac{1}{2}}\rho^{(2)}_{\text{1D-DASPT}|\text{blue}}&=Z_{l,k_0}Y_{l+\frac{1}{2},k_0+\frac{1}{2}}Y_{l+1,k_0}Z_{l+\frac{3}{2},k_0+\frac{1}{2}}\times\\
    &\hspace{3cm}Z_{L_x-\frac{3}{2},k_0+\frac{1}{2}}Y_{L_x-1,k_0}Y_{L_x-\frac{1}{2},k_0+\frac{1}{2}}Z_{L_x,k_0}\rho^{(2)}_{\text{1D-DASPT}|\text{blue}}
    \end{split}
\end{align}
\end{subequations}
To make the state $\rho^{(2)}_{\text{1D-DASPT}|\text{blue}}$ symmetric under the invertible symmetries, we add the projectors $Z_{l,k_0}Y_{l+\frac{1}{2},k_0+\frac{1}{2}}Y_{l+1,k_0}Z_{l+\frac{3}{2},k_0+\frac{1}{2}}=\pm 1$, and  $Z_{L_x-\frac{3}{2},k_0+\frac{1}{2}}Y_{L_x-1,k_0}Y_{L_x-\frac{1}{2},k_0+\frac{1}{2}}Z_{L_x,k_0}=\pm 1$.  The state with $Z_{l,k_0}Y_{l+\frac{1}{2},k_0+\frac{1}{2}}Y_{l+1,k_0}Z_{l+\frac{3}{2},k_0+\frac{1}{2}}=-1$ and $Z_{L_x-\frac{3}{2},k_0+\frac{1}{2}}Y_{L_x-1,k_0}Y_{L_x-\frac{1}{2},k_0+\frac{1}{2}}Z_{L_x,k_0}=-1$ has a strong non-invertible symmetry $\mathrm{D}^{(1)}$. 
\subsection{Other choices of interface cuts between $\rho_{\text{1D-DASPT}}$ and $\rho_{\text{blue}^{k_0;k_1}}$}\label{app:Otherinterfacecuts3}
\subsubsection{Interface cut at $(l,k_0)$ and $(L_x,k_0)$}
Let us consider an interface cut passing through $(l,k_0)$ and $(L_x,k_0)$. The interface state is
\begin{align}
    \rho^{(1)}_{\text{1D-DASPT}|\text{blue}^{k_0;k_1}}&=\prod_{i=0}^{l-1}\frac{(1+Z_{i,k_0}X_{i+\frac{1}{2},k_0+\frac{1}{2}}Z_{i+1,k_0})}{2}\prod_{i=l+1}^{L_x-2}\frac{(1-Y_{i,k_0}X_{i+\frac{1}{2},k_0+\frac{1}{2}}Y_{i+1,k_0})}{2}\frac{(1+\prod_iX_{i,k_0})}{2^{L_x}}\,.
\end{align}
Now we apply the symmetries on this state
\begin{subequations}
\begin{align}
    \prod_{i}X_{i,k_0}\rho^{(1)}_{\text{1D-DASPT}|\text{blue}^{k_0;k_1}}&=\rho^{(1)}_{\text{1D-DASPT}|\text{blue}^{k_0;k_1}}\\
    \begin{split}\prod_iX_{i+\frac{1}{2},k_0+\frac{1}{2}}\rho^{(1)}_{\text{1D-DASPT}|\text{blue}^{k_0;k_1}}&=Z_{l,k_0}X_{l+\frac{1}{2},k_0+\frac{1}{2}}Y_{l+1,k_0}\times Y_{L_x-1,k_0}X_{L_x-\frac{1}{2},k_0+\frac{1}{2}}Z_{L_x,k_0}\rho^{(1)}_{\text{1D-DASPT}|\text{blue}^{k_0;k_1}}\,.
    \end{split}
\end{align}
\end{subequations}
\red{When we add the projectors  $Z_{l,k_0}X_{l+\frac{1}{2},k_0+\frac{1}{2}}Y_{l+1,k_0}=\pm 1$, and $Y_{L_x-1,k_0}X_{L_x-\frac{1}{2},k_0+\frac{1}{2}}Z_{L_x,k_0}=\pm 1$ the interface state is symmetric under strong $\mathbb{Z}_2^{(r)}\times\mathbb{Z}_2^{(b)}$ symmetry. We know that the two states $\rho_{\text{1D-DASPT}}$ and $\rho_{\text{blue}^{k_0;k_1}}$ are in the same phase w.r.t the strong $\mathbb{Z}_2^{(r)}\times\mathbb{Z}_2^{(b)}$ via a symmetric two-way connecting finite-depth local channel.} However, these two are distinguished by non-invertible symmetry $\mathrm{D}^{(1)}$ since $\rho_{\text{1D-DASPT}}$ has strong $\mathrm{D}^{(1)}$ while $\rho_{\text{blue}^{k_0;k_1}}$ is not symmetric under $\mathrm{D}^{(1)}$.
\subsubsection{Interface cut at $(l+\frac{1}{2},k_0+\frac{1}{2})$ and $(L_x,k_0)$}
Now let us consider an interface cut passing through $(l+\frac{1}{2},k_0+\frac{1}{2})$ and $(L_x,k_0)$. The interface state is
\begin{align}
    \begin{split}\rho^{(2)}_{\text{1D-DASPT}|\text{blue}^{k_0;k_1}}&=\prod_{i=1}^{l-1}\frac{(1+Z_{i,k_0}X_{i+\frac{1}{2},k_0+\frac{1}{2}}Z_{i+1,k_0})}{2}\prod_{i=l+1}^{L_x-2}\frac{(1-Y_{i,k_0}X_{i+\frac{1}{2},k_0+\frac{1}{2}}Y_{i+1,k_0})}{2}\frac{(1+\prod_{i}X_{i,k_0})}{2^{L_x}}\,.
\end{split}
\end{align}
Now we apply the symmetries on this state
\begin{subequations}
\begin{align}
    \prod_iX_{i,k_0}\rho^{(2)}_{\text{1D-DASPT}|\text{blue}^{k_0;k_1}}&=\rho^{(2)}_{\text{1D-DASPT}|\text{blue}^{k_0;k_1}}\\
    \begin{split}\prod_iX_{i+\frac{1}{2},k_0+\frac{1}{2}}\rho^{(2)}_{\text{1D-DASPT}|\text{blue}^{k_0;k_1}}&=Z_{l,k_0}X_{l+\frac{1}{2},k_0+\frac{1}{2}}Y_{l+1,k_0}\times Y_{L_x-1,k_0}X_{L_x-\frac{1}{2},k_0+\frac{1}{2}}Z_{L_x,k_0}\rho^{(2)}_{\text{1D-DASPT}|\text{blue}^{k_0;k_1}}\,.
    \end{split}
\end{align}
\end{subequations}
\red{When we add the projectors  $Z_{l,k_0}X_{l+\frac{1}{2},k_0+\frac{1}{2}}Y_{l+1,k_0}=\pm 1$, and $Y_{L_x-1,k_0}X_{L_x-\frac{1}{2},k_0+\frac{1}{2}}Z_{L_x,k_0}=\pm 1$ the interface state is symmetric under strong $\mathbb{Z}_2^{(r)}\times\mathbb{Z}_2^{(b)}$ symmetry. We know that the two states $\rho_{\text{1D-DASPT}}$ and $\rho_{\text{blue}^{k_0;k_1}}$ are in the same phase w.r.t the strong $\mathbb{Z}_2^{(r)}\times\mathbb{Z}_2^{(b)}$ via a symmetric two-way connecting finite-depth local channel.}
\section{Commutative diagram} \label{app:commutative}
Let us consider a pure 2D state $\rho$. Now consider the single-site Pauli decoherence channel
\begin{align}
    \xi_i[\rho]=\frac{1}{2}\left(\rho+\mathfrak{X}_i\rho\mathfrak{X}_i\right)\,,\qquad \xi_{\mathcal{A}}[\rho]=\prod_{i\in \mathcal{A}}\xi_i[\rho]\,,
\end{align}
where $\mathcal{A}$ is a subset of the sites and $\mathfrak{X}_i$ be any Pauli operator $X$, $Y$ or $Z$. Now let us denote the trace operation on a subset of sites $\mathcal{B}$ by $\text{Tr}_{\mathcal{B}}$. Then we have
\begin{align}
    \text{Tr}_{\mathcal{B}}(\xi_{\mathcal{A}}[\rho])=\xi_{\mathcal{A}\cap\mathcal{B}^c}[\text{Tr}_{\mathcal{B}}(\rho)]\,.
\end{align}
This can be proven in the following steps
\begin{align}
    \text{Tr}_{\mathcal{B}}(\xi_{\mathcal{A}}[\rho])
    =&\red{\frac{1}{2^{-|\mathcal{A}|}}}\text{Tr}_{\mathcal{B}}\left(\sum_{\substack{\{i_l\},\\i_l\in\mathcal{A}}}\mathfrak{X}_{i_1}...\mathfrak{X}_{i_r}\rho\mathfrak{X}_{i_1}...\mathfrak{X}_{i_r}\right)\nonumber\\
    =&\red{\frac{1}{2^{-|\mathcal{A}|}}}\sum_{\{s\}_{\mathcal{B}}}\bra{\{s\}_{\mathcal{B}}}\sum_{\substack{\{i_l\},\\i_l\in\mathcal{A}}}\mathfrak{X}_{i_1}...\mathfrak{X}_{i_r}\rho\mathfrak{X}_{i_1}...\mathfrak{X}_{i_r}\ket{\{s\}_{\mathcal{B}}}\nonumber\\
    =&\red{\frac{1}{2^{-|\mathcal{A}\cap\mathcal{B}^c|}}}\sum_{\substack{\{i_l\},\\i_l\in\mathcal{A}\cap\mathcal{B}^c}}\mathfrak{X}_{i_1}...\mathfrak{X}_{i_s}\sum_{\{s\}_{\mathcal{B}}}\bra{\{s\}_{\mathcal{B}}}\rho\ket{\{s\}_{\mathcal{B}}}\mathfrak{X}_{i_1}...\mathfrak{X}_{i_s}\nonumber\\
    =&\hspace{0.2cm}\xi_{\mathcal{A}\cap\mathcal{B}^c}[\text{Tr}_{\mathcal{B}}(\rho)]\,.
\end{align}
In the second equality, we sum over the states in the basis of $\mathfrak{X}_i$ operators at each site.

The above derivation then implies the following commutative diagram
\begin{equation} \begin{tikzcd}
\rho_{\text{2D-pure}} \arrow{r}{\text{Tr}_{\text{B}}} \arrow[swap]{d}{\xi[\rho_{\text{2D-pure}}]} & \rho_{\text{1D-mixed}} \arrow{d}{\xi[\rho_{\text{1D-mixed}}]} \\
\rho_{\text{2D-mixed}} \arrow{r}{\text{Tr}_{\text{B}}}& \rho'_{\text{1D-mixed}}
\end{tikzcd}\,.
\end{equation}
\section{Some examples of 1D mixed-state phases via decohering pure states (with $\mathbb{Z}_2$ degrees of freedom)}\label{app:1Dexamples}

Here, we give examples of 1D mixed-states by decohering the corresponding 1D pure state. The resulting mixed-states in various cases below are related to the examples described in the main text and previous Appendices, where one traces over the bulk of 2D pure states. This corresponds to the top layer in the commutative diagram below. On the left end, we have the 2D pure state that we trace over to produce the 1D mixed-state in the middle. The same mixed-state can be obtained by decohering the 1D pure state at the right end.

\begin{equation} \begin{tikzcd}[row sep=huge, column sep = huge]
\rho_{\text{2D-pure}} \arrow{r}{\text{Tr}_{\text{B}}} \arrow[swap]{d}{\xi[\rho_{\text{2D-pure}}]} & \rho_{\text{1D-mixed}} \arrow{d}{\xi[\rho_{\text{1D-mixed}}]} & \rho_{\text{1D-pure}}\arrow[swap]{l}{\xi'[\rho_{\text{1D-pure}}]}\arrow{ld}{\xi\circ\xi'[\rho_{\text{1D-pure}}]}\\
\rho_{\text{2D-mixed}} \arrow{r}{\text{Tr}_{\text{B}}}& \rho'_{\text{1D-mixed}}
\end{tikzcd}
\end{equation}
\subsection{Example: 1D SWSSB by decohering the GHZ state}

Consider the GHZ state with two symmetries. One is the 0-form $\mathbb{Z}_2$ symmetry, $\eta = \prod_{i} X_i$. The other one is a ``1-form" symmetry $Z_i Z_j$, $i \neq j$. For periodic boundary conditions, the density matrix is given by,
\begin{equation}
    \begin{aligned}
    \rho_{GHZ} = \prod_{i} \frac{1+Z_i Z_j}{2} \cdot \frac{1+\eta}{2^N},
    \end{aligned}
\end{equation}
where $N$ is the number of qubits of the system. Consider the local Kraus operators, $K_i = \{1, X_i\}$. The decohered density matrix is thus given by
\begin{equation}
    \begin{aligned}
        \rho^{(1)} = \frac{1+\eta}{2^N}. \label{eq: swssb}
    \end{aligned}
\end{equation}

One can check that this density matrix still has the strong $\eta$ symmetry. However, the ``1-form" $Z_i Z_j$ symmetry becomes weak.
\begin{equation}
    \begin{aligned}
        Z_i Z_j \rho^{(1)} Z_i Z_j = \rho^{(1)}.
    \end{aligned}
\end{equation}
Different extremal points are labeled by the charge of the strong symmetry. We have
\begin{equation}
    \begin{aligned}
        \rho^{(2)} = Z_i \rho^{(1)} Z_i = \frac{1 - \eta}{2^N}.
    \end{aligned}
\end{equation}
They can be distinguished by applying the strong symmetry operator,
\begin{equation}
    \begin{aligned}
        \eta \rho^{(1)} = \rho^{(1)}, \quad \eta \rho^{(2)} = - \rho^{(2)}.
    \end{aligned}
\end{equation}
The above two density matrices, $\rho^{(1)}$ and $\rho^{(2)}$, form a locally indistinguishable set, and further can be proved to be an information convex set. One can also check that the R\'enyi-2 correlator of this state is non-vanishing. Therefore, this state is in the SWSSB phase.

We note that the form of $\rho^{(1)}$ appears in \eqref{eq:rhoA2Dcluster}, where the mixed-state $\rho^{\rm A}_{\text{1D-DASPT}}$ was obtained by tracing over the bulk of the 2D cluster state. The act of tracing over the bulk there provides an effective decoherence channel on the boundary.

Additionally, both density matrices have weak Kramers-Wannier symmetry $\mathsf{D}$, as in Eq.~(\ref{eq:D1lattice}),
\begin{equation}
    \begin{aligned}
        \mathsf{D} \rho^{(1)} = \rho^{(1)} \mathsf{D}, \quad \mathsf{D} \rho^{(2)} = \rho^{(2)} \mathsf{D}.
    \end{aligned}
\end{equation}

If we instead decohere in $\{1, Z_i\}$
 basis, then the resultant state is a mixture of $|0\cdots 0\rangle$  and $|1\cdots 1\rangle$.
 $\eta$ becomes a weak symmetry, and $Z_iZ_j$ is still a strong symmetry despite the fact that its support is two, unlike that of the global symmetry $\eta$. However, now that $\eta$ becomes a weak symmetry, and thus $Z_i$ will no longer be regarded as a charge, as a weak symmetry does not have a conserved charge despite ${\rm Tr}(\rho Z_i Z_j)=1$, even for $|i-j|\rightarrow \infty$. One expects that a small perturbation will break the equal weight mixture, thereby breaking the weak symmetry $\eta$. 
 \subsection{Example: 1D ASPT from decohering the cluster state in the Z basis}

Consider the $\mathbb{Z}_2 \times \mathbb{Z}_2$ cluster state. In periodic boundary conditions \red{with $2N$ sites}, the density matrix can be given as follows,
\begin{equation}
    \begin{aligned}
        \rho_{\mathrm{cl}} = \prod_{i \in e} \frac{1 + Z_{i-1}^o X_i^e Z_{i+1}^o}{2} \prod_{j \in o} \frac{1 + Z_{j-1}^e X_j^o Z_{j+1}^e}{2}. 
    \end{aligned}
\end{equation}
Consider the local Kraus operators $K_{i \in e} = \{1, Z_i^e\}$. After decoherence, we get,
\begin{equation}
    \begin{aligned}
        \rho_{\mathrm{cl}}' = \red{\frac{1}{2^N}}\prod_{j \in o} \frac{1 + Z_{j-1}^e X_j^o Z_{j+1}^e}{2}.
        \label{eq:ASPT}
    \end{aligned}
\end{equation}
This density matrix has a weak 0-form symmetry $\eta^e = \prod_{i \in e} X_i^e$, and a strong 0-form symmetry $\eta^o = \prod_{i \in o} X_i^o$. One can calculate the expectation value of the strong string order parameter,
\begin{equation}
    \begin{aligned}
        \mathrm{Tr} \left(Z^e X^o X^o ... X^o Z^e \rho_{\mathrm{cl}}'\right) = 1.
    \end{aligned}
\end{equation}
It's straightforward to see that the expectation value of the weak string order parameter is 1. In addition, one can also observe the weak symmetry localization,
\begin{equation}
    \begin{aligned}
          X^e_i X^e_{i+1} ... X^e_{j} \rho_{\mathrm{cl}}' X^e_{i} X^e_{i+1} ... X^e_{j} = Z^o_{i-1} Z^o_{j+1}\rho_{\mathrm{cl}}'Z^o_{i-1} Z^o_{j+1}.
    \end{aligned}
\end{equation}

Therefore, this density matrix is in the averaged SPT (ASPT) phases~\cite{ma2023average,Ma2025Topological}, classified by $H^2(\mathbb{Z}_2^{w},H^2(\mathbb{Z}_2^{s} , U(1)))$. Equivalently, the state can be given by the domain wall decoration between the strong $\mathbb{Z}_2^{s}$ symmetry and the weak $\mathbb{Z}_2^w$ symmetry.

Additionally, the density matrix has a Kramers-Wannier symmetry $\mathrm{D}^{(1)}$, as in Eq.~(\ref{eq:D1exp}),
\begin{align}
    \mathrm{D}^{(1)} \rho_{\mathrm{cl}}' = \rho_{\mathrm{cl}}' \mathrm{D}^{(1)}.
\end{align}

Comparing with the pure state case, which uses coherent proliferation of domain walls $\sum_{i\in {\rm even}} Z_{i-1}X_i Z_{i+1}$ to obtain $Z_2\otimes Z_2$ SPT, the mixed-state case uses incoherent proliferation: using the channel $\sum_{i\in {\rm even}} Z_{i-1}X_i Z_{i+1}[\,]$, where $A[\rho]\equiv A\rho A^\dagger$
.

\subsection{Example: Coexistence of SWSSB and ASPT (i.e., DASPT) by decohering the cluster state in the X basis} \label{sec:1DDASPT}

Starting from the $\mathbb{Z}_2 \times \mathbb{Z}_2$ cluster state. Now we consider the Kraus operators to be $K_{i \in e} = \{1, X_i^e\}$. The decohered density matrix is given by
\begin{align}
    \rho^{(1)}_{\mathrm{cl}}= \prod_{i \in e} \frac{1 + Z_{i-1}^o X_i^e Z_{i+1}^o}{2} \cdot \frac{1 + \eta^o}{2^N}.
\end{align}
Equivalently, the above density matrix can be obtained from tracing over the bulk of a 2d cluster state, as we studied in Sec.~\ref{sec:DASPTfrom2D}. One can notice that the first part of the density matrix resembles the density matrix of an ASPT phase, as we have shown in Eq.~\eqref{eq:ASPT}, while the second part resembles an SWSSB phase, as we have shown in Eq.~\eqref{eq: swssb}. This coexistence of SWSSB and ASPT can also be understood as domain wall decoration between the all $|+\rangle$ state on the first layer and a SWSSB density matrix on the second.

The symmetries of this density matrix are given as follows: A strong 0-form symmetry $\eta^e$, a strong 0-form symmetry $\eta^o$, and a weak ``1-form" symmetry $Z_i^o Z_j^o$. One can calculate the following properties,
\begin{subequations}
    \begin{align}
       & Z^o X^e ... X^e Z^o \rho_{\mathrm{cl}}^{(1)} = \rho_{\mathrm{cl}}^{(1)},\\  &Z^e X^o ... X^o Z^e \rho_{\mathrm{cl}}^{(1)} Z^e X^o ... X^o Z^e = \rho_{\mathrm{cl}}^{(1)},\\
       & \frac{\mathrm{Tr}(Z_i^o Z_j^o \rho_{\mathrm{cl}}^{(1)}Z^o_i Z^o_j \rho_{\mathrm{cl}}^{(1)})}{\mathrm{Tr}( (\rho_{\mathrm{cl}}^{(1)})^2)} = 1.
    \end{align}
\end{subequations}

Different extremal points can be labeled by the charge of the $\eta^o$ symmetry, and we obtain a second extremal point from the above state:
\begin{equation}
    \begin{aligned}
        \rho^{(2)}_{\mathrm{cl}} = Z_i^o \rho^{(1)}_{\mathrm{cl}} Z_i^o = \rho^{(1)}_{\mathrm{cl}}= \prod_{i \in e} \frac{1 + Z_{i-1}^o X_i^e Z_{i+1}^o}{2} \cdot \frac{1 - \eta^o}{2^N}.
    \end{aligned}
\end{equation}

$\rho^{(1)}_{\mathrm{cl}}$ and $\rho^{(2)}_{\mathrm{cl}}$ are locally indistinguishable due to the factor $1\pm \eta^{o}$, and $\mathrm{Tr}(\rho^{(1)}_{\mathrm{cl}} \rho^{(2)}_{\mathrm{cl}}) = 0$.

Additionally, there is a KW symmetry,
\begin{align}
    \mathrm{D}^{(1)} \rho^{(1)}_{\mathrm{cl}} = \rho^{(1)}_{\mathrm{cl}} \mathrm{D}^{(1)},\quad \mathrm{D}^{(1)} \rho^{(2)}_{\mathrm{cl}} = \rho^{(2)}_{\mathrm{cl}} \mathrm{D}^{(1)}
\end{align}

Therefore, this density matrix has both features of the ASPT phase and the SWSSB phase, and it was recently studied numerically in Ref.~\cite{Guo2025strong,Guo2025strong}. We note that the form of $\rho^{(1)}_{\rm cl}$ is identical to \eqref{eq:rhoA2Dcluster}, where the mixed-state $\rho^{\rm A}_{\text{1D-DASPT}}$ was obtained by tracing over the bulk of the 2D cluster state. The act of tracing over the bulk there provides an effective decoherence channel on the boundary.

In the open boundary condition case, the $\eta^o$ symmetry is explicitly broken. Therefore, the SWSSB does not exist anymore. Instead, the ASPT phase can still survive, for which both strong and weak symmetry can be localized~\cite{Ma2025Topological}. Therefore, one can still observe an ASPT phase in open boundary conditions.

\subsection{Example: Another DASPT by decohering the cluster state with nearest neighbor $YY$}
Here, we still start with the 1D pure cluster state, whose density matrix is
\begin{subequations}
\begin{align}
    \rho&=\prod_{i\in o}\left(\frac{1+Z_{i-1}^{e}X_{i}^{o}Z_{i+1}^e}{2}\right)\prod_{i\in e}\left(\frac{1+Z_{i-1}^oX_{i}^eZ_{i+1}^o}{2}\right)\,,
\end{align}
which is identical to 
\begin{align}
    \rho&=\prod_{i\in e}\left(\frac{1-Z_{i-2}^eY_{i-1}^oX_{i}^eY_{i+1}^oZ_{i+2}^e}{2}\right)\prod_{i\in o}\left(\frac{1+Z_{i-1}^eX_{i}^oZ_{i+1}^e}{2}\right)\,.
\end{align}
\end{subequations}
 But we now apply a different  decoherence channel $\xi[\rho]=\prod_i\xi_i[\rho]$ where $\xi_i[\rho]=\frac{1}{2}\left(\rho+Y_{i}^oY_{i+2}^o\rho Y_{i}^oY_{i+2}^o\right)$. This yields 
 \begin{align}
     \rho_{\text{odd}}^{(1)}&=\prod_{i\in e}\left(\frac{1-Z_{i-2}^eY_{i-1}^oX_{i}^eY_{i+1}^oZ_{i+2}^e}{2}\right)\left(\frac{1+\prod_{i\in o}X_{i}^o}{2^N}\right)
     \label{eq:rhoodd^1}
 \end{align}
 which can be seen identical to $\rho^{\rm A}_{\text{blue}}$ in \eqref{eq:rhoAblue}.

We mention yet another example that starting with a pure-state non-invertible SPT in \eqref{eq:pureNSPT}, we applied a decoherence channel that consists of nearest-neighbor $YY$'s to obtain the state in~\eqref{eq:rhoAbluek0k1}. 
\section{1D examples of mixed-state phases and non-invertible SWSSB}\label{app:1Dnon-invertibleSWSSB}

For the $\mathbb{Z}_2 \times \mathbb{Z}_2$ cluster state, consider the following Kraus operators, $K_i = \{1, X_i\}$ on \emph{both} sublattices, as opposed to just one sublattice considered in Appendix~\ref{sec:1DDASPT}. The decohered density matrix is given by
\begin{equation}
    \begin{aligned}
        \rho^{(1)} = \frac{1+\eta^e}{2^N} \frac{1 + \eta^o}{2^N}.
    \end{aligned}
\end{equation}
The symmetries of this density matrix are given as follows: A strong 0-form symmetry $\eta^e$, a strong 0-form symmetry $\eta^o$, a weak ``1-form" symmetry $Z_i^e Z_j^e$, a weak ``1-form" symmetry $Z_i^o Z_j^o$, and a Kramers-Wannier symmetry $\mathrm{D}^{(1)}$.

This density matrix is in the SWSSB phase for $\mathbb{Z}_2 \times \mathbb{Z}_2$ symmetry. Consequently, different extremal points (from the corresponding information-complete convex set) arise from the SWSSB and are labeled by the charges of the strong symmetry and are interconvertible by these charge operators. We thus have four distinct extremal points
\begin{equation}
    \begin{aligned}
        &\rho_{++} = \frac{1+\eta^e}{2^N} \frac{1 + \eta^o}{2^N}, \quad \rho_{+-} = \frac{1+\eta^e}{2^N} \frac{1 - \eta^o}{2^N}, \\
        &\rho_{-+} = \frac{1-\eta^e}{2^N} \frac{1 + \eta^o}{2^N}, \quad \rho_{--} = \frac{1-\eta^e}{2^N} \frac{1 - \eta^o}{2^N}.
    \end{aligned}
\end{equation}
The above matrices are locally indistinguishable and orthogonal to each other. In this section, we rescale $\mathrm{D}^{(1)}\longrightarrow 2 \mathrm{D}^{(1)}$ so that we have the fusion rule given in~\eqref{eq:D1fusion} (cf.~\eqref{eq:D1lattice} in the main text). The first density matrix only has weak $\mathrm{D}^{(1)}$ symemtry, while the other three have strong $\mathrm{D}^{(1)}$ symmetry with eigenvalue 0 (see Table~\ref{tab:repD8} for the strong charges). 

As Rep($D_8$) has five different representations, we expect to have two more extremal points (density matrices) that are charged under $\mathrm{D}^{(1)}$. To construct another $\mathrm{Rep}(D_8)$ SWSSB density matrix, we need to symmetrize the first density matrix $\rho_{++}$ under the KW action. Using that $\mathrm{D}^{(1)}\rho_{++}\sim \mathrm{D}^{(1)}$ and $\big(\mathrm{D}^{(1)} \big)^2\sim\rho_{++}$, we thus obtain
\begin{equation}
    \begin{aligned}
        \rho_{\rm KW_1} = \frac{1+\eta^e}{2^N} \frac{1 + \eta^o}{2^N} + \frac{2\mathrm{D}^{(1)}}{4^N}.
    \end{aligned}
\end{equation}
One can then verify that the resultant $\rho_{{\rm KW}_1}$ has charge 2 under $\mathrm{D}^{(1)}$
\begin{align}
    \mathrm{D}^{(1)} \rho_{\rm KW_1} = 2 \rho_{\rm KW_1}.
\end{align}
Here we have used the following facts:
\begin{equation}
    \begin{aligned}
        \mathrm{D}^{(1)} \eta_e = \eta_e \mathrm{D}^{(1)} = \mathrm{D}^{(1)},\ \mathrm{D}^{(1)} \eta_o = \eta_o \mathrm{D}^{(1)} = \mathrm{D}^{(1)},\ \big(\mathrm{D}^{(1)}\big)^2 = 1 + \eta_e + \eta_o + \eta_e \eta_o.
    \end{aligned}
    \label{eq:D1fusion}
\end{equation}

The other density matrix that carries nontrivial $\mathrm{D}^{(1)}$ charge is given by (using `-' instead of `+' in the construction of $\rho_{\mathrm{KW}_1}$)
\begin{align}
    \rho_{\mathrm{KW}_2} = \frac{1+\eta^e}{2^N} \frac{1 + \eta^o}{2^N} - \frac{2\mathrm{D}^{(1)}}{4^N}.
\end{align}
One can show that
\begin{align}
    \mathrm{D}^{(1)} \rho_{\mathrm{KW}_2} = -2 \rho_{\mathrm{KW}_2},
\end{align}
and that $\text{Tr}(\rho_{{\rm KW}_1}\rho_{{\rm KW}_2})=0$.
Thus, we have found 5 candidate density matrices supporting  $\mathrm{Rep}(D_8)$ SWSSB.

A similar study for the $\mathrm{Rep}(G)$ SWSSB on the qudit system can be found in Ref.~\cite{sun2025anomalousmatrixproductoperator,Schafer-Nameki:2025fiy}. 
$\mathrm{Rep}(D_8)$ SWSSB are labeled by the charge of the $\mathrm{Rep}(D_8)$ symmetry, which are classified by the following homomorphism,
\begin{align}
    [\omega] \in \mathrm{Hom} \left(\mathrm{Rep}(D_8), \mathbb{C}\right),
\end{align}
where $[\omega]$ represents different equivalent classes~\cite{ostrik2003module,ostrik2006module}. We show these classes in Table.~\ref{tab:repD8}. 

\begin{table}
\begin{center}
\begin{tabular}{||c | c c c||} 
 \hline
     & $\eta^e$ & $\eta^o$ & $\mathrm{D}^{(1)}$ \\ [0.5ex] 
 \hline\hline
 $\rho_{\rm KW_1}$ & 1 & 1 & 2 \\ 
 \hline
 $\rho_{+-}$ & 1 & -1 & 0 \\
 \hline
 $\rho_{-+}$ & -1 & 1 & 0 \\
 \hline
 $\rho_{--}$ & -1 & -1 & 0 \\
 \hline
 $\rho_{\mathrm{KW}_2}$ & 1 & 1 & -2 \\
 \hline
\end{tabular} 
\end{center}
\caption{This table illustrate the  $\mathrm{Rep}(D_8)$ symmetry charges of various density matrices.}
\label{tab:repD8}
\end{table}

For the two $\rho_{{\rm KW}}$'s we just constructed, we need to properly normalize them so that $\text{Tr}(\rho_{{\rm KW}})=1$.
This requires us to calculate $\text{Tr}(\rm \mathrm{D}^{(1)})$, which is the same as $\text{Tr}(\mathsf{V}\mathrm{D}^{(1)}\mathsf{V})$, where $\mathsf{V}\mathrm{D}^{(1)}\mathsf{V}$ is the Kennedy-Tasaki transformation that can be written in terms of an MPO~\cite{Meng:2024nxx} and $\mathsf{V}=\prod_{i}CZ_{i,i+1}$.

Now, consider the $2N$ spins with two spins per unit cell. Let us consider this effective site that contains two spins. The MPO tensor at the $j^{th}$ site is

    \begin{align}
   \mathcal{U}^j=\begin{pmatrix}
       \ket{++}_j\bra{++}+\ket{+-}_j\bra{+-} &\quad \ket{-+}_j\bra{-+}-\ket{--}_j\bra{--}\\
       \ket{-+}_j\bra{-+}+\ket{--}_j\bra{--} &\quad \ket{++}_j\bra{++}-\ket{+-}_j\bra{+-}
   \end{pmatrix} \,.
\end{align}
The matrix is in the virtual space, while the matrix element, which is an operator, acts on the physical space. Then, we have
\begin{align}
   \mathsf{V}\mathrm{D}^{(1)}\mathsf{V}=\text{Tr}_{\mathcal{H}_{\text{virt}}}\left(\mathcal{U}^{N}\mathcal{U}^{N-1}...\mathcal{U}^{1}\right)\,, 
\end{align}
where the trace is over the virtual space. We can explicitly evaluate the trace over the physical space of the operator $\mathsf{V}\mathrm{D}^{(1)}\mathsf{V}$, 
\begin{align}
    \text{Tr}_{\mathcal{H}_{\text{phy}}}\left(\mathsf{V}\mathrm{D}^{(1)}\mathsf{V}\right)&=\text{Tr}_{\mathcal{H}_{\text{phy}}}\left(\text{Tr}_{\mathcal{H}_{\text{virt}}}\left(\mathcal{U}^{N}\mathcal{U}^{N-1}...\mathcal{U}^{1}\right)\right)\nonumber\\
    &=\text{Tr}_{\mathcal{H}_{\text{virt}}}\left(\text{Tr}_{\mathcal{H}_{\text{phy}}}\left(\mathcal{U}^{N}\mathcal{U}^{N-1}...\mathcal{U}^{1}\right)\right)\nonumber\\
    &=\text{Tr}_{\mathcal{H}_{\text{virt}}}\left(\text{Tr}(\mathcal{U}^{N})\text{Tr}(\mathcal{U}^{N-1})...\text{Tr}(\mathcal{U}^{1})\right)\nonumber\\
    &=\text{Tr}_{\mathcal{H}_{\text{virt}}}\left(\begin{pmatrix}
        2 & 0\\
        2 & 0
    \end{pmatrix}^{N}\right)\nonumber\\
    &=\text{Tr}_{\mathcal{H}_{\text{virt}}}\left(\begin{pmatrix}
        2^N & 0\\
        2^N & 0
    \end{pmatrix}\right)\nonumber\\
    &=2^{N}\,.
\end{align}
Since $\mathsf{V}^2=1$, $\text{Tr}_{\mathcal{H}_{\text{phy}}}\left(\mathsf{V}\mathrm{D}^{(1)}\mathsf{V}\right)=\text{Tr}_{\mathcal{H}_{\text{phy}}}\left(\mathrm{D}^{(1)}\right)=2^N$.

Therefore, we have,
\begin{equation}
    \begin{aligned}
        \lim_{N \to \infty} \mathrm{Tr} (\rho_{\rm KW_1}) =  \lim_{N \to \infty} \mathrm{Tr} (\rho_{\mathrm{KW}_2}) = 1.
    \end{aligned}
\end{equation}
In the thermodynamic limit $N \to \infty$, the above five density matrices are locally indistinguishable, and they are orthogonal to each other. Therefore, they define the five extremal points of the information convex set for the $\mathrm{Rep}(D_8)$ SWSSB. 

Now we evaluate the following R\'enyi-2 correlators to show that these states are indeed SWSSB. Let us consider $Z_i^e$ and $Z_{i}^o$ to be the charged operators under the $\mathbb{Z}_2\times\mathbb{Z}_2$ symmetry and $(Z_{i-1}^eZ_{i+1}^e-X_i^o)$ to be the charged operator under the $\mathrm{D}^{(1)}$ symmetry. Now we evaluate the correlators for $\rho^{+-}$
\begin{align}
    &\frac{\text{Tr}\left(\rho^{+-}Z_i^eZ_j^e\rho^{+-}Z_i^eZ_j^e\right)}{\text{Tr}\left((\rho^{+-})^2\right)}=1\,,\qquad \frac{\text{Tr}\left(\rho^{+-}Z_i^oZ_j^o\rho^{+-}Z_i^oZ_j^o\right)}{\text{Tr}\left((\rho^{+-})^2\right)}=1\,,\nonumber\\
    &\frac{\text{Tr}\left(\rho^{+-}\left(Z_{i-1}^eZ_{i+1}^e-X_i^o\right)\left(Z_{j-1}^eZ_{j+1}^e-X_j^o\right)\rho^{+-}\left(Z_{i-1}^eZ_{i+1}^e-X_i^o\right)\left(Z_{j-1}^eZ_{j+1}^e-X_j^o\right)\right)}{\text{Tr}\left((\rho^{+-})^2\right)}=4\,.
\end{align}
R\'enyi-2 correlators for $\rho^{-+}$ and $\rho^{--}$ are similar. For $\rho_{\rm KW_1}$
\begin{align}
    &\frac{\text{Tr}\left(\rho_{\rm KW_1}Z_i^eZ_j^e\rho_{\rm KW_1}Z_i^eZ_j^e\right)}{\text{Tr}\left((\rho_{\rm KW_1})^2\right)}=\frac{1}{2}\frac{1+\text{Tr}\left[\frac{16\rm D^{(1)}}{4^N}\right]}{1+\text{Tr}\left[\frac{8\rm D^{(1)}}{4^N}\right]}\xrightarrow{N\rightarrow\infty}\frac{1}{2}\,,\nonumber\\
    &\frac{\text{Tr}\left(\rho_{\rm KW_1}(Z_{i-1}^eZ_{i+1}^e-X_i^o)(Z_{j-1}^eZ_{j+1}^e-X_j^o)\rho_{\rm KW_1}(Z_{i-1}^eZ_{i+1}^e-X_i^o)(Z_{j-1}^eZ_{j+1}^e-X_j^o)\right)}{\text{Tr}\left((\rho_{\rm KW_1})^2\right)}\nonumber\\
    &\hspace{3cm}=4\frac{1+\text{Tr}\left[\frac{8\rm D^{(1)}}{4^N}(1-Z_{i-1}^eX_i^oZ_{i+1}^e)(1-Z_{j-1}^eX_j^oZ_{j+1}^e)\right]}{1+\text{Tr}\left[\frac{8\rm D^{(1)}}{4^N}\right]}\xrightarrow{N\rightarrow\infty}4\,.
\end{align}
In the last line, we used 
\begin{align}
    \text{Tr}\left[\mathrm{D}^{(1)}Z_{i-1}^eX_i^oZ_{i+1}^e\right]&=\text{Tr}\left[\mathrm{V}\mathrm{D}^{(1)}Z_{i-1}^eX_i^oZ_{i+1}^e\mathrm{V}\right]\nonumber\\
    &=\text{Tr}\left[\mathrm{V}\mathrm{D}^{(1)}\mathrm{V}X_i^o\right]\nonumber\\
    &=\text{Tr}_{\mathcal{H}_{\text{phy}}}\left[\mathrm{V}\mathrm{D}^{(1)}\mathrm{V}X_i^o\right]\nonumber\\
    &=\text{Tr}_{\mathcal{H}_{\text{phy}}}\left[\text{Tr}_{\mathcal{H}_{\text{virt}}}\left(\mathcal{U}^{N}\mathcal{U}^{N-1}...\mathcal{U}^{1}\right)X_i^o\right]\nonumber\\
    &=\text{Tr}_{\mathcal{H}_{\text{virt}}}\left(\text{Tr}_{\mathcal{H}_{\text{phy}}}\left(\mathcal{U}^{N}\mathcal{U}^{N-1}...\mathcal{U}^{1}X_i^o\right)\right)\nonumber\\
    &=\text{Tr}_{\mathcal{H}_{\text{virt}}}\left(\text{Tr}(\mathcal{U}^{N})...\text{Tr}(U^iX_i^o)...\text{Tr}(\mathcal{U}^{1})\right)\nonumber\\
    &=\text{Tr}_{\mathcal{H}_{\text{virt}}}\left(\begin{pmatrix}
        2 & 0\\
        2 & 0
    \end{pmatrix}^{N-i}\begin{pmatrix}
        2 & 0\\
        -2 & 0
    \end{pmatrix}\begin{pmatrix}
        2 & 0\\
        2 & 0
    \end{pmatrix}^{i-1}\right)\nonumber\\
    &=\text{Tr}_{\mathcal{H}_{\text{virt}}}\left(\begin{pmatrix}
        2^N & 0\\
        2^N & 0
    \end{pmatrix}\right)\nonumber\\
    &=2^{N}\,,
\end{align}
and similarly $\text{Tr}\left[\mathrm{D}^{(1)}Z_{j-1}^oX_j^eZ_{j+1}^o\right]=2^N$.
We get similar results for $\rho_{\rm KW_2}$, indicating that the four states satisfy the R\'enyi-2 correlator criterion for SWSSB.

The above MPO analysis implies that the properly normalized duality operator $\mathrm{D}^{(1)}$ has zero trace in the thermodynamic limit. In general, the computation of this trace requires careful regularization. In cases where we have a TQFT with degenerate vacua, for example, at the end of a certain RG flow preserving categorical symmetry, the trace of a topological operator vanishes if the corresponding defect Hilbert space is empty~\cite{Chang:2018iay}. In 2D rational CFTs such as the Ising CFT and the tricritical Ising CFT, the trace of the duality operator vanishes when restricted to the primary states of the theory quantized on a circle \cite{DiFrancesco:1997nk}.

\section{ 1D $\mathbb{Z}_3^{(s)}\times\mathbb{Z}_3^{(w)}$ ASPT: two ASPTs and the interface between them}\label{app:interfaceZ3SPT}
\subsection{ASPTs from decohering pure state SPTs}
In this section, we consider the $\mathbb{Z}_3\times \mathbb{Z}_3$ cluster state and consider various decoherence channels. According to the classification of bosonic SPTs, in 1D $\mathbb{Z}_3\times\mathbb{Z}_3$ symmetric SPTs are classified by $H^2(\mathbb{Z}_3\times\mathbb{Z}_3,U(1))=\mathbb{Z}_3$. So there are two non-trivial SPTs. We consider one of them in periodic boundary conditions \red{with $2N$ sites}, concretely

\begin{align}
    \rho_{\text{SPT}_1}&=\prod_{i\in e}\left(\frac{1+Z_{i-1}^{o\dagger}X_i^eZ_{i+1}^{o}+Z_{i-1}^{o}X_i^{e\dagger}Z_{i+1}^{o\dagger}}{3}\right)\times\prod_{j\in o}\left(\frac{1+Z_{j-1}^{e}X_j^oZ_{j+1}^{e\dagger}+Z_{j-1}^{e\dagger}X_j^{o\dagger}Z_{j+1}^{e}}{3}\right)\,.
\end{align}
This state can be equivalently written as
\begin{align*}
    &\prod_{i\in o}CZ_{i,i-1}CZ^{\dagger}_{i,i+1}\ket{+}^{\otimes\Delta_e}\ket{+}^{\otimes\Delta_o}\,,\quad\text{with } \ket{+}=\frac{\left(\ket{0}+\ket{1}+\ket{2}\right)}{\sqrt{3}}\,,\quad \text{and }CZ_{i,i+1}=\sum_{a,b=0}^2\omega^{ab}\ket{a,b}\bra{a,b}\,.
\end{align*}
Now consider the local Kraus operators $K_{i\in e}=\{1, Z_i^e,Z_i^{e\dagger}\}$. After decoherence, we get the mixed-state
\begin{align}
    \rho_{\text{m-SPT}_1}=\red{\frac{1}{3^{N}}}\prod_{j\in o}\left(\frac{1+Z_{j-1}^{e}X_j^oZ_{j+1}^{e\dagger}+Z_{j-1}^{e\dagger}X_j^{o\dagger}Z_{j+1}^{e}}{3}\right)\,.
\end{align}
This density matrix has strong 0-form symmetry $\eta^o=\prod_{j\in o}X_{j}^o$ and weak 0-form symmetry $\eta^e=\prod_{j\in e}X_j^e$. One can calculate the expectation value of the strong string order parameter,
\begin{equation}
    \begin{aligned}
        \mathrm{Tr} \left(Z^e X^o X^o ... X^o Z^{e\dagger} \rho_{\text{m-SPT}_1}\right) = 1.
    \end{aligned}
\end{equation}
One can also calculate the expectation value of the weak string order parameter on the other sublattice,
\begin{equation}
    \begin{aligned}
         \mathrm{Tr} \left(Z^{o\dagger} X^e X^e ... X^e Z^o \rho_{\text{m-SPT}_1} Z^o X^{e\dagger} X^{e\dagger} ... X^{e\dagger} Z^{o\dagger}\right) = 1.
    \end{aligned}
\end{equation}
Now let us consider the other non-trivial SPT
\begin{align}
    \rho_{\text{SPT}_2}&=\prod_{i\in e}\left(\frac{1+Z_{i-1}^{o}X_i^eZ_{i+1}^{o\dagger}+Z_{i-1}^{o\dagger}X_i^{e\dagger}Z_{i+1}^{o}}{3}\right)\times\prod_{j\in o}\left(\frac{1+Z_{j-1}^{e\dagger}X_j^oZ_{j+1}^{e}+Z_{j-1}^{e}X_j^{o\dagger}Z_{j+1}^{e\dagger}}{3}\right)\,.
\end{align}
This state can be equivalently written as
\begin{align*}
    &\prod_{i\in o}CZ^{\dagger}_{i,i-1}CZ_{i,i+1}\ket{+}^{\otimes\Delta_e}\ket{+}^{\otimes\Delta_o}\,,\quad\text{with }\ket{+}=\frac{\left(\ket{0}+\ket{1}+\ket{2}\right)}{\sqrt{3}}\,,\quad \text{and }CZ_{i,i+1}=\sum_{a,b=0}^2\omega^{ab}\ket{a,b}\bra{a,b}\,.
\end{align*}
Now consider the local Kraus operators $K_{i\in e}=\{1, Z_i^e,Z_i^{e\dagger}\}$. After decoherence, we get the mixed-state
\begin{align}
    \rho_{\text{m-SPT}_2}=\red{\frac{1}{3^N}}\prod_{j\in o}\left(\frac{1+Z_{j-1}^{e\dagger}X_j^oZ_{j+1}^{e}+Z_{j-1}^{e}X_j^{o\dagger}Z_{j+1}^{e\dagger}}{3}\right)\,.
\end{align}
This density matrix has strong 0-form symmetry $\eta^o=\prod_{j\in o}X_{j}^o$ and weak 0-form symmetry $\eta^e=\prod_{j\in e}X_j^e$. One can calculate the expectation value of the strong string order parameter,
\begin{equation}
    \begin{aligned}
        \mathrm{Tr} \left(Z^{e\dagger} X^o X^o ... X^o Z^{e} \rho_{\text{m-SPT}_2}\right) = 1.
    \end{aligned}
\end{equation}
One can also calculate the expectation value of the weak string order parameter on the other sublattice,
\begin{equation}
    \begin{aligned}
         \mathrm{Tr} \left(Z^{o} X^e X^e ... X^e Z^{o\dagger} \rho_{\text{m-SPT}_2} Z^{o\dagger} X^{e\dagger} X^{e\dagger} ... X^{e\dagger} Z^{o}\right) = 1.
    \end{aligned}
\end{equation}
\subsection{Interface between the two $\mathbb{Z}_3^{(s)}\times\mathbb{Z}_3^{(w)}$ ASPTs}
According to the classification of mSPTs~\cite{ma2023average}, $H^1(\mathbb{Z}_3,H^1(\mathbb{Z}_3,U(1)))$, there are three mSPTs with $\mathbb{Z}_3$ strong and $\mathbb{Z}_3$ weak symmetry. One of the SPTs is the trivial SPT. To write down the other two non-trivial SPTs, we consider $N$ sites on a ring. We label them odd $o$ and even $e$. Let $X$ and $Z$ be $\mathbb{Z}_3$ shift and clock operators. The two non-trivial SPTs are 
\begin{subequations}
\begin{align}
    \rho_{\text{m-SPT}_1}=\prod_{j\in o}\left(\frac{1+Z_{j-1}^{e}X_j^oZ_{j+1}^{e\dagger}+Z_{j-1}^{e\dagger}X_j^{o\dagger}Z_{j+1}^{e}}{3}\right)\,\\
    \rho_{\text{m-SPT}_2}=\prod_{j\in o}\left(\frac{1+Z_{j-1}^{e\dagger}X_j^oZ_{j+1}^{e}+Z_{j-1}^{e}X_j^{o\dagger}Z_{j+1}^{e\dagger}}{3}\right)\,.
\end{align}
\end{subequations}
Now let us consider an interface between the two mSPTs: $\rho_{\text{m-SPT}_1}$ and $\rho_{\text{m-SPT}_2}$, where the two interfaces are put at $l$ and $2N-1$. Then the mixed-state with this interface is

\begin{align}
    \rho_{\text{m-SPT}_1|\text{m-SPT}_2}=\prod_{\substack{1\leq j<l\\
    j\in o}}\left(\frac{1+Z_{j-1}^{e}X_j^oZ_{j+1}^{e\dagger}+Z_{j-1}^{e\dagger}X_j^{o\dagger}Z_{j+1}^{e}}{3}\right)\prod_{\substack{l< j\leq 2N-1\\
    j\in o}}\left(\frac{1+Z_{j-1}^{e\dagger}X_j^oZ_{j+1}^{e}+Z_{j-1}^{e}X_j^{o\dagger}Z_{j+1}^{e\dagger}}{3}\right)\,.
\end{align}
On this mixed-state, we act with the strong symmetry and its action localizes at the two interfaces
\begin{align}
    \prod_{j\in o}X_j^o\,\rho_{\text{m-SPT}_1|\text{m-SPT}_2}=Z_{l-1}^eX_l^oZ_{l+1}^e Z_{2N-2}^{e\dagger}X_{2N-1}^oZ_{2N}^{e\dagger}\,\rho_{\text{m-SPT}_1|\text{m-SPT}_2}\equiv\mathbb{X}_l\,\mathbb{X}_{2N-1}\,\rho_{\text{m-SPT}_1|\text{m-SPT}_2}\,.
\end{align}

In particular, $\mathbb{X}_l$ and $\mathbb{X}_{2N-1}$ are charged under the weak symmetry $\prod_{j\in e}X_j^e$. This implies that, if we add the projection onto \red{$\frac{\left(1+\omega^q\,\mathbb{X}_l+\omega^{2q}\, \mathbb{X}_l^2\right)}{3}$ and $\frac{\left(1+\omega^{2q}\,\mathbb{X}_{2N-1}+\omega^{q}\, \mathbb{X}_{2N-1}^2\right)}{3}$ for $q=1,2$} to the interface state $\rho_{\text{m-SPT}_1|\text{m-SPT}_2}$ to preserve the strong symmetry $\prod_{j\in o}X_j^o$, we explicitly break the weak symmetry to nothing.
\end{document}